\documentclass[11pt]{article}

\newcommand{\be}{\begin{equation}}
\newcommand{\ee}{\end{equation}}
\usepackage{amsmath}
\usepackage{amssymb}
\usepackage{latexsym}
\usepackage{epsfig}
\usepackage{cite}

\usepackage[vcentermath]{youngtab}

\usepackage[makeroom]{cancel}

\usepackage{mathabx}

\def\hybrid{
        \topmargin -20pt
        \oddsidemargin 0pt
        \headheight 0pt \headsep 0pt
        \textwidth 6.25in % A4 paper
        \textheight 9.5in % A4 paper
        \marginparwidth .875in
        \parskip 5pt plus 1pt \jot = 1.5ex}

\hybrid

\newcommand{\RRR}{{\hbox{\rm R\kern-2.35mm R}}}

\def\ZZZ{{\hbox{ Z\kern-1.6mm Z}}}

\newcommand{\nin}[1] {\underline{\phantom{h}}\hskip-6pt {#1}}

      \def\D{{\cal D}}
      \def\H{{\cal H}}

\def\Ä{\varphi}  \def\¿{\varpi}	\def\Ï{\vartheta}

\def\Ç{\textstyle{Ç}}

\begin{document}

\begin{titlepage}
\rightline{December 2016} 
%\rightline{\today} 
\begin{center}
\vskip 2.5cm

{\Large \bf {Background Independent Double Field Theory at Order $\alpha'$\,:  \\[1.5ex]
Metric vs.~Frame-like Geometry }}

 %\vskip 0.5cm

  \vskip 2.0cm
 {\large {Olaf Hohm }}
 \vskip 1cm

{\em  \hskip -.1truecm Simons Center for Geometry and Physics, \\
Stony Brook University, \\
Stony Brook, NY 11794-3636, USA \vskip 5pt }

ohohm@scgp.stonybrook.edu\\

\vskip 2.5cm
{\bf Abstract}

\end{center}

\vskip 0.5cm

\noindent
\begin{narrower}

\baselineskip10pt

{\small

The higher derivative corrections in double field theory are revisited to first order in $\alpha'$. 
In first order perturbation theory around flat space,  
the gauge algebra is $\alpha'$ corrected, governed 
by two parameters $a, b$. One parameter choice corresponds to bosonic string theory, 
another to heterotic string theory. These results are generalized 
to second order in perturbation theory by a Noether procedure. 
Using consistency conditions derived from the Jacobi identity, 
it is shown that for general $a, b$ there 
is no generalized metric formulation. 
A manifestly background independent formulation 
is instead given by a frame formalism, 
with $\alpha'$-deformed frame transformations, as proposed by Marques and Nunez. 
Their construction is slightly generalized to an $\alpha'$-deformed 
$GL(D)\times GL(D)$ frame formalism. The perturbation theory around 
flat space is matched to second order with the results obtained by the Noether procedure. 
We also obtain 
a formulation based on the `non-symmetric' metric ${\cal E}$, the sum of metric and $b$-field. 
The transformations of ${\cal E}$ under $O(D,D)$ receive non-trivial 
$\alpha'$ corrections, unless $a=-b$, in which case there is a generalized metric formulation, 
in agreement with previous results in the literature.

}

\end{narrower}

\end{titlepage}

\tableofcontents

\section{Introduction} 

One of the most intriguing features of string theory is that already at the classical level 
it modifies general relativity by higher-derivative corrections and new symmetries or dualities. 
The classical target space equations, which are not known explicitly,
 are a generalization of the Einstein equations 
by an infinite number of higher-derivative corrections governed by the dimensionful 
parameter $\alpha'$. In addition, string theory features novel  duality symmetries, 
notably under the `T-duality' group $O(d,d;\mathbb{R})$ \cite{Meissner:1991zj,Maharana:1992my}, 
which from the point of view 
of standard geometry is a hidden symmetry that only emerges upon compactification on $d$-tori. 

Despite being central characteristics of string theory, until recently only very little was known about 
the interplay between the $\alpha'$ corrections and the $O(d,d;\mathbb{R})$ symmetry, 
which must be realized to all orders in $\alpha'$ \cite{Sen:1991zi,9109038,Hohm:2014sxa}. 
(See \cite{Meissner:1996sa,Kaloper:1997ux} for notable exceptions.) 
Double field theory, which is the framework that makes $O(d,d;\mathbb{R})$
manifest \textit{before} compactification by working on a suitably `doubled' 
spacetime \cite{Siegel:1993th,Hull:2009mi,Hohm:2010jy,Hohm:2010pp,Hohm:2010xe}, 
has been used recently to describe  $\alpha'$ corrections \cite{Hohm:2013jaa,Hohm:2014eba,Hohm:2014xsa,Marques:2015vua,Hohm:2015mka,Siegel:2015axg,Naseer:2016izx,Huang:2016bdd,Casali:2016atr,Hohm:2016lim,Lescano:2016grn}. 
This requires a deformation of certain structures in double field theory (DFT).  
Most importantly, the gauge symmetries are $\alpha'$-deformed, which 
in turn implies that the higher-derivative corrections are determined 
(at least partially) by a gauge symmetry principle.

An exactly duality and gauge invariant $\alpha'$-deformation 
was found in \cite{Hohm:2013jaa}. In terms of conventional geometric objects, this theory 
(referred to as HSZ theory) almost certainly encodes 
an infinite number of $\alpha'$ corrections. However, this deformation does not encode 
the most general higher-derivative corrections arising in bosonic or heterotic string theory, which were 
determined  in 
\cite{Hohm:2014eba,Hohm:2014xsa} perturbatively about flat space, using closed string field theory (CSFT) 
to cubic order in fields \cite{Kugo:1992md,KugoZwiebach}. 
More recently, in important work by Marques and Nunez \cite{Marques:2015vua}, 
the first order $\alpha'$ corrections were described by a deformation of the frame or vielbein formalism of DFT. 
Part of the motivation for the present paper was to clarify these results and to understand how they are
related to `metric-like' field variables. 
Reporting on results that were recently announced in \cite{Hohm:2016lge}, I will show that in general there are  
obstructions for a background independent 
formulation based on the so-called generalized metric. 
Instead, a formulation that is both manifestly invariant under all symmetries 
and background independent is given by a frame formalism, which  introduces   
pure gauge modes under a (generalized) local Lorentz symmetry. 
The need to work with a vielbein formalism is well-known from the problem of 
coupling fermions to gravity, but  to my knowledge the result discussed in this paper
is the first example of a purely bosonic theory that seems to \textit{require} a frame formulation. 
In the remainder of the introduction, I will sketch proofs for this conclusion and 
corroborating consistency checks.

Let us begin by reviewing some central structures of DFT in order to discuss where 
and how they need to be $\alpha'$-deformed. The fields depend on doubled coordinates $X^M=(\tilde{x}_i,x^i)$, 
where $M,N=1,\ldots, 2D$ are $O(D,D)$ indices, and $D$ denotes  the number of
spacetime dimensions. The associated doubled derivatives $\partial_M=(\tilde{\partial}^i,\partial_i)$
are subject to the `section constraint' or `strong constraint' 
 \be\label{SECTION}
  \eta^{MN}\partial_M\partial_NA \ \equiv \ \partial^M\partial_MA \ = \ 0\;, \qquad 
  \partial^MA\,\partial_MB \ = \ 0\;, \qquad 
  \eta_{MN} \ = \  \begin{pmatrix}   0 & {\bf 1}\\[0.5ex]
  {\bf 1} & 0 \end{pmatrix}\;, 
 \ee
for any fields $A, B$, where $\eta_{MN}$ denotes the $O(D,D)$ invariant metric. 
The most general solution to this constraint is given by $\tilde{\partial}^iA=0$  for all fields $A$, 
and all its $O(D,D)$ rotations. The metric tensor $g_{ij}$ and the Kalb-Ramond two-form 
$b_{ij}$ transform, to \textit{zeroth order} in $\alpha'$, according to a non-linear realization of $O(D,D)$. 
Combining $g$ and $b$ into the `non-symmetric' metric 
 \be
  {\cal E}_{ij} \ \equiv \ g_{ij} + b_{ij}\;, 
 \ee
the non-linear $O(D,D)$ action is given by 
 \be\label{calEtransf}
  {\cal E}'(X') \ = \ (a\,{\cal E}(X)+b)(c\,{\cal E}(X)+d)^{-1}\;, \qquad
  h \ \equiv \ \begin{pmatrix}   a & b\\[0.5ex]
  c & d \end{pmatrix} \ \in \ O(D,D)\;, 
 \ee   
where $X^{\prime M}=h^{M}{}_{N}X^N$. This is the familiar non-linear realization, typically 
realized on scalar fields in dimensionally reduced theories \cite{Maharana:1992my}, but here this notion is slightly generalized 
in that the coordinate argument $X$ is transformed as well. The non-linear $O(D,D)$ action can 
be linearized by passing to the `generalized metric' ${\cal H}_{MN}$, 
 \be\label{firstH}
  {\cal H}_{MN} \ = \  \begin{pmatrix}    g^{ij} & -g^{ik}b_{kj}\\[0.5ex]
  b_{ik}g^{kj} & g_{ij}-b_{ik}g^{kl}b_{lj}\end{pmatrix}\;,
 \ee
which is constrained to be $O(D,D)$ valued, satisfying ${\cal H}_{MK}\eta^{KL}{\cal H}_{LN}=\eta_{MN}$. 
The $O(D,D)$ action is now given by the covariant transformation 
 \be\label{HPRIME}
  {\cal H}_{MN}^{\prime}(X') \ = \ (h^{-1})^K{}_M\, (h^{-1})^L{}_N\, {\cal H}_{KL}(X)\;, 
 \ee 
which is equivalent to (\ref{calEtransf}), as we will review in more detail in sec.~5.1.

It should be emphasized that the $O(D,D)$ transformations (\ref{HPRIME}) or, equivalently, (\ref{calEtransf}) 
are genuine invariances of DFT thanks to the doubled coordinates transforming under $O(D,D)$.  
Although the strong constraint (\ref{SECTION}) implies that the fields depend only 
on $D$ coordinates, the solution of the constraint  does not need to be specified in order to define the theory.  
Since the constraint is manifestly $O(D,D)$ invariant, the theory is duality invariant. 
This invariance suffices in order to make manifest the emergence of $O(d,d)$ upon 
dimensional reduction on a torus $T^d$.\footnote{Here and in the following we consider the continuous group realized 
on the massless fields of the classical theory. The discrete group realized in the full string theory will be 
briefly discussed in the conclusion section.} Indeed, in this case the Kaluza-Klein ansatz 
implies that the fields are independent of the $d$ internal coordinates $y^m$, and 
since the fields are also independent of the dual coordinates $\tilde{y}_m$ the `unbroken' 
symmetry is $O(d,d)$. In contrast, picking a solution of the strong constraint from the start realizes 
only the `geometric subgroup' $GL(D,\mathbb{R})\ltimes \mathbb{R}^{\frac{1}{2}D(D-1)}$, 
consisting of `global' diffeomorphisms and antisymmetric shifts of the $b$-field. 
This symmetry is manifest in any theory written in terms covariant curvature tensors, 
the gauge invariant field strength $H={\rm d}b$ and 
their covariant derivatives. It becomes particularly important  
when including $\alpha'$ corrections to make the full $O(D,D)$ manifest, 
for there are numerous higher-derivative invariants that can be 
written with such conventional geometric objects, but only a small subset is actually duality invariant
and realized in string theory. Thus, DFT strongly constrains $\alpha'$ corrections.

In order to explain the obstacle for a generalized metric formulation that will be proved in this paper, 
we have to recall the perturbation theory around a constant background. 
Expanding ${\cal H}_{MN}$ around a constant $\bar{\cal H}_{MN}$, 
the fluctuation terms are constrained in order to preserve the constraint on ${\cal H}$. 
The proper expansion reads 
 \be\label{expandHAAAINntro}
  {\cal H}_{MN} \ = \ \bar{\cal H}_{MN} + h_{\,\nin{M}\bar{N}}+ h_{\,\nin{N}\bar{M}}
   -\tfrac{1}{2} h^{\,\nin{K}}{}_{\bar{M}}\, h_{\,\nin{K}\bar{N}} + \tfrac{1}{2} h_{\,\nin{M}}{}^{\bar{K}}\,
  h_{\,\nin{N}\bar{K}}  \ +  \ {\cal O}(h^3)\;, 
 \ee 
where we introduced projected
 $O(D,D)$ indices defined for a vector by $V_{\,\nin{M}}=P_M{}^{N}V_N$, $V_{\bar{M}}=\bar{P}_M{}^{N}V_N$, 
with the projectors 
 \be\label{projectorsINTRO}
  P_M{}^{N} \ =  \ \tfrac{1}{2}\big(\delta_M{}^{N} - \bar{\cal H}_M{}^N\big)\;, \qquad 
  \bar{P}_M{}^{N} \ =  \ \tfrac{1}{2}\big(\delta_M{}^{N} + \bar{\cal H}_M{}^N\big)\;, 
 \ee
satisfying $P^2=P$, $\bar{P}^2=\bar{P}$ and $P\bar{P}=0$ as a consequence of the constraint on the background 
generalized metric, $\bar{\cal H}_{M}{}^{K}\bar{\cal H}_{K}{}^{N}=\delta_M{}^{N}$. 
The fluctuation field  $h_{\,\nin{M}\bar{N}}$ has $D^2$ independent 
components and transforms covariantly under $O(D,D)$. 
The non-symmetric metric ${\cal E}_{ij}$ is unconstrained and so are its fluctuations, 
but in order for the fluctuation field to transform in a simple way under 
$O(D,D)$ a particular expansion is needed: 
 \be\label{calEExpansionIntro}
  {\cal E}_{ij} \ = \ E_{ij} + e_{ij} +\tfrac{1}{2}\, e_{i}{}^{k} e_{kj} \ + \ {\cal O}(e^3)\;, 
 \ee
where $E=G+B$ encodes the constant background metric $G$ and $B$-field.  
Indeed, the fluctuation field $e_{ij}$ naturally appears in CSFT,  
where it transforms under $O(D,D)$ precisely in such a way that the background independent field 
${\cal E}_{ij}$ transforms as required by (\ref{calEtransf}). 
Moreover, the two fields $h_{\,\nin{M}\bar{N}}$ and $e_{ij}$ are closely related, which can be 
made manifest by use of a constant background frame field ${E}_{A}{}^{M}$, 
with doubled tangent space indices $A=(a,\bar{a})$, as follows \cite{Hohm:2014xsa}:  
 \be\label{eandh}
  {E}_{a}{}^i\,{E}_{\,\bar{b}}{}^{\,j}\, e_{ij} \ = \ \tfrac{1}{2}\,{E}_{a}{}^{M}\, {E}_{\,\bar{b}}{}^{\,N} \, 
  h_{\,\nin{M}\bar{N}}\;, 
 \ee
provided the higher order terms in (\ref{expandHAAAINntro}), 
which are fixed by the constraint on ${\cal H}$ only up to field redefinitions of $h_{\,\nin{M}\bar{N}}$, 
are chosen appropriately. 

Let us now turn to the first $\alpha'$ correction. Starting from CSFT to cubic order in fields, 
it was shown in \cite{Hohm:2014xsa}
that the gauge algebra, $[\delta_{\xi_1},\delta_{\xi_2}]=\delta_{\xi_{12}}$, 
which to lowest order is given by the `C-bracket' for gauge parameters $\xi^M$, receives an 
$\alpha'$ correction and reads  
\be
\xi_{12}^{M} \ = \ \big[\xi_2,\xi_1\big]_{c}^M -\tfrac{1}{4}\,a\, K_{2}^{\,\nin{K}\,\nin{L}}\,\partial^M K_{1\,\nin{K}\,\nin{L}}
   +\tfrac{1}{4}\, b\, K_{2}^{\bar{K}\bar{L}} \,\partial^M K_{1 \bar{K}\bar{L}} - (1\leftrightarrow 2)\;, 
\ee
where $K_{1 MN}=2\partial_{[M}\xi_{1 N]}$, etc., and the explicit factors of $\alpha'$ are 
suppressed. More precisely, for $a=b=1$ this is the gauge algebra 
for bosonic string theory, but the deformation is consistent for any choice of $a, b$. In particular, 
for $a=1, b=0$ it corresponds to heterotic string theory, while for $a=1, b=-1$ it corresponds to the 
HSZ theory  \cite{Hohm:2013jaa}. This result holds to first order in 
perturbation theory, for which the fields ($h_{\,\nin{M}\bar{N}}$ or equivalently $e_{ij}$)
enter the gauge transformations linearly, while the gauge algebra is field independent. 
Using the Noether procedure, in sec.~2 this result is extended to second order in 
perturbation theory, for which the fields enter the gauge transformations quadratically 
and the gauge algebra is field dependent. 
We then ask whether there is a manifestly background independent formulation 
in terms of the full generalized metric ${\cal H}_{MN}$, with a 
field dependent gauge algebra. It turns out that for bosonic string theory 
there is a unique candidate gauge algebra in 
terms of  ${\cal H}_{MN}$, c.f.~(\ref{DFT+algebraFULL}) below, 
that is consistent with the second order results. 
It will then be shown in sec.~3, however, that this candidate expression does not define a consistent 
gauge algebra because it does not satisfy the Jacobi identity 
$\sum_{\rm cycl.} [[\delta_{\xi_1},\delta_{\xi_2}],\delta_{\xi_3}]=0$ when 
acting on fields.\footnote{The `C-bracket' algebra has a non-vanishing Jacobiator, but it represents a 
trivial parameter and therefore the Jacobi identity acting on fields is satisfied.} 
This proves that there is no generalized metric formulation for bosonic string theory to first order 
in $\alpha'$.

In sec.~4 we then 
turn to a frame or vielbein formalism in order to find a background independent 
formulation of the general first order $\alpha'$ corrections. 
We follow the proposal of Marques and Nunez \cite{Marques:2015vua}, 
which we clarify 
and slightly generalize by extending the ($\alpha'$-deformed) frame transformations to 
$GL(D)\times GL(D)$. The full frame field $E_{A}{}^{M}$, with tangent space indices $A=(a,\bar{a})$, 
is subject to local frame transformations that act as 
 \be\label{localFRAMEIntro}
  \delta_{\Lambda}E_{a}{}^{M} \ = \ \Lambda_{a}{}^{b} E_{b}{}^{M}+ \Sigma_{a}{}^{\bar{b}}(\Lambda,E)
  E_{\bar{b}}{}^{M}\;, 
 \ee  
and similarly for $E_{\bar{a}}{}^{M}$, where $\Lambda_{a}{}^{b}$ and $\Lambda_{\bar{a}}{}^{\bar{b}}$ 
are the $GL(D)\times GL(D)$ parameters, and 
   \be
  \Sigma_a{}^{\bar{b}} \ \equiv \ 
   \tfrac{a}{2}\, \D_{a}\Lambda_c{}^{d}\,\omega^{\bar{b}}{}_{d}{}^{c}
   +\tfrac{b}{2}\, \D^{\bar{b}}\Lambda_{\bar{c}}{}^{\bar{d}}\,\omega_{a\bar{d}}{}^{\bar{c}}\;. 
 \ee
Here ${\cal D}_A=E_A{}^{M}\partial_M$ and 
$\omega_{AB}{}^{C}$ 
are the $GL(D)\times GL(D)$ generalized spin connections, which are first order in derivatives of $E$
(c.f.~sec.~4.1).  
Thus, the second term in (\ref{localFRAMEIntro}) carries two derivatives and is of order $\alpha'$, 
thereby representing a deformation of the covariant frame rotations. 
These gauge transformations close to first order in $\alpha'$, 
which requires a deformation of both the generalized diffeomorphism algebra and of the 
$\mathfrak{gl}(D)\oplus \mathfrak{gl}(D)$ algebra, where 
in this paper we restrict ourselves exclusively to first order in $\alpha'$.
Let us also note that the structural form of the gauge transformations (\ref{localFRAMEIntro}) 
is analogous to the 
Green-Schwarz deformed local Lorentz transformations in $D=10$, $N=1$ string theory \cite{Green:1984sg}; 
indeed, they encode these transformations for a particular choice of $a, b$ 
\cite{Hohm:2014eba,Hohm:2014xsa,Marques:2015vua}. 
We will set up the perturbation theory around a constant background frame, following 
\cite{Siegel:1993th,Hohm:2011dz,Hohm:2015ugy}, and show how to identify 
the frame-like fluctuation fields with those in (\ref{eandh}) upon imposing appropriate 
gauge conditions. It is then shown that, up to parameter and field redefinitions, 
the results agree to second order in perturbation theory with those found in sec.~2 by the 
Noether method applied to the first order string theory results of \cite{Hohm:2014xsa}.
This gives an independent check that the $\alpha'$ deformed frame formalism is the proper 
background independent DFT formulation of the general first order $\alpha'$ corrections.

Thus, there is a satisfactory frame-like formulation of the first order $\alpha'$ corrections
in DFT, but it remains somewhat puzzling why a metric-like formulation should not exist in 
general. 
CSFT implies that DFT can be written in a manifestly $O(D,D)$ invariant way 
in terms of fluctuation fields $e_{ij}$ or, equivalently, $h_{\,\nin{M}\bar{N}}$, 
which is confirmed by the explicit results in this paper. 
What, then, prevents us from re-summing these fluctuations 
as in (\ref{expandHAAAINntro}) or (\ref{calEExpansionIntro}) in order to write the theory in terms 
of the background independent generalized metric ${\cal H}_{MN}$ or the background independent ${\cal E}_{ij}$?
It is certainly possible to write the $\alpha'$ corrections of bosonic string theory,  
and of the NS-NS sector of superstring theory, in terms of $g_{ij}$ and $b_{ij}$ and hence in 
terms of ${\cal E}_{ij}=g_{ij}+b_{ij}$. In order to resolve this apparent contradiction, I will show in sec.~5 that 
 in the $\alpha'$-deformed formalism the frame can indeed be parameterized in 
 terms of a field ${\cal E}_{ij}$, upon fixing an appropriate gauge, but the crux is that the ${\cal E}_{ij}$ 
 so defined \textit{does not} transform under $O(D,D)$ as in (\ref{calEtransf}). 
 Rather, the $O(D,D)$ transformation of ${\cal E}_{ij}$ is $\alpha'$-deformed. 
 Consequently, there is no generalized metric formulation, because that would imply an 
 undeformed $O(D,D)$ invariance. 
 More precisely, for general parameters $a, b$, the $O(D,D)$ transformations of $g_{ij}$ and $b_{ij}$
 are deformed in a non-trivial way, but for $a=-b$ we will see that the deformation is actually trivial and 
 hence removable by a redefinition, in agreement with the fact that for this case, corresponding 
 to HSZ theory \cite{Hohm:2013jaa}, there \textit{is} a generalized metric formulation 
 \cite{Hohm:2015mka}.

 The above conclusion can be summarized by saying that in string theory there is a conflict between manifest
 duality invariance and manifest background independence. 
 Indeed, in perturbation theory written in terms of $e_{ij}$ the $O(D,D)$ invariance 
 is manifestly realized to all orders in $\alpha'$ in the original sense, but then 
 the naive ${\cal E}_{ij}$ defined by (\ref{calEExpansionIntro}) is \textit{not} 
 the actual background independent variable in string theory to first order in $\alpha'$. 
 Conversely, writing the theory in a manifestly background independent way in terms 
 of a field ${\cal E}_{ij}$, the $O(D,D)$ symmetry is in general no longer 
 given by (\ref{calEtransf}) and not manifest. 
 In sec.~6, this observation will be used to give an independent proof for the no-go result 
 for a generalized metric formulation. Starting from the manifestly background independent
 frame formulation, I will prove that for the field $h_{\,\nin{M}\bar{N}}$ background independence only holds 
 --- in the sense that the theory is writable in terms of ${\cal H}_{MN}$ according to 
 (\ref{expandHAAAINntro}) ---  for $a=-b$, corresponding to HSZ theory \cite{Hohm:2013jaa}.

\section{Second order perturbation theory: Noether construction} \setcounter{equation}{0}
Here we determine the perturbative gauge structure of DFT around flat backgrounds 
to first order in $\alpha'$ but 
including two fields in the gauge transformations, to which we refer to as second order 
perturbation theory. This determines field-dependent terms in the gauge algebra. 
In the first subsection we review the results obtained from string field theory to 
first order in perturbation theory. In the second subsection we use the Noether method
to extend this to second order in perturbation theory. In the last subsection we 
determine the unique candidate expression for the background independent 
gauge algebra in terms of the generalized metric, 
of which it will be shown in the next section that it does not define a  consistent gauge algebra.

\subsection{Review of first order perturbation theory}

The gauge transformations for the generalized metric ${\cal H}_{MN}$ to zeroth order in $\alpha'$ 
are given by the generalized Lie derivative w.r.t.~the gauge parameter $\xi^M$,   
 \be\label{fullHGauge}
  \delta_{\xi}{\cal H}_{MN} \ = \ {\cal L}_{\xi}{\cal H}_{MN} \ \equiv \ 
  \xi^K\partial_K{\cal H}_{MN} + K_M{}^K {\cal H}_{KN}+K_N{}^K {\cal H}_{MK}\;, 
 \ee
where we defined 
 \be
  K_{MN} \ = \ \partial_M\xi_N - \partial_N\xi_M\;, 
 \ee
and indices are raised and lowered with the $O(D,D)$ invariant metric $\eta_{MN}$. 
Due to this form of the gauge transformations and the strong constraint, gauge parameters 
of the form $\xi^M=\partial^M\chi$ are trivial in that they do not generate a gauge transformation.  
The gauge transformations close, $[\delta_{\xi_1},\delta_{\xi_2}]=\delta_{\xi_{12}}$, 
according to the C-bracket 
 \be\label{CBRacket}
  \xi_{12}^M \ = \ \big[\xi_2,\xi_1\big]_{c}^M \ \equiv \ \xi_2^N\partial_N\xi_1^M - \tfrac{1}{2}\,\xi_{2N}\partial^M \xi_1^N
  -( 1 \leftrightarrow 2)\;, 
 \ee 
which is related to the generalized Lie derivative and the inner product $\langle\,, \rangle $ defined by the 
$O(D,D)$ metric $\eta$ by 
 \be\label{DCrelation}
  {\cal L}_{V}W \ = \ \big[V,W\big]_c + \tfrac{1}{2}\,\partial\, \big\langle V, W\big \rangle\;. 
 \ee  
The proof of closure is straightforward when using the following identity  
 \be\label{MASTER1}
  K_{12 MN} \ \equiv \ K_{MN}(\xi_{12}) \ = \ \xi_2^K\partial_KK_{1 MN}+K_{2M}{}^{K} K_{1KN} -( 1 \leftrightarrow 2)\;, 
 \ee
where $K_{1}\equiv K(\xi_1)$, etc., which can be verified by a quick direct computation employing the strong constraint.  
Taking the derivative of this equation we derive another useful identity,  
 \be\label{MASTER2}
 \begin{split}
  \partial_M K_{12KL} \ &= \ \xi_2^N\partial_N\partial_M K_{1KL}+K_{2M}{}^{N} \partial_N K_{1KL}
  -2\, K_{2[K}{}^{N} \partial_M K_{1L]N}
  -( 1 \leftrightarrow 2)\\[0.5ex]
  \ &\equiv \ {\cal L}_{\xi_2}\big(\partial_M K_{1KL}\big)-( 1 \leftrightarrow 2)\;, 
 \end{split}
 \ee      
where we used again the strong constraint and recognized on the right-hand side the 
generalized Lie derivative of an object with three indices. 
Let us finally note that the gauge transformation of the dilaton, 
 \be\label{dilatonvar}
  \delta_{\xi}\phi \ = \ \xi^N\partial_N\phi + \partial_N\xi^N \;, 
 \ee
will remain undeformed under $\alpha'$ corrections. Indeed, as will be established below, 
the deformations of the gauge algebra are such that the free index of $\xi_{12}^M$ 
is carried by a derivative, so that the corresponding transformation of $\phi$ is trivial 
by the strong constraint.

We now turn to the perturbative expansion (\ref{expandHAAAINntro}) around a flat background 
encoded in the constant background generalized metric $\bar{\cal H}_{MN}$, 
 \be\label{expandHAAA}
  {\cal H}_{MN} \ = \ \bar{\cal H}_{MN} + h_{\,\nin{M}\bar{N}}+ h_{\,\nin{N}\bar{M}}
   -\tfrac{1}{2} h^{\,\nin{K}}{}_{\bar{M}}\, h_{\,\nin{K}\bar{N}} + \tfrac{1}{2} h_{\,\nin{M}}{}^{\bar{K}}\,
  h_{\,\nin{N}\bar{K}}  \ +  \ {\cal O}(h^3)\;, 
 \ee 
where we recall the 
 projected $O(D,D)$ indices defined by $V_{\,\nin{M}}=P_M{}^{N}V_N$, $V_{\bar{M}}=\bar{P}_M{}^{N}V_N$, 
and similarly for any $O(D,D)$ tensor, with the projectors (\ref{projectorsINTRO}) defined in the 
introduction. 
 The independent fluctuation field is $h_{\,\nin{M}\bar{N}}$, which in agreement 
with our notation satisfies $P_{N}{}^{K}h_{\,\nin{M}\bar{K}} = \bar{P}_M{}^{K}h_{\,\nin{K}\bar{N}} = 0$.  
Since this field carries  
projected indices, it encodes the $D^2$ degrees of freedom corresponding to the metric and 
$b$-field fluctuations. It is easy to verify with (\ref{expandHAAA}) that, to second order in fields, ${\cal H}_{MN}$
satisfies the constraint ${\cal H}\eta{\cal H}=\eta$. The higher order terms in $h$ not displayed in (\ref{expandHAAA})
are needed to satisfy this constraint to all orders, as we will discuss in more detail later. 

Applying the gauge transformations (\ref{fullHGauge}) to (\ref{expandHAAA}) 
one may determine the gauge transformations of the fluctuation field $h_{\,\nin{M}\bar{N}}$, 
 \be
  \delta_{\xi} h_{\,\nin{M}\bar{N}} \ = \ 
  2(\partial_{\,\nin{M}}\xi_{\bar{N}}-\partial_{\bar{N}}\xi_{\,\nin{M}})
   + \xi^{P}\partial_P h_{\,\nin{M}\bar{N}}+K_{\,\nin{M}}{}^{\,\nin{K}}
   h_{\,\nin{K}\bar{N}} + K_{\bar{N}}{}^{\bar{K}} h_{\,\nin{M}\bar{K}}\;, 
 \ee
up to terms that are quadratic or of higher order in $h$.    
In the following we will use a notation, where the number of fields entering the 
gauge transformations or the gauge algebra is indicated in a square bracket $[ \,]$.  
We thus write 
 \be
  \begin{split}
   \delta_{\xi}^{[0]} h_{\,\nin{M}\bar{N}} \ &= \ 2(\partial_{\,\nin{M}}\xi_{\bar{N}}
   -\partial_{\bar{N}}\xi_{\,\nin{M}}) \ = \ 2K_{\,\nin{M}\bar{N}} \;, \\
   \delta_{\xi}^{[1]} h_{\,\nin{M}\bar{N}} \ &= \ \xi^{P}\partial_P h_{\,\nin{M}\bar{N}}+K_{\,\nin{M}}{}^{\,\nin{K}}
   h_{\,\nin{K}\bar{N}} + K_{\bar{N}}{}^{\bar{K}} h_{\,\nin{M}\bar{K}}\;. 
   \end{split}
 \ee
We also recall for later use the definition of the following linearized connections:  
\be\label{lincoNN}
   \Gamma_{\bar{M}\,\nin{N}\,\nin{K}} \ \equiv \ \partial_{\,\nin{N}}h_{\,\nin{K}\bar{M}} 
   -\partial_{\,\nin{K}}h_{\,\nin{N}\bar{M}}\;,\qquad 
    \Gamma_{\,\nin{M}\bar{N}\bar{K}} \ \equiv \ 
    \partial_{\bar{N}}h_{\,\nin{M}\bar{K}} -\partial_{\bar{K}}h_{\,\nin{M}\bar{N}}\;, 
 \ee 
which transform under the linearized gauge transformations as 
  \be\label{linnGaugeVar}  
   \delta_{\xi}^{[0]} \Gamma_{\bar{M}\,\nin{N}\,\nin{K}} \ = \ -2\partial_{\bar{M}}K_{\,\nin{N}\,\nin{K}}\;, \qquad 
  \delta_{\xi}^{[0]} \Gamma_{\,\nin{M}\bar{N}\bar{K}}  \ = \ 2\partial_{\,\nin{M}}K_{\bar{N}\bar{K}}\;. 
 \ee

Next let us review closure of the  gauge algebra in this perturbative scheme.   
Acting on the fluctuation field, closure reads  
  \be\label{flucclosure}
   \big[\delta_{\xi_1},\delta_{\xi_2}\big]h_{\,\nin{M}\bar{N}} \ = \ \delta_{\xi_{12}}h_{\,\nin{M}\bar{N}}\;. 
  \ee
This condition is trivial to zeroth order in fields, since the gauge variations $\delta_{\xi}^{[0]}$ are abelian. 
To first order in fields, closure requires 
 \be 
  \delta_{\xi_1}^{[0]}\big(\delta_{\xi_2}^{[1]}h_{\,\nin{M}\bar{N}}\big)-(1\leftrightarrow 2) \ = \ 
   \delta_{\xi_{12,c}}^{[0]} h_{\,\nin{M}\bar{N}}\;,
 \ee  
and it is instructive to verify this by a quick computation: 
  \be\label{lowest}
  \begin{split}
   \delta_{\xi_1}^{[0]}\big(\delta_{\xi_2}^{[1]}h_{\,\nin{M}\bar{N}}\big) -(1\leftrightarrow 2)
    \ &= \ 2\big(\xi_2^{K}\partial_K K_{1\,\nin{M}\bar{N}}+K_{2\,\nin{M}}{}^{\,\nin{K}}
   K_{1\,\nin{K}\bar{N}} + K_{2\bar{N}}{}^{\bar{K}} K_{1\,\nin{M}\bar{K}}\big)
   -(1\leftrightarrow 2) \\ 
     \ &= \  2\, K_{12\,\nin{M}\bar{N}} \ = \ \delta_{\xi_{12,c}}^{[0]}  h_{\,\nin{M}\bar{N}}  \end{split} 
  \ee 
using the antisymmetry in $(1\leftrightarrow 2)$ and the identity (\ref{MASTER1}) in the second line.

After having reviewed the perturbative gauge structure of the two-derivative theory, 
let us now turn to the first $\alpha'$ correction. We will follow the convention that 
the order of $\alpha'$ is indicated in round parenthesis $(\,)$, 
 \be
  \delta_{\xi} \ = \ \delta_{\xi}^{[0]} \ + \ \delta_{\xi}^{[1]} \ + \ \alpha'\, \delta_{\xi}^{[1](1)} \  + \ \cdots\;. 
 \ee
 When there is no 
explicit parenthesis the expression is to zeroth order in $\alpha'$. For instance, 
the C-bracket part can be written as $\xi_{12,c}\equiv \xi_{12}^{[0](0)}$. 
We recall that in this paper we
restrict ourselves to the first $\alpha'$ correction.    
In \cite{Hohm:2014xsa} it was shown that the first $\alpha'$ correction of the gauge transformations, 
linear in fields and 
hence corresponding to the cubic action, is given by  
   \be
  \delta_{\xi}^{[1](1)} h_{\,\nin{M}\bar{N}} \ = \ a\,
  \partial_{\,\nin{M}}K^{\,\nin{K}\,\nin{L}}\,\partial_{\,\nin{K}} 
   h_{\,\nin{L}\bar{N}}  -b\, \partial_{\bar{N}} K^{\bar{K}\bar{L}}\,\partial_{\bar{K}} h_{\,\nin{M}\bar{L}}\;, 
  \ee 
or, in terms of the linearized connections (\ref{lincoNN}), 
 \be\label{linCOnnFORM}
   \delta_{\xi}^{[1](1)} h_{\,\nin{M}\bar{N}} \ = \ \tfrac{a}{2}\,
  \partial_{\,\nin{M}}K^{\,\nin{K}\,\nin{L}}\,\Gamma_{\bar{N}\,\nin{K}\,\nin{L}} 
   -\tfrac{b}{2}\, \partial_{\bar{N}} K^{\bar{K}\bar{L}}\,\Gamma_{\,\nin{M}\bar{K}\bar{L}}\;, 
  \ee 
with two free parameters $a, b$, which in line with the notation of \cite{Hohm:2014xsa} 
we sometimes re-parameterize as    
  \be
  a \ = \ \gamma^+ + \gamma^-\;, \qquad
  b \ = \ \gamma^+ - \gamma^-\;. 
 \ee
For $a=b=1$ or $\gamma^+=1, \gamma^-=0$ these gauge transformations follow from bosonic closed 
string field theory  \cite{Hohm:2014xsa}, but the deformation of the gauge structure is consistent for 
arbitrary parameters. Indeed, for $\gamma^+=\gamma^- = \frac{1}{2}$ the deformation corresponds 
to heterotic string theory, while for $\gamma^+=0, \gamma^-=1$ we obtain the  $\alpha'$
corrections of the HSZ theory \cite{Hohm:2013jaa}. 
 
Let us verify that the above gauge transformations are consistent by proving closure. 
Taking into account in (\ref{flucclosure}) terms to zeroth order in fields 
and first order in $\alpha'$, the closure condition reads  
\be\label{firstorderalpha'cond}
 \delta_{\xi_1}^{[0]}\big(\delta_{\xi_2}^{[1](1)} h_{\,\nin{M}\bar{N}}\big)-(1\leftrightarrow 2) 
   \ = \ \delta_{\xi_{12}^{[0](1)}}^{[0]} h_{\,\nin{M}\bar{N}}\;, 
\ee   
where we allow for a deformation $\xi_{12}^{[0](1)}$ of the gauge algebra, which is of first order in $\alpha'$ 
and zeroth order in fields.  
In order to simplify the presentation it is convenient to use a notation in which the antisymmetrization in 
$(1\leftrightarrow 2)$ is left implicit. Thus, in the remainder of this section we leave out the 
term $-(1\leftrightarrow 2)$ at the end of all equations. 
We then compute for the left-hand side of (\ref{firstorderalpha'cond}), using (\ref{linnGaugeVar}) 
and (\ref{linCOnnFORM}), 
 \be
  \begin{split}
     \delta_{\xi_1}^{[0]}\big(\delta_{\xi_2}^{[1](1)} h_{\,\nin{M}\bar{N}}\big) 
    \ &= \ -a\, \partial_{\,\nin{M}}K_2^{\,\nin{K}\,\nin{L}}\,\partial_{\bar{N}} K_{1\,\nin{K}\,\nin{L}}
   -b\, \partial_{\bar{N}} K_2^{\bar{K}\bar{L}}\,\partial_{\,\nin{M}} K_{1\bar{K}\bar{L}}\\
   \ &= \ \partial_{\,\nin{M}}\big(-\tfrac{1}{2}a K_{2}^{\,\nin{K}\,\nin{L}}\,\partial_{\bar{N}} K_{1\,\nin{K}\,\nin{L}}
   -\tfrac{1}{2} b K_{1\bar{K}\bar{L}} \,\partial_{\bar{N}} K_2^{\bar{K}\bar{L}}\big)\\
   &\qquad -\partial_{\bar{N}}\big(-\tfrac{1}{2} a K_{2}^{\,\nin{K}\,\nin{L}}\,\partial_{\,\nin{M}} K_{1\,\nin{K}\,\nin{L}}
   -\tfrac{1}{2}b  K_{1\bar{K}\bar{L}} \,\partial_{\, \nin{M}} K_2^{\bar{K}\bar{L}}\big)\\
   \ &\equiv \ 2\,\partial^{}_{\,\nin{M}}\xi^{[0](1)}_{12 \bar{N}}-2\, \partial^{}_{\bar{N}}\xi^{[0](1)}_{12\,\nin{M}}\;, 
  \end{split}
 \ee    
where we wrote out in the last step the desired right-hand side of  (\ref{firstorderalpha'cond}). 
We infer that closure holds for 
 \be\label{zerothalpha'algebra}
  \xi_{12}^{[0](1)M} \ = \ -\tfrac{1}{4}a\, K_{2}^{\,\nin{K}\,\nin{L}}\,\partial^M K_{1\,\nin{K}\,\nin{L}}
   +\tfrac{1}{4} b\, K_{2}^{\bar{K}\bar{L}} \,\partial^M K_{1 \bar{K}\bar{L}} \;. 
 \ee  
This concludes our review of first order perturbation theory in presence of the first $\alpha'$ correction.

\subsection{Gauge structure in second order perturbation theory} 

Our goal in this subsection is to extend the above results to second order perturbation theory, 
i.e., to determine the gauge transformations including two fields, denoted 
in our above notation as $\delta_{\xi}^{[2](1)} h_{\,\nin{M}\bar{N}}$. 
This corresponds to the quartic action around a flat background. While for bosonic string theory 
in principle this could be derived from closed string field theory, such a computation would be 
forbiddingly tedious, in addition to conceptual subtleties that arise in string field theory beyond cubic level. 
Therefore, we will instead use a Noether construction, in which we systematically determine 
the quadratic terms in the gauge transformations and the linear terms in the gauge algebra by demanding closure. 
Note, in particular, that to this order we see the first appearance of \textit{field-dependent} 
structures in the gauge algebra. This determines uniquely the gauge structure of second order 
perturbation theory, up to the same two free parameters $a, b$ discussed above. 
The dilaton will play no role in this discussion, and its gauge transformation remains undeformed. 
  
We begin by writing out the closure condition (\ref{flucclosure}) to the desired order, 
which includes all terms to first order in $\alpha'$ and linear  in $h$, 
 \be 
  \begin{split}
   & \delta_{\xi_1}^{[1]}\big(\delta_{\xi_2}^{[1](1)} h_{\,\nin{M}\bar{N}}\big)
   -\delta_{\xi_2}^{[1](1)}\big(\delta_{\xi_1}^{[1]} h_{\,\nin{M}\bar{N}}\big)
   +\delta_{\xi_1}^{[0]}\big(\delta_{\xi_2}^{[2](1)} h_{\,\nin{M}\bar{N}}\big)-(1\leftrightarrow 2) \\
   &\quad \ = \ \big(\delta_{\xi_{12}^{[1](1)}}^{[0]} +\delta_{\xi_{12}^{[0](1)}}^{[1]} 
   +\delta_{\xi_{12,c}}^{[1](1)} \big)h_{\,\nin{M}\bar{N}}\;. 
  \end{split}
 \ee    
 We reorder this equation as follows: 
  \be\label{reorderedREL}
   \begin{split}
      {\cal E}_{\,\nin{M}\bar{N}} \ &\equiv \ \delta_{\xi_1}^{[1]}\big(\delta_{\xi_2}^{[1](1)} h_{\,\nin{M}\bar{N}}\big)
       -\delta_{\xi_2}^{[1](1)}\big(\delta_{\xi_1}^{[1]} h_{\,\nin{M}\bar{N}}\big) -(1\leftrightarrow 2) \\
      & \qquad -\delta_{\xi_{12,c}}^{[1](1)}h_{\,\nin{M}\bar{N}}-\delta_{\xi_{12}^{[0](1)}}^{[1]} h_{\,\nin{M}\bar{N}} \\[0.5ex]
      \ &= \ \delta_{\xi_{12}^{[1](1)}}^{[0]}h_{\,\nin{M}\bar{N}}
      -\delta_{\xi_1}^{[0]}\big(\delta_{\xi_2}^{[2](1)} h_{\,\nin{M}\bar{N}}\big)-(1\leftrightarrow 2) \;, 
   \end{split}
  \ee    
which is now written so that the terms defining ${\cal E}_{\,\nin{M}\bar{N}}$ are all computable from the formulas 
of the previous subsection. The Noether procedure 
is then to find a deformation of $\xi_{12}$ to first order in $h$ and a deformation of $\delta_{\xi}$ 
to second order in $h$ so that in (\ref{reorderedREL}) the left-hand side equals the right-hand side.

We compute for the first two terms on the left-hand side: 
  \be
  \begin{split}
   \delta_{\xi_1}^{[1]}\big(\delta_{\xi_2}^{[1](1)} h_{\,\nin{M}\bar{N}}\big)
   \ = \ &\,a\, \partial_{\,\nin{M}} K_2^{\,\nin{K}\,\nin{L}}\,\partial_{\,\nin{K}}
   \big(  \xi_1^{P}\partial_P h_{\,\nin{L}\bar{N}}+K_{1\,\nin{L}}{}^{\,\nin{P}}
   h_{\,\nin{P}\bar{N}} + K_{1\bar{N}}{}^{\bar{P}}h_{\,\nin{L}\bar{P}}\big)   \\[0.5ex]
   &\,-b\, \partial_{\bar{N}}K_2^{\bar{K}\bar{L}}\,\partial_{\bar{K}}
   \big( \xi_1^{P}\partial_P h_{\,\nin{M}\bar{L}}+K_{1\,\nin{M}}{}^{\,\nin{P}}
   {h_{\,\nin{P}\bar{L}}} + K_{1\bar{L}}{}^{\bar{P}} h_{\,\nin{M}\bar{P}} \big)\;, 
   \end{split}  
  \ee
 and 
  \be
  \begin{split}
   -\delta_{\xi_2}^{[1](1)}\big(\delta_{\xi_1}^{[1]} h_{\,\nin{M}\bar{N}}\big) \ = \ &\, 
    -\delta_{\xi_2}^{[1](1)}\big(\xi_1^{P}\partial_P h_{\,\nin{M}\bar{N}}+K_{1\,\nin{M}}{}^{\,\nin{P}}
   h_{\,\nin{P}\bar{N}} + K_{1\bar{N}}{}^{\bar{P}} h_{\,\nin{M}\bar{P}}\big)\\[0.5ex]
   \ = \ &\, -\xi_1^P\partial_P\big(a\,  \partial_{\,\nin{M}}K_2^{\,\nin{K}\,\nin{L}}\,\partial_{\,\nin{K}} 
   h_{\,\nin{L}\bar{N}}  - b\, \partial_{\bar{N}} K_2^{\bar{K}\bar{L}}\,{\partial_{\bar{K}} h_{\,\nin{M}\bar{L}}}\big)\\[0.5ex]
   &\, -K_{1\,\nin{M}}{}^{\,\nin{P}}\big(a\, \partial_{\,\nin{P}}K_2^{\,\nin{K}\,\nin{L}}\,\partial_{\,\nin{K}} 
   h_{\,\nin{L}\bar{N}}  - b\, \partial_{\bar{N}} K_2^{\bar{K}\bar{L}}\,\partial_{\bar{K}} h_{\,\nin{P}\bar{L}}\big)\\[0.5ex]
   &\, -K_{1\bar{N}}{}^{\bar{P}}\big( a\,\partial_{\,\nin{M}}K_2^{\,\nin{K}\,\nin{L}}\,\partial_{\,\nin{K}} 
   h_{\,\nin{L}\bar{P}}  -b\,  \partial_{\bar{P}} K_2^{\bar{K}\bar{L}}\,\partial_{\bar{K}} h_{\,\nin{M}\bar{L}}\big)\;. 
  \end{split} 
  \ee     
Various terms cancel between the two structures.   
Next, for the $\xi_{12}$ terms on the left-hand side of  (\ref{reorderedREL}) we compute 
 \be
 \begin{split}
  -&
  \delta_{\xi_{12,c}}^{[1](1)}h_{\,\nin{M}\bar{N}} \ = \ -a\, \partial_{\,\nin{M}} K_{12}^{\,\nin{K}\,\nin{L}}\,\partial_{\,\nin{K}}
  h_{\,\nin{L}\bar{N}} +b\,  \partial_{\bar{N}} K_{12}^{\bar{K}\bar{L}}\,\partial_{\bar{K}} h_{\,\nin{M}\bar{L}}\\
  \ &= \ a\, \big(-\xi_2^P\partial_P\partial_{\,\nin{M}} K_{1}^{\,\nin{K}\,\nin{L}}
  - K_{2\,\nin{M}}{}^{P} \partial_P K_1^{\,\nin{K}\,\nin{L}}
   - K_{2}^{\,\nin{K}}{}_{P}\partial_{\,\nin{M}} K_1^{P\,\nin{L}}
  - K_2{}^{\,\nin{L}}{}_{P}\,\partial_{\,\nin{M}} K_1^{\,\nin{K}P}\big)  
  \partial_{\,\nin{K}} h_{\,\nin{L}\bar{N}} \\
  &\qquad +b\, \big(\xi_2^P\partial_P\partial_{\bar{N}}K_1^{\bar{K}\bar{L}}
  + {K_{2\bar{N}}{}^{P}\partial_P K_1^{\bar{K}\bar{L}}}
  +K_2{}^{\bar{K}}{}_{P}\partial_{\bar{N}} K_1^{P\bar{L}}+ K_2{}^{\bar{L}}{}_{P} \partial_{\bar{N}} K_1^{\bar{K}P}\big)
  \partial_{\bar{K}} h_{\,\nin{M}\bar{L}}\;, 
 \end{split} 
 \ee 
where we used the identity (\ref{MASTER2}).  
Moreover, 
 \be\label{notransport}
  -\delta_{\xi_{12}^{[0](1)}}^{[1]} h_{\,\nin{M}\bar{N}} \ = \ 
  -\xi_{12}^{[0](1)P}\partial_P h_{\,\nin{M}\bar{N}} - K_{12}^{[0](1)}{}_{\,\nin{M}}{}^{\,\nin{K}}h_{\,\nin{K}\bar{N}}
  - K_{12}^{[0](1)}{}_{\bar{N}}{}^{\bar{K}} h_{\,\nin{M}\bar{K}}\;, 
 \ee
for which we have to use that from (\ref{zerothalpha'algebra}) 
 \be
 \begin{split}
  K_{12}^{[0](1)}{}_M{}^{N} 
   \ = \ \, -\tfrac{1}{2}a\, \partial_M K_2^{\,\nin{K}\,\nin{L}}\partial^N K_{1\,\nin{K}\,\nin{L}} 
   +\tfrac{1}{2} b\, \partial_M K_2^{\bar{K}\bar{L}} \partial^N K_{1\bar{K}\bar{L}} \;, 
  \end{split}
 \ee  
while the transport term in (\ref{notransport}) vanishes. This gives in total 
 \be
 \begin{split}
   -\delta_{\xi_{12}^{[0](1)}}^{[1]} h_{\,\nin{M}\bar{N}} \ = \ 
   &\, \tfrac{1}{2}a \partial_{\,\nin{M}} K_2^{\,\nin{P}\,\nin{Q}}\partial^{\,\nin{K}} K_{1\,\nin{P}\,\nin{Q}} \, h_{\,\nin{K}\bar{N}}
   -\tfrac{1}{2}b  \partial_{\,\nin{M}} K_2^{\bar{P}\bar{Q}} \partial^{\,\nin{K}} K_{1\bar{P}\bar{Q}} 
   \, h_{\,\nin{K}\bar{N}}\\
   &\, +\tfrac{1}{2}a \partial_{\bar{N}} K_2^{\,\nin{P}\,\nin{Q}}\partial^{\bar{K}} K_{1\,\nin{P}\,\nin{Q}} \, h_{\,\nin{M}\bar{K}}
     -\tfrac{1}{2}b   \partial_{\bar{N}} K_2^{\bar{P}\bar{Q}} \partial^{\bar{K}} K_{1\bar{P}\bar{Q}} 
   \, h_{\,\nin{M}\bar{K}}\;. 
 \end{split}
 \ee  
We can now collect all terms of the left-hand side of (\ref{reorderedREL}): 
 \be\label{calEbareh} 
 \begin{split}
  {\cal E}_{\,\nin{M}\bar{N}}
  \ = \ &\, a\,\partial_{\,\nin{M}} K_2^{\,\nin{K}\,\nin{L}}\,\partial_{\,\nin{K}}
        K_{1\bar{N}}{}^{\bar{P}}\, h_{\,\nin{L}\bar{P}}
        - b\,
         \partial_{\bar{N}}K_2^{\bar{K}\bar{L}}\,\partial_{\bar{K}}K_{1\,\nin{M}}{}^{\,\nin{P}}\,h_{\,\nin{P}\bar{L}}\\
     &\, -\tfrac{1}{2} b \, \partial_{\,\nin{M}} K_2^{\bar{P}\bar{Q}} \partial^{\,\nin{K}} K_{1\bar{P}\bar{Q}} 
     \,h_{\,\nin{K}\bar{N}}
     +\tfrac{1}{2}a\, \partial_{\bar{N}} K_2^{\,\nin{P}\,\nin{Q}}\partial^{\bar{K}} K_{1\,\nin{P}\,\nin{Q}} \, 
     h_{\,\nin{M}\bar{K}} \\
 &\, -a\, K_{2\,\nin{M}}{}^{\bar{P}}\partial_{\bar{P}}K_1^{\,\nin{K}\,\nin{L}}\,
  \partial_{\,\nin{K}} h_{\,\nin{L}\bar{N}} + b\, K_{2\bar{N}}{}^{\,\nin{P}}\,\partial_{\,\nin{P}} K_1^{\bar{K}\bar{L}}\,
  \partial_{\bar{K}} h_{\,\nin{M}\bar{L}}\\
  &\,+a\, \partial_{\,\nin{M}}K_2^{\,\nin{K}\bar{P}}\,K_{1}{}^{\,\nin{L}}{}_{\bar{P}}\, \Gamma_{\bar{N}\,\nin{K}\,\nin{L}}
  -b\,  \partial_{\bar{N}}K_2^{\,\nin{P}\bar{K}}\, K_{1\,\nin{P}}{}^{\bar{L}}\, \Gamma_{\,\nin{M}\bar{K}\bar{L}}
   \\
  &\, +a\, \partial_{\,\nin{M}}K_2^{\,\nin{K}\,\nin{L}}\, K_{1\,\nin{K}}{}^{\bar{P}}\,\partial_{\bar{P}} h_{\,\nin{L}\bar{N}}
  -b\, \partial_{\bar{N}}K_2^{\bar{K}\bar{L}}\, K_{1\bar{K}}{}^{\,\nin{P}}\,\partial_{\,\nin{P}} h_{\,\nin{M}\bar{L}}\;, 
  \end{split}
 \ee 
where we used the linearized connections (\ref{lincoNN}) to simplify some terms. 
Moreover, the terms were organized into those with bare $h$ in the first two lines 
and with $\partial h$ in the final three lines.

Our next challenge is to write ${\cal E}_{\,\nin{M}\bar{N}}$ as a total $\delta^{[0]}$ variation of the unknown 
$\delta^{[2](1)}h$ and as a `curl' in $\partial_{\,\nin{M}}$ and $\partial_{\bar{N}}$, 
corresponding to the deformation of the gauge algebra. 
A direct computation shows that various terms can be combined into a total variation,  
and we find 
 \be\label{zwischenresultat}
 \begin{split}
  {\cal E}_{\,\nin{M}\bar{N}} \ = \ &\,\delta_{\xi_2}^{[0]}\big(-\tfrac{1}{2} a\,  
  h_{\,\nin{M}}{}^{\bar{P}}\partial_{\bar{P}}K_1^{\,\nin{K}\,\nin{L}}\,
  \partial_{\,\nin{K}} h_{\,\nin{L}\bar{N}} -\tfrac{1}{2}b\,  h^{\,\nin{P}}{}_{\bar{N}} \,\partial_{\,\nin{P}} K_1^{\bar{K}\bar{L}}\,
  \partial_{\bar{K}} h_{\,\nin{M}\bar{L}} \\
  &\qquad -\tfrac{1}{2}a\, \partial_{\,\nin{M}}K_1^{\,\nin{K}\,\nin{L}}\, h_{\,\nin{K}}{}^{\bar{P}}\,
  \partial_{\bar{P}} h_{\,\nin{L}\bar{N}}
  -\tfrac{1}{2}b\, \partial_{\bar{N}}K_1^{\bar{K}\bar{L}}\, h^{\,\nin{P}}{}_{\bar{K}} \,\partial_{\,\nin{P}} h_{\,\nin{M}\bar{L}}\big)
  \\
  &\qquad + a\, \partial_{\,\nin{M}}K_2^{\,\nin{K}\,\nin{L}}\,\partial_{\bar{N}} K_{1\,\nin{K}}{}^{\bar{P}}
  h_{\,\nin{L}\bar{P}}
  + b\, \partial_{\bar{N}}K_2^{\bar{K}\bar{L}}\,\partial_{\,\nin{M}}K_{1}{}^{\,\nin{P}}{}_{\bar{K}}
  \, h_{\,\nin{P}\bar{L}} \\
  &\qquad +a\, \partial_{\,\nin{M}}K_2^{\,\nin{K}\bar{P}}\,K_{1}{}^{\,\nin{L}}{}_{\bar{P}}\, \Gamma_{\bar{N}\,\nin{K}\,\nin{L}}
  - b\, \partial_{\bar{N}}K_2^{\,\nin{P}\bar{K}}\, K_{1\,\nin{P}}{}^{\bar{L}}\, \Gamma_{\,\nin{M}\bar{K}\bar{L}}\;. 
 \end{split}
 \ee 
We will now show that the remaining terms in the third and fourth line can be interpreted as a 
field-dependent deformation of the gauge algebra, up to further terms in $\delta^{[2](1)}h$. 
To this end it is convenient to start with an ansatz for the gauge algebra that follows from 
the cubic algebra determined in \cite{Hohm:2014xsa}, as reviewed in the previous subsection, 
by promoting the background generalized metric to a full generalized metric, writing 
\be\label{HKKalgebra}
  \xi_{12}^{\prime M} \ = \ \tfrac{1}{4}\,{\cal H}^{KL} K_{2 K}{}^{P}\partial^M K_{1LP} -(1\leftrightarrow 2)\;. 
\ee  
Using the 
expansion (\ref{expandHAAA}) we can read off the order $h$ terms for the bosonic string, i.e., 
for $a=b=1$. Moreover, we should expect terms 
including $\partial h$. All in all, we use 
 \be
 \begin{split}
  \xi_{12}^{M[1](1)} \ = \ &\, \tfrac{1}{4}a\, h^{\,\nin{K}\bar{L}} K_{2\,\nin{K}}{}^{\,\nin{P}}
  \partial^M K_{1\bar{L}\,\nin{P}}
  +\tfrac{1}{4}b \,  h^{\,\nin{K}\bar{L}} K_{2\,\nin{K}}{}^{\bar{P}}\partial^M K_{1\bar{L}\bar{P}} \\
  &\, +\tfrac{1}{4} a\, h^{\,\nin{L}\bar{K}} K_{2\bar{K}}{}^{\,\nin{P}}\partial^M K_{1\,\nin{L}\,\nin{P}}
  +\tfrac{1}{4}b \,  h^{\,\nin{L}\bar{K}}K_{2\bar{K}}{}^{\bar{P}}\partial^M K_{1\,\nin{L}\bar{P}}\\
  &\, -\tfrac{1}{4}a\,\partial^Mh^{\,\nin{K}\bar{L}}\, K_{2\,\nin{K}}{}^{\,\nin{P}} K_{1\bar{L}\,\nin{P}}
  +\tfrac{1}{4}b   \, \partial^Mh^{\,\nin{K}\bar{L}}\, K_{2\,\nin{K}}{}^{\bar{P}} K_{1\bar{L}\bar{P}}\;, 
 \end{split} 
 \ee 
where we restored arbitrary parameters $a, b$ and 
set the coefficients in the last line to the values that momentarily will be 
fixed by the closure computation.  
Computing  $\delta_{\xi_{12}^{[1](1)}}^{[0]}h$ with this ansatz and using (\ref{zwischenresultat}) 
we find 
\be\label{zwischenresultat44}
 \begin{split}
  {\cal E}_{\,\nin{M}\bar{N}} \ = \ &\,\delta_{\xi_2}^{[0]}\big(-\tfrac{1}{2} a\,
  h_{\,\nin{M}}{}^{\bar{P}}\partial_{\bar{P}}K_1^{\,\nin{K}\,\nin{L}}\,
  \partial_{\,\nin{K}} h_{\,\nin{L}\bar{N}} -\tfrac{1}{2}b\,  h^{\,\nin{P}}{}_{\bar{N}} \,\partial_{\,\nin{P}} K_1^{\bar{K}\bar{L}}\,
  \partial_{\bar{K}} h_{\,\nin{M}\bar{L}} \\
  &\qquad -\tfrac{1}{2}a\, \partial_{\,\nin{M}}K_1^{\,\nin{K}\,\nin{L}}\, h_{\,\nin{K}}{}^{\bar{P}}\,
  \partial_{\bar{P}} h_{\,\nin{L}\bar{N}}
  -\tfrac{1}{2}b\,  \partial_{\bar{N}}K_1^{\bar{K}\bar{L}}\, h^{\,\nin{P}}{}_{\bar{K}} \,\partial_{\,\nin{P}} h_{\,\nin{M}\bar{L}} \\
  &\qquad 
  -\tfrac{1}{2}a\,\partial_{\,\nin{M}} K_1^{\,\nin{K}\bar{P}}\,h^{\,\nin{L}}{}_{\bar{P}}
  \,\Gamma_{\bar{N}\,\nin{K}\,\nin{L}} + \tfrac{1}{2}b\,
  \partial_{\bar{N}}K_1^{\,\nin{P}\bar{K}} \, h_{\,\nin{P}}{}^{\bar{L}}\,
  \Gamma_{\,\nin{M}\bar{K}\bar{L}}\big)
  \\[0.5ex]
    &\,+ \delta_{\xi_{12}^{[1](1)}}^{[0]}h_{\,\nin{M}\bar{N}} \\
  &\; - \partial_{\,\nin{M}}h^{\,\nin{K}\bar{L}}\Big(a\, K_{2\,\nin{K}}{}^{\,\nin{P}}\partial_{\bar{N}} 
    K_{1\bar{L}\,\nin{P}}
  +b\, K_{2\bar{L}}{}^{\bar{P}}\partial_{\bar{N}} K_{1\,\nin{K}\bar{P}}\Big)\\
  &\; +\partial_{\bar{N}}h^{\,\nin{K}\bar{L}}\Big(
  a\, K_{2\,\nin{K}}{}^{\,\nin{P}}\partial_{\,\nin{M}}
    K_{1\bar{L}\,\nin{P}}
  +b\, K_{2\bar{L}}{}^{\bar{P}}\partial_{\,\nin{M}} K_{1\,\nin{K}\bar{P}}\Big) \;. \end{split}
 \ee 
The final two lines are also rewritable as total $\delta^{[0]}$ variations: 
\be\label{zwischenresultat454}
 \begin{split}
   {\cal E}_{\,\nin{M}\bar{N}} \ = \ &\,\delta_{\xi_2}^{[0]}\big(-\tfrac{1}{2} a\, 
  h_{\,\nin{M}}{}^{\bar{P}}\partial_{\bar{P}}K_1^{\,\nin{K}\,\nin{L}}\,
  \partial_{\,\nin{K}} h_{\,\nin{L}\bar{N}} -\tfrac{1}{2}b\,  h^{\,\nin{P}}{}_{\bar{N}} \,\partial_{\,\nin{P}} K_1^{\bar{K}\bar{L}}\,
  \partial_{\bar{K}} h_{\,\nin{M}\bar{L}} \\[0.5ex]
  &\qquad -\tfrac{1}{2}a\, \partial_{\,\nin{M}}K_1^{\,\nin{K}\,\nin{L}}\, h_{\,\nin{K}}{}^{\bar{P}}\,
  \partial_{\bar{P}} h_{\,\nin{L}\bar{N}}
  -\tfrac{1}{2}b\,  \partial_{\bar{N}}K_1^{\bar{K}\bar{L}}\, 
  h^{\,\nin{P}}{}_{\bar{K}} \,\partial_{\,\nin{P}} h_{\,\nin{M}\bar{L}} \\[0.5ex]
  &\qquad 
  -\tfrac{1}{2}a\, \partial_{\,\nin{M}} K_1^{\,\nin{K}\bar{P}}\,h^{\,\nin{L}}{}_{\bar{P}}
  \,\Gamma_{\bar{N}\,\nin{K}\,\nin{L}} + \tfrac{1}{2}b\, 
  \partial_{\bar{N}}K_1^{\,\nin{P}\bar{K}} \, h_{\,\nin{P}}{}^{\bar{L}}\,
  \Gamma_{\,\nin{M}\bar{K}\bar{L}}\big)
  \\[1.5ex]
    &\,+ \delta_{\xi_{12}^{[1](1)}}^{[0]}h_{\,\nin{M}\bar{N}} \\[1ex]   &\; +
   \delta^{[0]}_{\xi_2}\big(-\tfrac{a}{2}\partial_{\,\nin{M}} h^{\,\nin{P}}{}_{\bar{K}}\,\partial_{\bar{N}}h^{\,\nin{Q}\bar{K}}
   K_{1\,\nin{P}\,\nin{Q}} 
   +\tfrac{b}{2}
   \partial_{\,\nin{M}} h_{\,\nin{K}}{}^{\bar{P}}\,\partial_{\bar{N}} h^{\,\nin{K}\bar{Q}} K_{1\bar{P}\bar{Q}}\big)  \;, 
  \end{split}
 \ee 
as one may verify by a direct computation. 
Thus, we have succeeded in rewriting ${\cal E}_{\,\nin{M}\bar{N}}$ as a total $\delta^{[0]}$ variation 
and a modification of the gauge algebra, where we recall again the the $(1\leftrightarrow 2)$
antisymmetrization is left implicit.

Summarizing, we have established closure for 
  \be\label{FINALGAUGE}
  \begin{split}
   \delta_{\xi}^{[2](1)} h_{\,\nin{M}\bar{N}} \ = \ & \, -\tfrac{1}{4} a\, 
  h_{\,\nin{M}}{}^{\bar{P}}\partial_{\bar{P}}K^{\,\nin{K}\,\nin{L}}\,
  \Gamma_{\bar{N}\,\nin{K}\,\nin{L}}
  -\tfrac{1}{4} b\, h^{\,\nin{P}}{}_{\bar{N}} \,\partial_{\,\nin{P}} K^{\bar{K}\bar{L}}\,
  \Gamma_{\,\nin{M}\bar{K}\bar{L}} \\[0.5ex]
  &\, -\tfrac{1}{2}a\, \partial_{\,\nin{M}}K^{\,\nin{K}\,\nin{L}}\, h_{\,\nin{K}}{}^{\bar{P}}\,
  \partial_{\bar{P}} h_{\,\nin{L}\bar{N}}
  -\tfrac{1}{2}b\,  
  \partial_{\bar{N}}K^{\bar{K}\bar{L}}\, h^{\,\nin{P}}{}_{\bar{K}} \,\partial_{\,\nin{P}} h_{\,\nin{M}\bar{L}}\\[0.5ex]
  & \, 
  -\tfrac{1}{2}a\, \partial_{\,\nin{M}} K^{\,\nin{K}\bar{P}}\,h^{\,\nin{L}}{}_{\bar{P}}
  \,\Gamma_{\bar{N}\,\nin{K}\,\nin{L}} + \tfrac{1}{2}b\, \partial_{\bar{N}}K^{\,\nin{P}\bar{K}} \, h_{\,\nin{P}}{}^{\bar{L}}\,
  \Gamma_{\,\nin{M}\bar{K}\bar{L}}\\[0.5ex]
  &\, -\tfrac{1}{2}a\, \partial_{\,\nin{M}} h^{\,\nin{P}}{}_{\bar{K}}\,\partial_{\bar{N}}h^{\,\nin{Q}\bar{K}}
   K_{\,\nin{P}\,\nin{Q}} 
   +\tfrac{1}{2}b\, 
   \partial_{\,\nin{M}} h_{\,\nin{K}}{}^{\bar{P}}\,\partial_{\bar{N}} h^{\,\nin{K}\bar{Q}} K_{\bar{P}\bar{Q}}\;, 
 \end{split}
 \ee
using the linearized connections (\ref{lincoNN}), with the following modification of the gauge algebra 
  \be\label{Finalalgebra} 
 \begin{split}
  \xi_{12}^{M[1](1)} \ = \ &\, \tfrac{1}{4}a\,
   h^{\,\nin{K}\bar{L}} K_{2\,\nin{K}}{}^{\,\nin{P}}\partial^M K_{1\bar{L}\,\nin{P}}
  +\tfrac{1}{4} b\, h^{\,\nin{K}\bar{L}} K_{2\,\nin{K}}{}^{\bar{P}}\partial^M K_{1\bar{L}\bar{P}} \\
  &\, +\tfrac{1}{4} a\, h^{\,\nin{L}\bar{K}} K_{2\bar{K}}{}^{\,\nin{P}}\partial^M K_{1\,\nin{L}\,\nin{P}}
  +\tfrac{1}{4} b\, h^{\,\nin{L}\bar{K}}K_{2\bar{K}}{}^{\bar{P}}\partial^M K_{1\,\nin{L}\bar{P}}\\
  &\, -\tfrac{1}{4}a \,\partial^Mh^{\,\nin{K}\bar{L}}\, K_{2\,\nin{K}}{}^{\,\nin{P}} K_{1\bar{L}\,\nin{P}}
  +\tfrac{1}{4}b  \, \partial^Mh^{\,\nin{K}\bar{L}}\, K_{2\,\nin{K}}{}^{\bar{P}} K_{1\bar{L}\bar{P}}\;.
 \end{split} 
 \ee 
We have thus shown that the gauge structure can be extended to second order in perturbation 
theory of arbitrary choices for the parameters $a, b$. 
For our discussion in the next subsection, we need the special case 
$a=b=1$ corresponding to bosonic string theory:   
   \be\label{DFT+2algebra}
 \begin{split}
  \xi_{12}^{M[1](1)} \ = \ &\, \tfrac{1}{4}\, h^{\,\nin{K}\bar{L}} K_{2\,\nin{K}}{}^{P}\partial^M K_{1\bar{L}P}
  +\tfrac{1}{4}\, h^{\,\nin{L}\bar{K}} K_{2\bar{K}}{}^{P}\partial^M K_{1\,\nin{L}P}\\
  &\, -\tfrac{1}{4}\,\partial^Mh^{\,\nin{K}\bar{L}}\, K_{2\,\nin{K}}{}^{\,\nin{P}} K_{1\bar{L}\,\nin{P}}
  +\tfrac{1}{4} \, \partial^Mh^{\,\nin{K}\bar{L}}\, K_{2\,\nin{K}}{}^{\bar{P}} K_{1\bar{L}\bar{P}}
 \;, 
 \end{split}
 \ee 
where in the first line we combined terms with summations over projected indices into 
terms with summation over unprojected indices. 

It should be emphasized that the above expression for the gauge algebra is only well-defined 
up to parameter redefinitions and the addition of trivial parameters of the form $\xi^M=\partial^M\chi$, 
as these do not affect the gauge transformations. Therefore, it is not yet evident that the above 
\textit{field-dependent} deformation of the algebra is non-trivial and cannot be removed by 
such redefinitions. Indeed, for the special case $a=-b$ the algebra (\ref{Finalalgebra}) is equivalent to the 
field-independent gauge algebra of the HSZ theory constructed in \cite{Hohm:2013jaa}. 
In contrast, for $a=b$ the above deformation is not removable, as will be confirmed in subsequent sections. 
The expression (\ref{DFT+2algebra}) is such that it can be promoted to a 
background independent candidate formula in terms of the generalized metric, to which we turn 
in the next subsection.

\subsection{Candidate gauge algebra in terms of generalized metric} 

We now aim to constrain the gauge structure of a putative background independent generalized metric 
formulation of bosonic string theory to first order in $\alpha'$, using the above perturbative results. 
Given that the gauge algebra is intrinsically field dependent (in the sense that the field dependence 
cannot be removed by parameter redefinitions and/or the addition of trivial parameters), 
it follows that the gauge algebra for a generalized metric formulation must be field dependent, too. 
We will show that there is a unique expression that is compatible with the above result  (\ref{DFT+2algebra}). 

To this end, let us discuss the expansion of a generalized metric around a constant background 
in a little more detail than in (\ref{expandHAAA}). Specifically, including higher order terms in the expansion, 
we have 
 \be\label{fullHexpan}
 \begin{split}
  {\cal H}_{MN} \ = \ \bar{\cal H}_{MN} &+ h_{\,\nin{M}\bar{N}}+ h_{\,\nin{N}\bar{M}}
  -\tfrac{1}{2} h^{\,\nin{K}}{}_{\bar{M}}\, h_{\,\nin{K}\bar{N}} + \tfrac{1}{2} h_{\,\nin{M}}{}^{\bar{K}}\,
  h_{\,\nin{N}\bar{K}}\\
  &-\tfrac{1}{8} h^{\,\nin{K}}{}_{\bar{M}}\, h_{\,\nin{K}}{}^{\bar{L}} \,h^{\,\nin{P}}{}_{\bar{L}} \,h_{\,\nin{P}\bar{N}}
  +\tfrac{1}{8} h_{\,\nin{M}}{}^{\bar{K}}\, h^{\,\nin{L}}{}_{\bar{K}}\, h_{\,\nin{L}}{}^{\bar{P}}\, h_{\,\nin{N}\bar{P}}
  \  + \  {\cal O}(h^6)\;, 
 \end{split}
 \ee 
in terms of the independent fluctuation $h_{\,\nin{M}\bar{N}}$. 
This expansion is such that the generalized metric satisfies the constraints 
${\cal H}_M{}^{K}{\cal H}_{KN}=\eta_{MN}$ and $\eta^{MN}{\cal H}_{MN}=0$. 
Note that there are no cubic or quintic terms and that all higher-order terms in $h$ carry the `diagonal' 
index projections $\nin{M} \nin{N}$ or $\bar{M}\bar{N}$. Without loss of generality,  
this can be assumed for the complete series expansion. Indeed, if we had a term of odd power in $h$, 
it is easy to see that it would have the index structure $\nin{M}\bar{N}$ or $\nin{N}\bar{M}$ and hence 
be removable by a 
field redefinition of $h_{\,\nin{M}\bar{N}}$. However, below we will encounter 
field variables that are such redefinitions of $h_{\,\nin{M}\bar{N}}$. 

This observation also implies that there is no subtlety when translating the closure condition 
on the full ${\cal H}_{MN}$, which has an infinite series expansion, to the closure condition on $h_{\,\nin{M}\bar{N}}$. 
Indeed, contracting (\ref{fullHexpan}) with the constant background projectors $P$ and $\bar{P}$, we obtain 
  \be
  h_{\,\nin{M}\bar{N}} \ = \  P_{M}{}^{K} \,   \bar{P}_{N}{}^{L} \,  {\cal H}_{KL}\;. 
 \ee  
Thus, the closure condition $[\delta_{\xi_1},\delta_{\xi_2}]{\cal H}_{MN}=\delta_{\xi_{12}}{\cal H}_{MN}$ 
immediately yields the closure condition on the fluctuation field, 
$[\delta_{\xi_1},\delta_{\xi_2}]h_{\nin{M}\bar{N}}=\delta_{\xi_{12}}h_{\nin{M}\bar{N}}$.

Our goal is now to find a candidate expression for the gauge algebra in terms of ${\cal H}$ that reduces 
to the perturbative result (\ref{DFT+2algebra}) upon expanding according to (\ref{fullHexpan}). 
For the terms of the structural form $h K \partial K$ there is nothing left to do, for these 
terms have already been obtained from the background independent (\ref{HKKalgebra}).   
For the terms of the form $\partial h\,K_2\,K_1$ the following structure 
reproduces the right terms:  
 \be\label{secondHansatz}
 \begin{split}
   {\cal H}^{K}{}_{R}\,\partial^M {\cal H}^{RL}\,{\cal H}^{PQ}\,K_{2KP}\,K_{1LQ} \ = \ &\;
  (\bar{P}-P)^K{}_{R}\,\partial^M(h^{\,\nin{R}\bar{L}}+h^{\,\nin{L}\bar{R}})
  (\bar{P}-P)^{PQ} K_{2KP} \,K_{1LQ} \\
  \ = \ &\; (\partial^Mh^{\,\nin{L}\bar{K}}-\partial^M h^{\,\nin{K}\bar{L}})
  (K_{2K}{}^{\bar{P}} \,K_{1L\bar{P}}-K_{2K}{}^{\,\nin{P}}\, K_{1L\,\nin{P}}) \\
    \ = \ &\;  -2\, \partial^M h^{\,\nin{K}\bar{L}}K_{2\,\nin{K}}{}^{\bar{P}} \,K_{1\bar{L}\bar{P}}
  \ + \ 2\, \partial^M h^{\,\nin{K}\bar{L}}K_{2\,\nin{K}}{}^{\,\nin{P}}\, K_{1\bar{L}\,\nin{P}}\;, 
 \end{split}
 \ee  
where we used the (implicit) antisymmetry in $(1\leftrightarrow 2)$.  
Comparing with (\ref{DFT+2algebra}) and recalling (\ref{HKKalgebra}) we 
conclude that the following expression is compatible with the perturbative gauge algebra, 
  \be\label{DFT+algebraFULL}
  \xi_{12}^{(1) M} \ = \ \tfrac{1}{4}\,{\cal H}^{KL} K_{2 K}{}^{P}\partial^M K_{1LP}
  \ - \ \tfrac{1}{8}\, {\cal H}^{K}{}_{R}\,\partial^M {\cal H}^{RL}\,{\cal H}^{PQ}\,K_{2KP}\,K_{1LQ} 
  -(1\leftrightarrow 2)\;. 
 \ee  
 
We close this section by arguing that this is the unique expression consistent with 
the perturbative gauge structure. 
First note that the expressions 
here are such that `integrating by parts' with $\partial^M$, using the freedom of adding trivial parameters of the 
form $\partial^M\chi$, does not allow us to change the form of the algebra. 
This follows from the antisymmetry in $(1\leftrightarrow 2)$ and the constraint on ${\cal H}$. 
Thus, the above is the unique writing of these algebra terms. 

Could one add further terms to (\ref{DFT+algebraFULL}) that do not contribute to 
second order perturbation theory but that may be relevant to yet higher order? 
To address this question, let us state our assumptions more explicitly: 
First, we recall that the gauge transformations of the dilaton did not receive 
$\alpha'$ corrections to first and second order in perturbation theory and that the
gauge algebra is independent of the dilaton. Moreover, 
the same holds in the exact construction of \cite{Hohm:2013jaa}. 
We hence assume that the dilaton gauge transformations remain undeformed and that 
the gauge algebra does not depend on the dilaton to all orders, which will be confirmed 
independently in later sections. 
This then implies that the expression for $\xi_{12}^{M}$ must carry the free index on 
a derivative, so that the undeformed dilaton gauge transformation (\ref{dilatonvar}) is consistent 
with a deformed gauge algebra 
by the strong constraint. 
Similarly, we assume that the form of the trivial parameter is not 
$\alpha'$-corrected and thus still of the form $\xi^M=\partial^M\chi$. 
We therefore demand that the gauge algebra expression 
vanishes when $\xi_1$ and $\xi_2$ are trivial in this sense. This is satisfied if that expression is 
written in terms of $K_{MN}(\xi_{1,2})=2\partial_{[M}\xi_{1,2 N]}$.\footnote{Strictly speaking, by the strong constraint 
this condition is also satisfied for an algebra expression containing a bare gauge parameter 
in the form $\xi_{1,2}^N\partial_N$.  
The perturbative Noether procedure makes it fairly clear, however, that bare gauge parameters do not arise at order $\alpha'$. Therefore, we will assume that the 
gauge algebra does not contain bare gauge parameters. Again, this will be confirmed independently 
by the results in sec.~4.}

To classify such terms it is helpful to recall that the bosonic string has the $\mathbb{Z}_2$ 
symmetry sending $b$ to $-b$. As reviewed in \cite{Hohm:2014xsa}, this symmetry 
is realized in $O(D,D)$ invariant expressions iff the number of $\eta$ used to contract indices is 
even. The terms in (\ref{DFT+algebraFULL}), which have one ${\cal H}$ and three ${\cal H}$,  
are $\mathbb{Z}_2$ even, and it is easy to see by inspection that 
they are the unique $\mathbb{Z}_2$ invariant structures with these numbers of ${\cal H}$.
Could we write  a term with five ${\cal H}$? Such a structure would take the schematic form 
 \be\label{ansatzALG}
  \xi_{12}^M \ \sim  \ {\cal H}^{\bullet\bullet}\ {\cal H}^{\bullet\bullet}\ {\cal H}^{\bullet\bullet}\ {\cal H}^{\bullet\bullet}
  \ \partial^M{\cal H}^{\bullet\bullet}\, K_{2\bullet\bullet}\, K_{1\bullet\bullet}\;. 
 \ee 
Let us first verify that this ansatz is $\mathbb{Z}_2$ even. 
The $K_2 K_1$ term is $\mathbb{Z}_2$ even, and we need to contract four lower and 
ten upper indices, which requires three $\eta$. Together with the 
single $\eta$  contained in $\partial^M$, this indeed leads to a $\mathbb{Z}_2$ even term. 
Note that a single term of the structural form (\ref{ansatzALG}) would contribute to first order in fields 
and thus be inconsistent with the perturbative results, but one might imagine that there is a linear 
combination of several such terms that do not contribute to first order. 
However, it is easy to see that no term of the above structure exists.  
We first note that we cannot write a term with contracted indices on two bare ${\cal H}$, 
because by the constraint on ${\cal H}$ that would give a 
Kronecker delta, reducing to the previous case of three ${\cal H}$. 
It is impossible, however, to write a term without such a contraction, because 
even contracting the two indices on $\partial^M{\cal H}^{\bullet\bullet}$ with two indices on bare ${\cal H}$, 
we still have six indices left on bare ${\cal H}$ fields that can only be contracted with the four indices on 
$K_2K_1$. 
Similarly, it is easy to see that an algebra ansatz with the derivative on one of the $K$ factors does not exist. 
We conclude that there are no further terms that we could add to the gauge algebra, 
and therefore (\ref{DFT+algebraFULL}) is the unique candidate expression for the full background independent 
gauge algebra in terms of the generalized metric.

\section{No-go theorem for generalized metric formulation of bosonic string theory}\setcounter{equation}{0}
In this section we prove that the unique candidate expression for the gauge algebra does \textit{not} in fact 
define a consistent gauge algebra to first order in $\alpha'$.\footnote{The results of this section were partly 
obtained in collaboration with Ashoke Sen and Barton Zwiebach.} 
To this end we prove that the expression is not compatible with 
the Jacobi identity for the putative gauge variations $\delta_{\xi}$. 
This is a necessary consistency condition that, as we will see, can be tested using only the gauge transformations 
to zeroth order in $\alpha'$.

\subsection{Consistency conditions from Jacobi identity}
We begin by deriving consistency conditions from the Jacobi identity for the gauge variations. 
Consider fields $\phi_i$ subject to a gauge symmetry parameterized by $\xi$, 
 \be
  \delta_\xi   \phi_i  \ = \  R_i ( \xi | \phi )\;, 
 \ee
where $R_i$ is linear in $\xi$. We assume off-shell closure,\footnote{This is sufficient for the discussion in this 
paper, because if we had a generalized metric 
formulation with only on-shell closure, this would imply only on-shell closure for the perturbative transformations,
in contradiction with the off-shell closure found in sec.~2.}
so that 
  \be\label{Fclosure}
   \bigl[ \,\delta_{\xi_1} \,, \, \delta_{\xi_2} \, \bigr] 
   \ = \ \delta_{F(\xi_1, \xi_2 ; \phi) } \;,  
 \ee   
for some function $F$ that is antisymmetric in its first two arguments and that we allow to be field-dependent. 
Writing out the variations on the left-hand-side, this implies  
\be\label{explicitcomm}
  {\delta R_i (\xi_2 | \phi) \over \delta\phi_k}  \, R_k (\xi_1| \phi) 
  \  -  \ {\delta R_i (\xi_1 | \phi) \over \delta\phi_k}  \, R_k (\xi_2| \phi)  
  \ = \  R_i ( F(\xi_1, \xi_2;\phi) | \phi) \;. 
\ee
Any infinitesimal symmetry variations must satisfy the Jacobi identity,  
\be
\sum_{\rm cyclic}  \big[ \, \delta_{\xi_1} , \, \big[ \, \delta_{\xi_2} , \, \delta_{\xi_3}\, \big]\, \big] \ = \ 0 \,, 
\ee
which yields immediately with (\ref{Fclosure}), 
 \be\label{Jacob}
\sum_{\rm cyclic}  \big[ \, \delta_{\xi_1} , \, \delta_{F(\xi_2, \xi_3 ; \phi) } \, \big]   \ = \ 0 \,. 
\ee
Let us work out this commutator, using the linearity of $R_i$ in its first argument, 
 \be
 \begin{split}
  &\big[ \, \delta_{\xi_1} , \, \delta_{F(\xi_2, \xi_3 ; \phi) } \, \big] \phi_i \ = \
  \delta_{\xi_1} R_{i}(F(\xi_2,\xi_3;\phi)|\phi) - \delta_{F(\xi_2,\xi_3;\phi)}R_i(\xi_1|\phi) \\[0.5ex]
  &\; \ = \ R_i(\delta_{\xi_1}F(\xi_2,\xi_3;\phi)|\phi) 
    +      {\delta R_i (F(\xi_2,\xi_3;\phi) | \phi) \over \delta\phi_k}  \, R_k (\xi_1| \phi)
     -   {\delta R_i (\xi_1 | \phi) \over \delta\phi_k}  \, R_k (F(\xi_2,\xi_3;\phi) | \phi)  \\[0.5ex] 
   &\; \ = \   R_i(\delta_{\xi_1}F(\xi_2,\xi_3;\phi)|\phi)
   + R_i(F(\xi_1,F(\xi_2,\xi_3;\phi);\phi)|\phi)\;, 
  \end{split}
  \ee 
where we used (\ref{explicitcomm}) in the last step. Introducing the notation $\delta_\xi = \delta (\xi)$ for better readability, 
we have thus shown: 
 \be
   \big[ \, \delta_{\xi_1} \,, \, \delta_{F(\xi_2, \xi_3 ; \phi) } \, \big] \ = \  
   \delta( \delta_{\xi_1} \hskip-1pt F (\xi_2, \xi_3;\phi)  
 \phantom{\bigl(} +\    F (\xi_1,  F (\xi_2, \xi_3;\phi) ; \phi )  )\;. 
 \ee
With (\ref{Jacob}) it follows that under the cyclic sum the effective gauge parameter 
appearing here must act trivially on fields, i.e., 
 \be
\label{vmgg}
 \sum_{\rm cyclic} \delta_{\xi_1} \hskip-1pt F (\xi_2, \xi_3;\phi)  
  \  + \    F (\xi_1,  F (\xi_2, \xi_3;\phi) ; \phi )  \ = \ \hbox{trivial gauge parameter}\;. 
\ee  
This is a necessary consistency condition for any field-dependent gauge algebra $F(\xi_1,\xi_2,\phi)$. 

Next, we specialize this condition to the generalized metric expression and perform an $\alpha'$ expansion 
(recalling our notation that round parenthesis indicate the powers in $\alpha'$),  
 \be\label{JACOBIATOR}
 \begin{split}
  J(\xi_1,\xi_2,\xi_3) \ &\equiv \  \sum_{\rm cycl}\delta_{\xi_1}^{(0)}F^{(1)}(\xi_2,\xi_3;{\cal H})+
  F^{(0)}(\xi_1,F^{(1)}(\xi_2,\xi_3;{\cal H}))
  +F^{(1)}(\xi_1,F^{(0)}(\xi_2,\xi_3);{\cal H}) \\
  \ &= \  \hbox{trivial gauge parameter}\;, 
 \end{split}
 \ee
where we used that $F^{(0)}$ is field independent and hence does not transform under
gauge transformations. We also recall that a trivial parameter is still a total $\partial^M$ derivative. 
Note that the Jacobiator can be computed from the candidate expression $F$ for the gauge algebra,
using only the gauge transformations to zeroth order in $\alpha'$.

\subsection{Proof that the Jacobiator is non-trivial}

We now compute the Jacobiator (\ref{JACOBIATOR}) for the unique candidate expression (\ref{DFT+algebraFULL}) 
for the gauge algebra and show that it is non-trivial. Recalling also the zeroth-order C-bracket part we have 
 \be
  \begin{split}
   F^{(0)}(\xi_1,\xi_2)^M \ &= \ \big[\xi_2,\xi_1\big]_{c}^M\;, \\
   F^{(1)}(\xi_1,\xi_2;{\cal H})^M \ &= \
   \tfrac{1}{4}\,{\cal H}^{KL} K_{2 K}{}^{P}\partial^M K_{1LP}\\
  &\quad\;  -  \tfrac{1}{8}\, {\cal H}^{K}{}_{R}\,\partial^M {\cal H}^{RL}\,{\cal H}^{PQ}\,K_{2KP}\,K_{1LQ} 
  -(1\leftrightarrow 2) \;. 
  \end{split}
 \ee  
It is convenient to organize the computation in terms of `covariant' contributions given by the generalized 
Lie derivative ${\cal L}_{\xi}$, plus additional `non-covariant' terms. One uses, for instance, that the 
gauge transformation of $\partial {\cal H}$ reads 
 \be
  \delta_{\xi}^{(0)}(\partial_M{\cal H}_{KL}) \ = \ {\cal L}_{\xi}(\partial_M{\cal H}_{KL}) 
  + \partial_M K_K{}^{P} \, {\cal H}_{PL}+ \partial_M K_L{}^{P} \, {\cal H}_{KP}\;, 
 \ee   
as can be quickly verified with (\ref{fullHGauge}).  
We can then compute the first term on the r.h.s.~of (\ref{JACOBIATOR}), leaving here 
and in the rest of this section the cyclic sum implicit, 
 \be
 \begin{split}
  \delta_{\xi_1}^{(0)}F^{(1)}(\xi_2,\xi_3;{\cal H})^M \ = \  & 
  \tfrac{1}{2}\,({\cal L}_{\xi_1}{\cal H}^{KL}) K_{3 K}{}^{P}\partial^M K_{2LP}
    -  \tfrac{1}{4}\, {\cal L}_{\xi_1}({\cal H}^{K}{}_{R}\,\partial^M {\cal H}^{RL}\,{\cal H}^{PQ})\,K_{3KP}\,K_{2LQ} \\
    & +\tfrac{1}{4}\,\partial^MK_1^{KL}\,{\cal H}^{PQ} K_{3KP}\, K_{2LQ}
    -\tfrac{1}{4}\,{\cal H}^{KR}{\cal H}^{LS}\partial^MK_{1RS}\,{\cal H}^{PQ}\,K_{3KP}\, K_{2LQ}\,,
  \end{split}  
 \ee
where the terms in the second line come from the non-covariant variation of $\partial\H$ and we used the constraint 
on ${\cal H}$.  
Using the cyclic sum and relabeling indices, this can be rewritten as 
\be\label{delta0F1}
 \begin{split}
  &\delta_{\xi_1}^{(0)}F^{(1)}(\xi_2,\xi_3;{\cal H})^M \ = \ 
  \tfrac{1}{2}\,({\cal L}_{\xi_1}{\cal H}^{KL}) K_{3 K}{}^{P}\partial^M K_{2LP}
    -  \tfrac{1}{4}\, {\cal L}_{\xi_1}({\cal H}^{K}{}_{R}\,\partial^M {\cal H}^{RL}\,{\cal H}^{PQ})\,K_{3KP}\,K_{2LQ} \\
    &\qquad  +\tfrac{1}{4}\,{\cal H}^{KL}\, \partial^MK_3^{PQ}\, K_{2KP}\, K_{1LQ}
    -\tfrac{1}{12}\,{\cal H}^{KR}{\cal H}^{LS}{\cal H}^{PQ} \partial^M\big(K_{1RS}\,K_{2LQ}\, K_{3KP}\big)\;. 
  \end{split}  
 \ee
Next, we compute the second term on the r.h.s.~of (\ref{JACOBIATOR}), 
 \be\label{F0ofF1}
 \begin{split}
  F^{(0)}(\xi_1,F^{(1)}(\xi_2,\xi_3;{\cal H}))^M \ &= \ -\big[\xi_1,F^{(1)}(\xi_2,\xi_3;\H)\big]_{C}^M \\
  \ &= \ -{\cal L}_{\xi_1}F^{(1)}(\xi_2,\xi_3;\H)^M
  +\tfrac{1}{2}\,\partial^M\big(\xi_1^NF^{(1)}(\xi_2,\xi_3;\H)_N\big)\;, 
 \end{split}
 \ee 
using the relation (\ref{DCrelation}) between the C-bracket and the generalized Lie derivative. 
Note that this relation holds for arbitrary $F^{(1)}$. 
Finally, for the third term in (\ref{JACOBIATOR}) we compute 
 \be
  \begin{split}
   &F^{(1)}\big(\xi_1, F^{(0)}\big(\xi_2,\xi_3\big);\H\big) \ = \ \tfrac{1}{4}\,{\cal H}^{KL}K([\xi_3,\xi_2]_C)_K{}^{P}
   \partial^MK_{1LP} \\
   &\quad \, -\tfrac{1}{4}\,\H^{KL}K_{1K}{}^{P}\partial^M K([\xi_3,\xi_2]_C)_{LP} 
    -\tfrac{1}{4}\,{\cal H}^{K}{}_{R}\,\partial^M\H^{LR} \H^{PQ}K([\xi_3,\xi_2]_C)_{KP}K_{1LQ}\;, 
  \end{split}
 \ee   
using in the second line that the $\partial \H$ part of $F^{(1)}$ is automatically antisymmetric in its two arguments. 
Next we use the two relations (\ref{MASTER1}), (\ref{MASTER2}), which also imply 
\be 
 K([\xi_3,\xi_2]_c)_K{}^{P} \ = \ {\cal L}_{\xi_3}K_{2K}{}^{P} + K_{2K}{}^{Q} K_{3Q}{}^{P} 
 -(2\leftrightarrow 3)\;, 
\ee 
to find after a short computation 
 \be
 \begin{split}
   F^{(1)}\big(\xi_1, F^{(0)}\big(\xi_2,\xi_3\big);\H\big) \ = \  &\,
   \tfrac{1}{2}\,\H^{KL}{\cal L}_{\xi_3}(K_{2K}{}^{P}\,\partial^MK_{1LP}) 
   -\tfrac{1}{2}\,\H^{KL}K_{3}^{PQ} K_{2KQ}\,\partial^MK_{1LP}\\
   &\, -\tfrac{1}{4}\,{\cal H}^{K}{}_{R}\,\partial^M{\cal H}^{LR} {\cal H}^{PQ}{\cal L}_{\xi_3}\big(K_{2KP} K_{1LQ}\big)\\
   &\, +\tfrac{1}{2}\,{\cal H}^{K}{}_{R}\,\partial^M{\cal H}^{LR} {\cal H}^{PQ} K_{3P}{}^{S} K_{2KS} K_{1LQ}\;. 
  \end{split}
  \ee 
Combining this with (\ref{delta0F1}), the generalized Lie derivative terms add up to the Lie derivative 
of $F^{(1)}$, and we obtain  
 \be
 \begin{split}
  &\delta_{\xi_1}^{(0)}F^{(1)}(\xi_2,\xi_3;{\cal H})^M+F^{(1)}\big(\xi_1, F^{(0)}\big(\xi_2,\xi_3\big);\H\big) \\
  & \ = \ {\cal L}_{\xi_1}F^{(1)}(\xi_2,\xi_3;\H)^M 
  +\tfrac{1}{4}\,{\cal H}^{KL}\, \partial^MK_3^{PQ}\, K_{2KP}\, K_{1LQ} \\
   &\qquad  -\tfrac{1}{12}\,{\cal H}^{KR}{\cal H}^{LS}{\cal H}^{PQ} \partial^M\big(K_{1RS}\,K_{2LQ}\, K_{3KP}\big)\\
  &\qquad   -\tfrac{1}{2}\,\H^{KL}K_{3}^{PQ} K_{2KQ}\,\partial^MK_{1LP}
  +\tfrac{1}{2}\,{\cal H}^{K}{}_{R}\,\partial^M{\cal H}^{LR} {\cal H}^{PQ} K_{3P}{}^{S} K_{2KS} K_{1LQ}\;. 
 \end{split}
 \ee
Finally, combining this with (\ref{F0ofF1}), the generalized Lie derivatives cancel, and we get for the Jacobiator 
 \be
  \begin{split}
    J^M \ &= \   \tfrac{1}{2}\,\partial^M\big(\xi_1^NF^{(1)}(\xi_2,\xi_3;\H)_N\big) \\
  &\qquad +\tfrac{1}{4}\, {\cal H}^{KL}\,\partial^M\big(K_3{}^{PQ}K_{2KP} K_{1LQ}\big)
  -\tfrac{1}{12}\,{\cal H}^{KR}{\cal H}^{LS}{\cal H}^{PQ}\, \partial^M\big(K_{1RS}\,K_{2LQ}\, K_{3KP}\big)\\
  &\qquad +\tfrac{1}{2}\,{\cal H}^{K}{}_{R}\,\partial^M{\cal H}^{LR} {\cal H}^{PQ} K_{3P}{}^{S} K_{2KS} K_{1LQ}\;, 
 \end{split}
 \ee 
where we used that the  $ KK\partial K$ structures combine into a total derivative under the cyclic sum. 
It is convenient 
to slightly rewrite this result by using the freedom to add trivial parameters and hence  `integrating by parts', 
\be\label{FINALJACOB}
  \begin{split}
    J^M \ &= \ \partial^M\big(\, \tfrac{1}{2}\,\xi_1^NF^{(1)}(\xi_2,\xi_3;\H)_N
      +  \tfrac{1}{4}\, {\cal H}^{KL}\,K_3{}^{PQ}K_{2KP} K_{1LQ} \\
       &\qquad \qquad -\tfrac{1}{12}\,{\cal H}^{KR}{\cal H}^{LS}{\cal H}^{PQ}\, K_{1RS}\,K_{2LQ}\, K_{3KP}\big)\\[0.5ex]
       &\qquad -\tfrac{1}{4}\, \partial^M{\cal H}^{KL}\,K_3{}^{PQ}K_{2KP} K_{1LQ}
       +\tfrac{1}{4}\,\partial^M{\cal H}^{KR}\, \H^{LS}\H^{PQ}\, K_{1RS}\,K_{2LQ}\, K_{3KP}\\
       &\qquad +\tfrac{1}{2}\,{\cal H}^{K}{}_{R}\,\partial^M{\cal H}^{LR} {\cal H}^{PQ} K_{3P}{}^{S} K_{2KS} K_{1LQ}\;. 
  \end{split}
 \ee    
The terms in the first and second line are total derivatives and hence trivial parameters, but 
the terms in the third and fourth line are not. 
Indeed, these terms are not secretly zero as can be confirmed by expanding around a background 
to first order in $h$, which yields 
 \be
  J^M \ =  \ -2\,\partial^Mh^{\,\nin{K}\bar{L}}\, K_{3}^{\bar{P}\,\nin{Q}}\, K_{2\,\nin{K}\bar{P}}\,K_{1\bar{L}\,\nin{Q}}
  +{\cal O}(h^2)\;, 
 \ee
where we dropped the trivial total derivative terms.   
Therefore, the Jacobi identity on fields is not satisfied and the candidate expression $F$ does not 
define a consistent gauge algebra. Since this was the unique expression for the gauge algebra in terms of the 
generalized metric that is consistent with the known perturbative result for bosonic string theory, 
it follows that there is no pure generalized metric formulation for the bosonic string, as we wanted to prove. 

It is reassuring to verify that the above no-go result is consistent with the result in second order 
perturbation theory established above. Since the candidate expression for the gauge algebra to 
that  order agrees with the gauge algebra that was found with the Noether procedure
and hence must be consistent, the failure of the Jacobi identity can only show up to third order in 
perturbation theory. 
To second order in perturbation theory, in the Jacobiator only the field independent 
terms enter. Such terms originate from the terms in the 
first and second line of (\ref{FINALJACOB}), upon keeping only constant background structures, and these 
terms are indeed trivial. In contrast, the terms in the third and fourth line 
all have $\H$ under the derivative, and so only enter to order $h$,  
corresponding to the Jacobiator to third order in 
perturbation theory.

\section{Background independent frame formulation}\setcounter{equation}{0}

In the previous section we have proved that a generalized metric formulation 
cannot give a background independent description of the general 
$\alpha'$ corrections in string theory. In this section it is shown that a frame formulation 
found by Marques and Nunez, which we clarify and slightly generalize to allow for deformed $GL(D)\times GL(D)$ 
frame transformations, provides the general background independent formulation to first  
order in $\alpha'$. After reviewing the conventional frame formalism to zeroth order in $\alpha'$,  
in the first subsection we give a self-contained discussion of the $\alpha'$-deformed 
frame formalism. In the second subsection we discuss the 
perturbation theory around flat backgrounds and show that it produces precisely, 
up to trivial parameter and field redefinitions, the results of second-order perturbation theory 
obtained above. As a consistency check we verify in the final subsection that the Jacobiator 
is consistent to third order in perturbation theory, showing explicitly that the obstacle 
found for the gauge algebra in terms of the generalized metric is circumvented in the 
frame formulation.

\subsection{$\alpha'$-deformed $GL(D)\times GL(D)$ frame formulation} 
We begin by reviewing the frame formalism with $GL(D)\times GL(D)$ local frame transformations 
to zeroth order in $\alpha'$ as developed by Siegel in \cite{Siegel:1993th}
and related to the original DFT in \cite{Hohm:2010xe}. 
(See also \cite{Hohm:2011si,Hohm:2012mf,Jeon:2011cn,Deser:2016qkw} 
for further investigations of the geometry of DFT.) 
The basic fields are the dilaton density and the frame field $E_{A}{}^{M}$, 
which is a vector under generalized diffeomorphisms, as in (\ref{fullHGauge}), 
and transforms under local frame transformations, 
 \be\label{localFRAME}
  \delta_{\Lambda}E_{A}{}^{M} \ = \ \Lambda_A{}^{B} E_{B}{}^{M}\;. 
 \ee
Here the flat index splits as $A=(a,\bar{a})$, where unbarred indices refer to the 
left $GL(D)$ and barred indices refer to the right $GL(D)$. Accordingly, the matrix 
$\Lambda_A{}^{B}$ is block-diagonal, with entries $\Lambda_a{}^b$ and $\Lambda_{\bar{a}}{}^{\bar{b}}$.  
The frame field is subject to the constraint that the 
tangent space metric, which is used to raise and lower flat indices and 
obtained by flattening the indices of the $O(D,D)$ metric,  is 
block-diagonal: 
  \be\label{TANGENTmetric}
  {\cal G}_{AB} \ \equiv \ E_{A}{}^{M} \,E_{B}{}^{N}\, \eta_{MN}  \ = \     \begin{pmatrix}    {\cal G}_{ab} & 0\\[0.5ex]
  0 & {\cal G}_{\bar{a}\bar{b}} \end{pmatrix} \;. 
 \ee 
This is a $GL(D)\times GL(D)$ covariant constraint.   

Next, let us briefly review the generalized connections for the local frame transformations. 
Working with the flattened partial derivative 
\be
 {\cal D}_{A} \ \equiv \ E_{A}{}^{M}\partial_M\;, 
\ee 
we may define covariant derivatives acting, say, on a vector, 
 \be
  \nabla_AV_B \ = \ {\cal D}_{A}V_{B}+\omega_{AB}{}^{C} V_{C}\;, \qquad
  \nabla_AV^B \ = \ {\cal D}_{A}V^B-\omega_{AC}{}^{B}V^C\;, 
\ee
with the generalized spin connection components $\omega_{AB}{}^{C}$ 
transforming as 
 \be\label{ConnTRans}
  \delta_{\Lambda}\omega_{AB}{}^{C} \ = \ -\nabla_A\Lambda_{B}{}^{C} + \Lambda_A{}^{D}\,\omega_{DB}{}^{C}\;. 
 \ee 
Not all connections can be determined 
in terms of the frame field and dilaton upon imposing covariant constraints. However, 
for our purposes here we only need to consider the following connection components 
 \be\label{connections}
  \omega_{a\bar{b}}{}^{\bar{c}} \ = \ -\Omega_{a\bar{b}}{}^{\bar{c}}\;, \qquad
   \omega_{\bar{a}b}{}^{c} \ = \ -\Omega_{\bar{a}b}{}^{c}\;, 
 \ee
which are uniquely determined in terms of the generalized coefficients of anholonomy, 
defined by 
  \be
  \big[ {E}_{A},{E}_{B}\big]_{c}^M \ \equiv \ \Omega_{AB}{}^{C}{E}_{C}{}^{M}\;, 
 \ee   
where the C-bracket (\ref{CBRacket}) is evaluated for the frame field viewed as a generalized vector.

We now turn to the $\alpha'$-deformed gauge transformations. 
In this we will generalize the construction in \cite{Marques:2015vua}
by enlarging the local frame transformations from a doubled local Lorentz group 
to a local $GL(D)\times GL(D)$ symmetry. 
The generalized diffeomorphisms are undeformed and given by generalized 
Lie derivatives as in (\ref{fullHGauge}), but  the local frame transformations 
are deformed. They take the same form as (\ref{localFRAME}), but with the off-diagonal entries of 
the transformation matrix being non-zero, expressed in terms of 
 higher derivatives of the diagonal components, 
 \be\label{DeformedMAtrix}
  \Lambda_A{}^{B} \  = \  \begin{pmatrix}    \Lambda_a{}^b & \Sigma_a{}^{\bar{b}}(\Lambda,E)\\[0.5ex]
  \Sigma_{\bar{a}}{}^{b}(\Lambda,E) & \Lambda_{\bar{a}}{}^{\bar{b}} \end{pmatrix}    \;, 
 \ee
where $\Sigma$ is defined in terms of derivatives of the gauge parameters $\Lambda_a{}^b$ and 
${\Lambda}_{\bar{a}}{}^{\bar{b}}$
and the generalized connections (\ref{connections}):
 \be\label{SigmaFORM}
  \Sigma_a{}^{\bar{b}} \ \equiv \ -\Sigma^{\bar{b}}{}_a \ \equiv \ 
   \tfrac{a}{2}\, \D_{a}\Lambda_c{}^{d}\,\omega^{\bar{b}}{}_{d}{}^{c}
   +\tfrac{b}{2}\, \D^{\bar{b}}\Lambda_{\bar{c}}{}^{\bar{d}}\,\omega_{a\bar{d}}{}^{\bar{c}}\;, 
 \ee
and flat indices are still raised and lowered with the tangent space metric 
(\ref{TANGENTmetric}). An important consistency condition is that the frame transformations 
of ${\cal G}_{AB}$ are undeformed, which is guaranteed due to $\Sigma_{a\bar{b}}+\Sigma_{\bar{b}a}=0$. 
Thus, it is still consistent to impose ${\cal G}_{a\bar{b}}=0$. 
Also note that these deformed transformations can be viewed as a generalization of the 
transformations required by the Green-Schwarz mechanism in that also here the local 
`Lorentz' transformations receive higher-derivative terms involving the Lorentz connections.  
In this paper we will not attempt to construct an invariant action, which would generalize the 
action given in \cite{Marques:2015vua}, as the gauge structure is already 
sufficient for our present purposes. 
 
Next, let us prove closure of these deformed frame transformations, which curiously  
requires a deformation both of the (generalized) diffeomorphism algebra and of the 
$GL(D)\times GL(D)$ algebra. To this end we need to use (\ref{ConnTRans}) to compute the 
local frame transformation of $\Sigma_{a}{}^{\bar{b}}$. Setting momentarily $b=0$ for simplicity, we have 
  \be 
  \begin{split}
   \delta_{\Lambda_1}^{(0)}\Sigma_{a}{}^{\bar{b}}(\Lambda_2,E) &- (1\leftrightarrow 2) \ = \ 
   \tfrac{a}{2}\,\Lambda_{1a}{}^{e}\,{\cal D}_{e}\Lambda_{2c}{}^{d}\,\omega^{\bar{b}}{}_{d}{}^{c} \\
   & +\tfrac{a}{2}\,{\cal D}_{a}\Lambda_{2c}{}^{d}\big(-{\cal D}^{\bar{b}}\Lambda_{1d}{}^{c}
   -\Lambda_{1\bar{e}}{}^{\bar{b}}\,\omega^{\bar{e}}{}_{d}{}^{c}
   +\Lambda_{1d}{}^{e}\,\omega^{\bar{b}}{}_{e}{}^{c} 
   -\Lambda_{1e}{}^{c}\,\omega^{\bar{b}}{}_{d}{}^{e} \big)- (1\leftrightarrow 2) \;. 
  \end{split}
 \ee  
Using the antisymmetry in $(1\leftrightarrow 2)$ it is easy to see that the final two terms 
in the last line can be combined into one, then including the total derivative ${\cal D}_a(\Lambda_2\Lambda_1)$. 
We can thus write  the result as 
 \be\label{deltaSigma} 
  \begin{split}
   \delta_{\Lambda_1}^{(0)}\Sigma_{a}{}^{\bar{b}}(\Lambda_2,E) - (1\leftrightarrow 2) \ = \ &\,
   \Lambda_{1a}{}^{e}\, \Sigma_{e}{}^{\bar{b}}(\Lambda_2,E) 
   -\Lambda_{1\bar{e}}{}^{\bar{b}}\,\Sigma_{a}{}^{\bar{e}}(\Lambda_2, E) 
   +\Sigma_{a}{}^{\bar{b}}(\Lambda_2\Lambda_1,E)\\
   &\, -\tfrac{a}{2}\,{\cal D}_{a}\Lambda_{2c}{}^{d}\,{\cal D}^{\bar{b}}\Lambda_{1d}{}^{c}
   -\tfrac{b}{2}\,{\cal D}^{\bar{b}}\Lambda_{2\bar{c}}{}^{\bar{d}}\,{\cal D}_{a}\Lambda_{1\bar{d}}{}^{\bar{c}}
   - (1\leftrightarrow 2)\;, 
  \end{split}
 \ee  
where we reintroduced an arbitrary parameter $b$, for which the computation proceeds in complete parallel.  
We can now work out the closure condition to first order in $\alpha'$ on, say, $E_{a}{}^{M}$: 
 \be\label{closureSTEP}
 \begin{split}
  \big[\delta_{\Lambda_1},\delta_{\Lambda_2}\big]E_{a}{}^{M} \ = \ &\, 
  \delta_{\Lambda_1}\big(\Lambda_{2a}{}^{b} E_{b}{}^{M}+\Sigma_{a}{}^{\bar{b}}(\Lambda_2,E)E_{\bar{b}}{}^{M}\big)
  - (1\leftrightarrow 2) \\
  \ = \ &\,\Lambda_{2a}{}^{b}\big(\Lambda_{1b}{}^{c} E_{c}{}^{M}
  +\Sigma_{b}{}^{\bar{e}}(\Lambda_1,E)E_{\bar{e}}{}^{M}\big) \\
  &\, +\delta_{\Lambda_1}^{(0)}\Sigma_{a}{}^{\bar{b}}(\Lambda_2,E) E_{\bar{b}}{}^{M} 
  +\Sigma_{a}{}^{\bar{b}}(\Lambda_2,E)\Lambda_{1\bar{b}}{}^{\bar{e}} E_{\bar{e}}{}^{M} 
  - (1\leftrightarrow 2)
  \\
  \ = \ &\, [\Lambda_2,\Lambda_1]_a{}^b E_{b}{}^{M}
  + \Sigma_{a}{}^{\bar{b}}([\Lambda_2,\Lambda_1],E) E_{\bar{b}}{}^{M}\\
  &\, -\tfrac{a}{2}\,{\cal D}_a\Lambda_{2c}{}^{d}\,{\cal D}^{\bar{b}}\Lambda_{1d}{}^{c} \, E_{\bar{b}}{}^{M}
  +\tfrac{b}{2}\, {\cal D}_{a}\Lambda_{2\bar{c}}{}^{\bar{d}}\,{\cal D}^{\bar{b}}
  \Lambda_{1\bar{d}}{}^{\bar{c}}\, E_{\bar{b}}{}^{M} - (1\leftrightarrow 2)\;. 
 \end{split}
 \ee 
Here we used (\ref{deltaSigma}). Specifically, the terms in the first line of (\ref{deltaSigma}) were needed to produce 
the  desired local frame transformation with parameter $\Lambda_{12}\equiv [\Lambda_2,\Lambda_1]$, 
while the terms in the second line yielded the extra contribution in the last line of (\ref{closureSTEP}). 
Moreover, we ignored the order $\alpha'$ variation of $\Sigma$ as such terms are of order $\alpha^{\prime\, 2}$.

In order to establish closure we have to show that the remaining terms in the last line of (\ref{closureSTEP}) 
can be interpreted  as a deformation of the $GL(D)\times GL(D)$ algebra
and/or the generalized diffeomorphism algebra. Indeed, for the following $\alpha'$ correction 
of the diffeomorphism algebra 
 \be\label{xiDEFor}
  \xi_{12}^{(1)M} \ = \ \tfrac{a}{4}\,\Lambda_{2c}{}^{d}\, \partial^M\Lambda_{1d}{}^{c}
  -\tfrac{b}{4}\, \Lambda_{2\bar{c}}{}^{\bar{d}}\, \partial^M\Lambda_{1\bar{d}}{}^{\bar{c}}-(1\leftrightarrow 2)\;, 
 \ee
and the following $\alpha'$ correction of the $GL(D)\times GL(D)$ algebra  
  \be\label{Lambda12prime}
  \begin{split}
   \Lambda^{(1)}_{12a}{}^{b} \ &= \ \tfrac{a}{2}\, {\cal D}_a\Lambda_{2c}{}^{d}\,{\cal D}^b\Lambda_{1d}{}^{c}
   -\tfrac{b}{2}\,{\cal D}_{a}\Lambda_{2\bar{c}}{}^{\bar{d}}\,{\cal D}^b\Lambda_{1\bar{d}}{}^{\bar{c}}
   -(1\leftrightarrow 2)\;, \\
    \Lambda^{(1)}_{12\bar{a}}{}^{\bar{b}} \ &= \ \tfrac{a}{2}\, {\cal D}_{\bar{a}}\Lambda_{2c}{}^{d}\,
    {\cal D}^{\bar{b}}\Lambda_{1d}{}^{c}
   -\tfrac{b}{2}\,{\cal D}_{\bar{a}}\Lambda_{2\bar{c}}{}^{\bar{d}}\,{\cal D}^{\bar{b}}\Lambda_{1\bar{d}}{}^{\bar{c}}
   -(1\leftrightarrow 2)\;, 
  \end{split} 
  \ee 
the terms in the second line  of (\ref{closureSTEP}) are reproduced.  To verify this, we compute 
 \be\label{effDIFF}
 \begin{split}
  \delta_{\xi_{12}^{(1)}}E_{a}{}^{M}  
  \ &= \ 
  \big(\partial^M\xi_{12N}^{(1)} - \partial_N\xi_{12}^{(1)M}\big) E_{a}{}^{N} \\
  \ &= \ \big(\tfrac{a}{2}\,\partial^M\Lambda_{2c}{}^{d}\,\partial_N\Lambda_{1d}{}^{c} 
  -\tfrac{b}{2}\,\partial^M\Lambda_{2\bar{c}}{}^{\bar{d}}\,\partial_N\Lambda_{1\bar{d}}{}^{\bar{c}}\big)E_{a}{}^{N}
  -(1\leftrightarrow 2) \\
  \ &= \ \tfrac{a}{2}\,\partial^M\Lambda_{2c}{}^{d}\,{\cal D}_a\Lambda_{1d}{}^{c} 
  -\tfrac{b}{2}\,\partial^M\Lambda_{2\bar{c}}{}^{\bar{d}}\,{\cal D}_a\Lambda_{1\bar{d}}{}^{\bar{c}}
  -(1\leftrightarrow 2) \\
  \ &= \ \big(\tfrac{a}{2}\,{\cal D}^B\Lambda_{2c}{}^{d}\,{\cal D}_a\Lambda_{1d}{}^{c} 
  -\tfrac{b}{2}\,{\cal D}^B\Lambda_{2\bar{c}}{}^{\bar{d}}\,{\cal D}_a\Lambda_{1\bar{d}}{}^{\bar{c}}\big)E_{B}{}^{M}
  -(1\leftrightarrow 2) \;, 
 \end{split}
 \ee 
where we used that the generalized Lie derivative does not yield a transport term since in (\ref{xiDEFor})
the free index is carried by a derivative. 
For the frame transformation of (\ref{Lambda12prime}) we have 
 \be\label{EffLOR}
    \delta_{\Lambda_{12}^{(1)}}E_{a}{}^{M} \ = \ \Lambda^{(1)}_{12a}{}^{b} E_{b}{}^{M} \ = \ 
    \big(\tfrac{a}{2}\, {\cal D}_a\Lambda_{2c}{}^{d}\,{\cal D}^b\Lambda_{1d}{}^{c}
   -\tfrac{b}{2}\,{\cal D}_{a}\Lambda_{2\bar{c}}{}^{\bar{d}}\,{\cal D}^b\Lambda_{1\bar{d}}{}^{\bar{c}}\big) E_{b}{}^{M}
   -(1\leftrightarrow 2) \;. 
 \ee  
It is straightforward to see that the sum of  (\ref{effDIFF}) and (\ref{EffLOR}) yields 
precisely the terms in the last line  of (\ref{closureSTEP}). This proves closure 
of the frame transformations on $E_{a}{}^{M}$, while 
closure on $E_{\bar{a}}{}^{M}$ follows similarly. 
Summarizing, we have proved closure of the frame transformations according to 
 \be
  \big[\delta_{\Lambda_1},\delta_{\Lambda_2}\big] \ = \ \delta_{\Lambda_{12}} + \delta_{\xi_{12}^{(1)}}\;, \quad 
  \text{where} \qquad \Lambda_{12} \ = \ \big[\Lambda_2,\Lambda_1\big]+\Lambda_{12}^{(1)}\;, 
 \ee 
with the $\alpha'$ corrected gauge algebras (\ref{xiDEFor}) and (\ref{Lambda12prime}). 

Finally, let us compute the mixed commutator between local frame transformations and generalized 
diffeomorphisms. To this end it is convenient to first establish 
 \be
  \delta_{\xi}\Sigma_{a}{}^{\bar{b}}(\Lambda, E) \ = \ {\cal L}_{\xi}\Sigma_{a}{}^{\bar{b}}(\Lambda, E) 
  -\Sigma_{a}{}^{\bar{b}}(\xi^N\partial_N\Lambda, E)\;. 
 \ee
This follows from (\ref{SigmaFORM}) by using that the spin connections transform as scalars 
under diffeomorphisms. We also recall that in such closure computations gauge parameters 
like $\Lambda$ are not to be varied. 
Closure can now be verified by a quick computation:  
 \be
 \begin{split}
   \big[\delta_{\xi},\delta_{\Lambda}\big]E_a{}^{M} \ = \ &\, \delta_{\xi}\big(\Lambda_a{}^b E_{b}{}^{M}
   +\Sigma_{a}{}^{\bar{b}}(\Lambda, E) E_{\bar{b}}{}^{M}\big)
   -\delta_{\Lambda}\big({\cal L}_{\xi}E_{a}{}^M\big) \\
   \ = \ &\, \Lambda_a{}^b {\cal L}_{\xi}E_b{}^M 
   + {\cal L}_{\xi}\big(\Sigma_{a}{}^{\bar{b}}(\Lambda, E) E_{\bar{b}}{}^{M}\big)
   -\Sigma_{a}{}^{\bar{b}}(\xi^N\partial_N\Lambda, E)E_{\bar{b}}{}^{M}\\
   &\, - \,{\cal L}_{\xi}\big(\Lambda_a{}^b E_b{}^M+\Sigma_{a}{}^{\bar{b}}(\Lambda,E)E_{\bar{b}}{}^{M}\big)\\
   \ = \ &\, -\xi^N\partial_N\Lambda_a{}^b\,E_b{}^M +\Sigma_a{}^{\bar{b}}(-\xi^N\partial_N\Lambda, E)\;. 
 \end{split}
 \ee   
The last line equals an $\alpha'$-deformed frame transformation.  
We have thus established closure and shown that this part of the algebra is undeformed: 
 \be
  \big[\delta_{\xi},\delta_{\Lambda}\big]E_a{}^{M} \ = \ \delta_{\Lambda'} E_{a}{}^{M}\;, \qquad
  \Lambda'_a{}^{b} \ = \ -\xi^N\partial_N\Lambda_a{}^{b}\;, 
 \ee
and analogously for $E_{\bar{a}}{}^{M}$.

\subsection{Second order perturbation theory around flat backgrounds}

We now develop the perturbation theory around constant backgrounds, following the original treatment 
given in \cite{Siegel:1993th} and investigated in \cite{Hohm:2011dz,Hohm:2015ugy}, 
but including the $\alpha'$-deformation discussed above. 
In principle, the extension to curved backgrounds is straightforward, as in \cite{Hohm:2015ugy}, 
but for our present purposes it is sufficient to restrict to flat backgrounds. 
The constant background frame is denoted by $\bar{E}_{A}{}^{M}$ and we expand the full frame field as 
 \be\label{backgroundExp}
  E_{A}{}^{M} \ = \ \bar{E}_{A}{}^{M} - h_{A}{}^{B}\bar{E}_{B}{}^{M}\;, 
 \ee
introducing the fluctuation  $h$ with flat indices. This expansion is taken to be exact.
 
Let us first briefly discuss the gauge symmetries in this perturbative framework, 
as detailed in \cite{Siegel:1993th,Hohm:2011dz,Hohm:2015ugy} to zeroth order 
in $\alpha'$. 
The generalized diffeomorphisms and local frame transformations act on the fluctuation field as 
 \be\label{gaugeFLuc}
 \begin{split}
  \delta h_{AB} \ = \ &\,K_{AB} + \xi^CD_C h_{AB}  + K_{B}{}^{C} h_{AC} 
    \ - \ \Lambda_{AB}+\Lambda_{A}{}^{C} h_{CB}\;, 
 \end{split}
\ee
where now flattening and unflattening is done with the background frame, and all indices are raised and lowered with the background tangent space metric. Moreover, we use the notation  
 \be
  K_{AB} \ \equiv \ D_A\xi_B - D_B\xi_A\;, \qquad D_A \ \equiv \ \bar{E}_{A}{}^{M}\partial_M\;. 
 \ee
Let us recall that the above form of the gauge transformations includes the $\alpha'$-deformation, 
in the form of the off-diagonal gauge parameters $\Lambda_{a\bar{b}}= \Sigma_{a\bar{b}}(\Lambda, E)$.  
To lowest order in fields and $\alpha'$, we infer that $\delta_{\Lambda}h_{ab}=-\Lambda_{ab}$ and 
$\delta_{\Lambda}h_{\bar{a}\bar{b}}=-\Lambda_{\bar{a}\bar{b}}$ and therefore that the diagonal components 
of the fluctuation field are pure gauge. We can thus impose
 \be
  h_{ab}  \ = \ h_{\bar{a}\bar{b}} \ = \ 0\;, 
 \ee
which we take to be exact gauge fixing conditions. This implies with ${\cal G}_{a\bar{b}}=0$ that 
$h_{a\bar{b}}=- h_{\bar{b}a}$. The above gauge condition  requires compensating 
gauge transformations, because a generalized diffeomorphism with parameter $\xi$ in general 
will not preserve the gauge. With (\ref{gaugeFLuc}) one verifies that the compensating 
gauge transformations are parameterized by 
 \be\label{compLambda}
  \begin{split}
   \Lambda_{ab} \ &= \ K_{ab} + K_{b}{}^{\bar{c}} \, h_{a\bar{c}} \ + \ {\cal O}(\alpha')\;, \\
   \Lambda_{\bar{a}\bar{b}} \ &= \ K_{\bar{a}\bar{b}} - K_{\bar{b}}{}^{{c}}\, h_{c\bar{a}} \ + \ {\cal O}(\alpha')\;,   
  \end{split}
 \ee 
up to terms of higher order in $\alpha'$. 

Next, we determine the $\alpha'$-corrected gauge transformations of $h_{a\bar{b}}$.  
The $\alpha'$ corrections to the compensating gauge transformations are determined for $h_{ab}$ according to (\ref{gaugeFLuc}) by 
 \be 
  \delta^{(1)} h_{ab} \ = \   -\Lambda_{ab}^{(1)} - \Sigma_{a}{}^{\bar{c}}(\Lambda^{(0)},h)\, h_{b\bar{c}} \ = \ 0 \;, 
 \ee
and similarly for $h_{\bar{a}\bar{b}}$.  Thus, the compensating parameter takes the 
form $ \Lambda_{ab}^{(1)}  =  -\Sigma_{a}{}^{\bar{c}}(\Lambda^{(0)},h)h_{b\bar{c}}$. 
The gauge transformation for the physical $h_{a\bar{b}}$ is then determined by  
 \be
  \delta^{(1)} h_{a\bar{b}} \ = \ -\Sigma_{a\bar{b}} \ + \ \Lambda^{(1)}_{a}{}^{c} h_{c\bar{b}}\;.  
 \ee
Since $\Sigma_{a}{}^{\bar{b}}$ starts at linear order in $h$, the compensating parameter 
$\Lambda_{ab}^{(1)}$ is quadratic in $h$ and  thus in the gauge transformation for $h_{a\bar{b}}$  
it contributes only to third order in $h$. Hence for second order perturbation theory 
the compensating transformation of ${\cal O}(\alpha')$ is immaterial, and 
the gauge transformations are determined from 
 \be\label{alpha'perturbativedeformed}
  \delta^{(1)} h_{a\bar{b}} \ = \ -\Sigma_{a\bar{b}} \ =  \ 
  -\tfrac{a}{2}\,\D_a\Lambda^{cd} \,\omega_{\bar{b}dc} - \tfrac{b}{2}\, \D_{\bar{b}}\Lambda^{\bar{c}\bar{d}}\,
  \omega_{a\bar{d}\bar{c}}\;,  
 \ee
upon eliminating $\Lambda$ according to (\ref{compLambda}) and expanding to second 
order in $h$. Here we used that one can 
freely raise and lower indices with the constant background tangent space metric.

In the remainder of this section we determine the $\alpha'$-deformed perturbative gauge transformations 
explicitly up to second order in fields in order to prove that the result is equivalent to that obtained
by the Noether method in sec.~2. We begin by expanding the generalized spin connections in (\ref{alpha'perturbativedeformed}) up to second order in fields.  Using (\ref{connections}), one finds 
 \be\label{ExplConne}
  \begin{split}
   \omega_{\bar{b}dc}  \  &= \ \Gamma_{\bar{b}dc}  - h_{c}{}^{\bar{d}}\,\Gamma_{d\bar{b}\bar{d}}
   -h_{d}{}^{\bar{c}}\, D_{\bar{c}}h_{c\bar{b}}\;, \\
   \omega_{a\bar{d}\bar{c}} \ &= \ -\Gamma_{a\bar{d}\bar{c}} - h^{d}{}_{\bar{c}}\, \Gamma_{\bar{d}ad}
   -h^{d}{}_{\bar{d}}\,D_dh_{a\bar{c}}\;, 
  \end{split}
 \ee  
where we defined the connections to first order: 
 \be
  \Gamma_{a\bar{b}\bar{c}} \ \equiv  \ D_{\bar{b}}h_{a\bar{c}} - D_{\bar{c}}h_{a\bar{b}}\;, 
  \qquad 
  \Gamma_{\bar{a}bc} \ \equiv  \ D_bh_{c\bar{a}}-D_{c}h_{b\bar{a}}\;. 
 \ee 
This, together with the compensating parameters (\ref{compLambda}), needs to be inserted into (\ref{alpha'perturbativedeformed}). 
To first order in fields we read off 
 \be\label{frametransper}
  \delta_{\xi}^{(1)[1]} h_{a\bar{b}} \ = \ \tfrac{a}{2}\, D_aK^{cd}\,\Gamma_{\bar{b}cd} 
   - \tfrac{b}{2}\, D_{\bar{b}}K^{\bar{c}\bar{d}}\,\Gamma_{a\bar{c}\bar{d}}\;. 
 \ee
This is in precise agreement with (\ref{linCOnnFORM}), as can be seen by using the technique 
introduced in \cite{Hohm:2014xsa} that allows one to covert flat indices into projected $O(D,D)$ indices 
by means of a background frame field. Specifically, with the background frame field $\bar{E}_{A}{}^{M}$
and its inverse we have the identifications 
 \be\label{hhdictionary}
  h_{\,\nin{M}\bar{N}} \ = \ 2\, \bar{E}_{M}{}^{a} \bar{E}_{N}{}^{\bar{b}} \, h_{a\bar{b}}\;, \qquad
  h_{a\bar{b}} \ = \ \tfrac{1}{2}\, \bar{E}_{a}{}^{M} \bar{E}_{\bar{b}}{}^{N}\,h_{\,\nin{M}\bar{N}}\;, 
 \ee
while gauge parameters and  derivatives  are converted in the obvious fashion, 
$\xi_{\,\nin{M}}=E_{M}{}^{a}\xi_{a}$, etc. 
It then follows immediately that (\ref{frametransper}) agrees with (\ref{linCOnnFORM}). 

Let us now turn to second order perturbation theory, where we include terms up to 
quadratic order in $h$. 
In order to simplify the algebra we set $b=0$, which can be done without loss of 
generality because the $\mathbb{Z}_2$ action mapping unbarred and barred indices into 
each other allows one to reconstruct the full result (see \cite{Hohm:2014xsa} for more details).  
Using (\ref{compLambda}), (\ref{alpha'perturbativedeformed}) and (\ref{ExplConne}),  
we thus start from 
\be
 \begin{split}
  \delta_{\xi}^{(1)} h_{a\bar{b}} \ = \ -\tfrac{a}{2}(D_a-h_{a}{}^{\bar{c}}D_{\bar{c}})(K^{cd}+K^{d}{}_{\bar{d}}\, h^{c\bar{d}})
   (\Gamma_{\bar{b}dc}  - h_{c}{}^{\bar{e}}\,\Gamma_{d\bar{b}\bar{e}}
   -h_{d}{}^{\bar{e}}\, D_{\bar{e}}h_{c\bar{b}})\;, 
  \end{split}
 \ee  
and collect all terms up to quadratic order in $h$, 
 \be\label{DVRRTGR}
 \begin{split}
  \delta_{\xi}^{(1)[2]} h_{a\bar{b}} \ = \   -\tfrac{a}{2}\big[&-D_aK^{cd}\,h_{c}{}^{\bar{e}} \,\Gamma_{d\bar{b}\bar{e}}
  -D_aK^{cd}\,h_{d}{}^{\bar{e}} \, D_{\bar{e}}h_{c\bar{b}}
  +D_aK^{d\bar{d}}\,h^{c}{}_{\bar{d}}\,\Gamma_{\bar{b}dc} \\ 
  &+K^{d\bar{d}} D_ah^{c}{}_{\bar{d}}\,\Gamma_{\bar{b}dc} 
  -h_{a}{}^{\bar{c}} \,D_{\bar{c}}K^{cd}\,\Gamma_{\bar{b}dc}\, \big]\;. 
 \end{split}
 \ee 
Upon converting flat into curved indices and using the dictionary (\ref{hhdictionary}) between $h_{a\bar{b}}$ 
and $h_{\,\nin{M}\bar{N}}$, this is equivalent to 
 \be\label{hFORMgaugetr}
  \begin{split}
  \delta_{\xi}^{(1)[2]} h_{\,\nin{M}\bar{N}}  =    -\tfrac{a}{4}\big[&-\partial_{\,\nin{M}}K^{\,\nin{K}\,\nin{L}}\,
  h_{\,\nin{K}}{}^{\bar{P}} \,\Gamma_{\,\nin{L}\bar{N}\bar{P}}
  -\partial_{\,\nin{M}}K^{\,\nin{K}\,\nin{L}}\,h_{\,\nin{L}}{}^{\bar{P}} \, \partial_{\bar{P}}h_{\,\nin{K}\bar{N}}
  +\partial_{\,\nin{M}}K^{\,\nin{K}\bar{L}}\,h^{\,\nin{P}}{}_{\bar{L}}\,\Gamma_{\bar{N}\,\nin{K}\,\nin{P}} \\
  &+K^{\,\nin{K}\bar{L}} \partial_{\,\nin{M}}h^{\,\nin{P}}{}_{\bar{L}}\,\Gamma_{\bar{N}\,\nin{K}\,\nin{P}} 
  +h_{{\,\nin{M}}}{}^{\bar{P}} \,\partial_{\bar{P}}K^{\,\nin{K}\,\nin{L}}\,\Gamma_{\bar{N}\,\nin{K}\,\nin{L}}\,\big]\;. 
 \end{split}
 \ee     
The global factor of 2 relative to (\ref{DVRRTGR}) is due to (\ref{hhdictionary}) and the fact that we have 
on the left-hand side one $h$, but on the right-hand side two. 
We can also determine the gauge algebra to this order in perturbation theory,  which follows from 
(\ref{xiDEFor}) upon inserting (\ref{compLambda}). Note that for this computation the ${\cal O}(\alpha')$
contributions to the compensating parameters are not needed as these would only contribute to 
order $\alpha'^{\,2}$. We compute 
 \be 
 \begin{split}
 \xi_{12}^M \ &= \ \tfrac{a}{4}\Lambda_{2}{}^{cd}\partial^M\Lambda_{1dc} \ = \ 
 \tfrac{a}{4}(K_2^{cd} + K_2^{d\bar{e}}\, h^{c}{}_{\bar{e}})\,\partial^M
 (K_{1dc} + K_{1c\bar{d}}\, h_{d}{}^{\bar{d}}) \\
 \ &= \ -\tfrac{a}{4}K_2^{cd}\partial^M K_{1cd} +\tfrac{a}{4}K_2^{cd}\partial^MK_{1c\bar{d}}\, h_{d}{}^{\bar{d}}
 +\tfrac{a}{4}K_2^{cd}K_{1c\bar{d}}\, \partial^M h_{d}{}^{\bar{d}}
 +\tfrac{a}{4}h^{c}{}_{\bar{e}}\,K_2^{d\bar{e}}\partial^M K_{1dc} \\
 \ &= \ -\tfrac{a}{4}K_2^{cd}\partial^M K_{1cd} 
 +\tfrac{a}{4}h^{d\bar{d}}\, K_{2d}{}^{c}\, \partial^MK_{1\bar{d}c}
 +\tfrac{a}{4}h^{c\bar{e}}\,K_{2\bar{e}}{}^{d}\, \partial^M K_{1cd}
 +\tfrac{a}{4}\partial^M h^{d\bar{d}}\,K_{2d}{}^{c}K_{1\bar{d}c} \;. 
 \end{split}
 \ee
Rewriting this algebra in terms of $h_{\,\nin{M}\bar{N}}$, we obtain 
 \be\label{algebrafromFRAME}
 \begin{split}
  \xi_{12}^M \ = \  &\, -\tfrac{a}{4}K_2^{\,\nin{K}\,\nin{L}}\partial^M K_{1\,\nin{K}\,\nin{L}} 
 +\tfrac{a}{8}h^{\,\nin{K}\bar{L}}\, K_{2\,\nin{K}}{}^{\,\nin{P}}\, \partial^MK_{1\bar{L}\,\nin{P}}
 +\tfrac{a}{8}h^{\,\nin{L}\bar{K}}\,K_{2\bar{K}}{}^{\,\nin{P}}\, \partial^M K_{1\,\nin{L}\,\nin{P}} \\
 &\, +\tfrac{a}{8}\partial^M h^{\,\nin{K}\bar{L}}\,K_{2\,\nin{K}}{}^{\,\nin{P}}K_{1\bar{L}\,\nin{P}}\;, 
 \end{split} 
 \ee
where we used (\ref{hhdictionary}). 

Let us now verify that the above gauge transformations and gauge algebra are equivalent to 
those found in second order perturbation theory in sec.~2 by the Noether method.  
We have to show that they are equal up to field and parameter redefinitions and the addition of 
trivial gauge parameters. 
In order to compare the two results, we denote 
the transformations and algebra derived in this section from the frame formalism by hats, 
and those determined in sec.~2 without hats. Comparing then (\ref{FINALGAUGE}) and (\ref{Finalalgebra}) 
with (\ref{hFORMgaugetr}) and (\ref{algebrafromFRAME}), recalling $b=0$,  we read off 
 \be\label{NoetherFrameClash}
  \begin{split}
   \widehat{\delta}_{\xi}^{[2](1)}h_{\,\nin{M}\bar{N}}  \ = \ \,& \delta_{\xi}^{[2](1)}h_{\,\nin{M}\bar{N}}
   +\tfrac{a}{4}\,\partial_{\,\nin{M}}K^{\,\nin{K}\,\nin{L}}\,h_{\,\nin{K}}{}^{\bar{P}}\partial_{\bar{P}}h_{\,\nin{L}\bar{N}}\\
   &+\tfrac{a}{4}\,\partial_{\,\nin{M}}K^{\,\nin{K}\bar{P}}\,h^{\,\nin{L}}{}_{\bar{P}}\Gamma_{\bar{N}\,\nin{K}\,\nin{L}}
   +\tfrac{a}{2}\,\partial_{\,\nin{M}}h^{\,\nin{P}}{}_{\bar{K}}\,\partial_{\bar{N}}h^{\,\nin{Q}\bar{K}}\,K_{\,\nin{P}\,\nin{Q}}\\
   &-\tfrac{a}{4}\,K^{\,\nin{K}\bar{L}}\,\partial_{\,\nin{M}}h^{\,\nin{P}}{}_{\bar{L}}\,\Gamma_{\bar{N}\,\nin{K}\,\nin{P}}
   +\tfrac{a}{4}\,\partial_{\,\nin{M}}K^{\,\nin{K}\,\nin{L}}\,h_{\,\nin{K}}{}^{\bar{P}}\,\Gamma_{\,\nin{L}\bar{N}\bar{P}}\,, \\[0.5ex]
   \widehat{\xi}_{12}^{M[1](1)} \ = \ &\, \xi_{12}^{M[1](1)}  +\Delta^M(h,\xi)
   \;, 
  \end{split}
 \ee  
where the difference between the two gauge algebras is  
\be\label{DeltaDiff}
 \Delta^M(h,\xi) \ = \  - \tfrac{a}{8}\, h^{\,\nin{K}\bar{L}}\,K_{2\,\nin{K}}{}^{\,\nin{P}}
   \partial^MK_{1\bar{L}\,\nin{P}}
   -\tfrac{a}{8}\, h^{\,\nin{L}\bar{K}} K_{2\bar{K}}{}^{\,\nin{P}}\partial^MK_{1\,\nin{L}\,\nin{P}}
   +\tfrac{3}{8}a\,\partial^Mh^{\,\nin{K}\bar{L}}\, K_{2\,\nin{K}}{}^{\,\nin{P}} K_{1\bar{L}\,\nin{P}} \;. 
 \ee
 
Consider now the parameter redefinition   
 \be
  \xi^{\prime M} \ = \ \xi^M  +   F^M(h,\xi)\;,  
 \ee
where $F$ is of order $\alpha'$,  linear in $\xi$ and quadratic in $h$, so that it modifies the gauge transformations 
to the appropriate order. We do not allow terms linear in $h$, because that would also affect the 
transformations $\delta^{[1](1)}h$ that we matched already. 
This parameter redefinition yields the modified gauge transformations 
 \be\label{paramredEffect}
  \delta_{\xi'} h_{\nin{M}\bar{N}} \  = \  \delta_{\xi} h_{\nin{M}\bar{N}}
 \ + \ 2(\partial_{\,\nin{M}}F_{\bar{N}}(h,\xi) - \partial_{\bar{N}}F_{\,\nin{M}}(h,\xi)) \  + \  {\cal O}(h^3)\;. 
 \ee
It is easy to see that closure then holds for $\delta_{\xi'}h$ with the modified algebra 
  \be
  \xi_{12}^{\prime M} \ = \ \xi_{12}^M \ + \ \big(\delta_{\xi_1}^{[0]}F^M(h,\xi_2) - (1\leftrightarrow 2) \big) \;. 
 \ee  
Thus, any two expressions for the gauge algebra that differ by a total $\delta_{\xi}^{[0]}$ variation 
as above are equivalent under parameter redefinitions. 
It turns out that in order to show that the extra terms on the right-hand side of the gauge 
algebra in (\ref{NoetherFrameClash}) are trivial, we also need to invoke trivial parameters, 
which take the form of a total derivative. 
Finally, it is straightforward to verify that (\ref{DeltaDiff}) can be written as 
 \be\label{paramREDEF}
  \Delta^M(h,\xi) \ =  \ -\tfrac{a}{8}\delta_{\xi_1}^{[0]}\big(\partial^Mh^{\,\nin{K}\bar{L}} K_{2\,\nin{K}}{}^{\,\nin{P}}
  h_{\nin{P}\bar{L}}\big)
  +\partial^M\big(-\tfrac{a}{8} h^{\,\nin{L}\bar{K}} K_{2\bar{K}}{}^{\,\nin{P}} K_{1\,\nin{L}\,\nin{P}}\big)\;, 
 \ee 
which proves that both algebras are identical up to a parameter redefinition and a trivial parameter. 
For the gauge transformations in (\ref{NoetherFrameClash}) one may confirm by a direct computation 
that 
 \be
 \begin{split}
  \widehat{\delta}_{\xi}h_{\,\nin{M}\bar{N}} \ = \ &\,\delta_{\xi}h_{\,\nin{M}\bar{N}}
  \ + \ \delta_{\xi}^{[0]}\big(\tfrac{a}{8}\,\partial_{\,\nin{M}}h^{\,\nin{K}\bar{L}} 
  h^{\,\nin{P}}{}_{\bar{L}}\,\Gamma_{\bar{N}\,\nin{K}\,\nin{P}}\big)\\[0.5ex]
  &\, +2\,\partial_{\,\nin{M}}\big(-\tfrac{a}{8}\,\partial_{\bar{N}}h^{\,\nin{K}\bar{L}} K_{\,\nin{K}}{}^{\,\nin{P}} 
  h_{\,\nin{P}\bar{L}}\, \big)
  -2\,\partial_{\bar{N}}\big(-\tfrac{a}{8}\,\partial_{\,\nin{M}}h^{\,\nin{K}\bar{L}} K_{\,\nin{K}}{}^{\,\nin{P}} 
  h_{\,\nin{P}\bar{L}}\, \big)\;. 
 \end{split} 
 \ee 
The extra term in the first line is removable by a field redefinition, which does not affect the 
gauge algebra. The terms in the second line 
are removable by the same parameter redefinition as in (\ref{paramREDEF}), as follows with (\ref{paramredEffect}). 
This completes our proof that the gauge transformations and gauge algebra found here 
are equivalent to those determined by the Noether method.

\subsection{Third order perturbation theory: proof that the Jacobiator is trivial}

In the previous subsection we have shown that to second order in perturbation theory 
the gauge structures determined by the Noether method agree, up to trivial parameter and 
field redefinitions,  with those following from 
the background independent $\alpha'$-deformed frame formalism, thereby giving an independent 
consistency check for this formalism. In sec.~3 we identified an obstruction for a generalized metric formulation 
that arises to third order in perturbation theory by showing that the unique candidate gauge algebra 
in terms of the generalized metric does \textit{not} lead to a trivial Jacobiator
and thus cannot define a consistent gauge algebra. 
Therefore, it is reassuring and an important consistency test 
that for the frame formalism there is no such obstruction to third order, as we verify now. 

We start by determining the gauge algebra to third order in perturbation theory, i.e., 
we determine the terms quadratic in $h$ in the gauge algebra. 
These are again obtained by inserting (\ref{compLambda}) into the gauge algebra (\ref{xiDEFor}), 
 \be
 \begin{split}
  F^{M(1)[2]}(\xi_1,\xi_2) \ \equiv \ \xi_{12}^{M(1)[2]} \ = \ 
  &\,\tfrac{a}{4}\, h^{c}{}^{\bar{e}} h_{d}{}^{\bar{d}} K_2{}^{d\bar{e}}\,\partial^M K_{1c\bar{d}}
  +\tfrac{a}{4}\, h^{c}{}_{\bar{e}}\,\partial^M h_{d}{}^{\bar{d}} K_{2}^{d\bar{e}} K_{1c\bar{d}}\\
  &\, -\tfrac{b}{4}\,h^{a\bar{b}} h^{c\bar{d}} K_{2 a\bar{d}} \, \partial^M K_{1c\bar{b}}
  -\tfrac{b}{4}\,h^{a\bar{b}}\,\partial^Mh^{c\bar{d}} K_{2a\bar{d}}\,  K_{1c\bar{b}}\;, 
 \end{split} 
 \ee 
where we renamed the algebra expression in order to streamline our discussion below.   
Converting into $O(D,D)$ indices, we obtain 
 \be\label{3rdorderalgebra}
  \begin{split}
   F^{M(1)[2]}(\xi_1,\xi_2) \ = \ &\, \tfrac{a}{16}\,h^{\,\nin{K}\bar{L}} h^{\,\nin{P}\bar{Q}}\,
   K_{2\,\nin{P}\bar{L}}\,\partial^MK_{1\,\nin{K}\bar{Q}} \ - \ \tfrac{b}{16}\, h^{\,\nin{L}\bar{K}} h^{\,\nin{Q}\bar{P}}\,
   K_{2\,\nin{L}\bar{P}}\,\partial^MK_{1\,\nin{Q}\bar{K}}\\
    & \, + \tfrac{a}{16}\,h^{\,\nin{K}\bar{L}} \partial^M h^{\,\nin{P}\bar{Q}}\,K_{2\,\nin{P}\bar{L}}\,K_{1\,\nin{K}\bar{Q}}
    \ - \ \tfrac{b}{16}\,h^{\,\nin{L}\bar{K}} \partial^M h^{\,\nin{Q}\bar{P}}\,K_{2\,\nin{L}\bar{P}}\,K_{1\,\nin{Q}\bar{K}}\;. 
  \end{split}
 \ee   
Moreover, from the construction of the frame formalism it follows that 
the gauge algebra is exact, with no higher order terms beyond quadratic order in $h$.

We now compute the Jacobiator for (\ref{3rdorderalgebra}) and show that it is trivial. 
As a warm-up we first verify this to second order,  using the gauge algebra linear in $h$, 
for which the Jacobiator is field independent. 
We start from the expression (\ref{JACOBIATOR}) for the Jacobiator and 
expand in the number of fields (suppressing the cyclic sum here and in the following), 
 \be
  J^{[0]} \ = \  \delta_{\xi_1}^{[0]}F^{(1)[1]}(\xi_2,\xi_3;h) + F^{(0)[0]}(\xi_1,F^{(1)[0]}(\xi_2,\xi_3))
  +F^{(1)[0]}(\xi_1,F^{(0)[0]}(\xi_2,\xi_3))\;. 
 \ee  
We set again $b=0$, 
which can be done without loss of generality, and use the algebra in the form (\ref{Finalalgebra}), i.e.,  
 \be
 \begin{split}
  F^{M[1](1)}(\xi_1,\xi_2) \ = \ &\, \tfrac{1}{4}a\,
   h^{\,\nin{K}\bar{L}} K_{2\,\nin{K}}{}^{\,\nin{P}}\partial^M K_{1\bar{L}\,\nin{P}}
  +\tfrac{1}{4} a\, h^{\,\nin{L}\bar{K}} K_{2\bar{K}}{}^{\,\nin{P}}\partial^M K_{1\,\nin{L}\,\nin{P}}
  \\
  &\, -\tfrac{1}{4}a \,\partial^Mh^{\,\nin{K}\bar{L}}\, K_{2\,\nin{K}}{}^{\,\nin{P}} K_{1\bar{L}\,\nin{P}} \;, 
 \end{split} 
 \ee 
and the C-bracket (\ref{CBRacket}) for $F^{(0)[0]}$ and (\ref{zerothalpha'algebra}) for $F^{(1)[0]}$. 
An explicit computation then gives after some algebra 
 \be
  \begin{split}
   J^{[0]M} \ = \ \partial^M\Big(-\tfrac{a}{8}\,\xi_1^K K_{3}^{\,\nin{P}\,\nin{Q}}\,
   \partial_KK_{2\,\nin{P}\,\nin{Q}}-\tfrac{a}{4}\,K_1^{\,\nin{P}\,\nin{Q}} K_{3\,\nin{P}}{}^{\bar{K}} K_{2\,\nin{Q}\bar{K}}
   -\tfrac{a}{12}\, K_{1}^{\,\nin{P}\,\nin{Q}} K_{3\,\nin{P}}{}^{\,\nin{K}} K_{2\,\nin{Q}\,\nin{K}}\Big)\;. 
 \end{split}
 \ee
Thus, the Jacobiator is a total derivative and hence trivial, as we wanted to prove. 

To third order perturbation theory we expand the Jacobiator
 (\ref{JACOBIATOR}) as 
  \be
   \begin{split}
    J^{[1]}(\xi_1,\xi_2,\xi_3) \ = \ &\,\delta_{\xi_1}^{[0]}F^{(1)[2]}(\xi_2,\xi_3;h)
    +\delta_{\xi_1}^{[1]}F^{(1)[1]}(\xi_2,\xi_3;h) \\[0.5ex]
    &\,+F^{(0)[0]}(\xi_1,F^{(1)[1]}(\xi_2,\xi_3 ;h))
    +F^{(1)[1]}(\xi_1,F^{(0)[0]}(\xi_2,\xi_3);h)\;, 
   \end{split}
  \ee  
which yields the terms linear in $h$. Using (\ref{3rdorderalgebra}) a straightforward 
but somewhat tedious computation shows that this is indeed trivial. 
To this end, one may freely integrate by parts and discard total derivatives. 
This confirms the validity of the gauge algebra to 
third order in the perturbation theory based on the frame-like formalism.

\section{Formulation in terms of non-symmetric metric}\setcounter{equation}{0}
In previous sections it was proved that there is no generalized metric formulation for 
bosonic string theory when including $\alpha'$ corrections. Rather, a background independent 
and manifestly $O(D,D)$ invariant formulation is given by a frame or vielbein formalism 
with $\alpha'$-deformed local frame transformations. In order to make both 
background independence and $O(D,D)$ invariance manifest, we are forced to introduce 
unphysical, pure gauge degrees of freedom (that are not present in closed string field theory).
This result is puzzling, because for bosonic string theory (and more generally for the bosonic 
sectors of superstring theory) it is always possible to write the $\alpha'$ corrections in terms
of $g_{ij}$ and $b_{ij}$ and hence in terms of ${\cal E}_{ij}\equiv g_{ij}+b_{ij}$, without 
additional pure gauge degrees of freedom. In this section we clarify this by showing how 
a formulation in terms of ${\cal E}_{ij}$ can be obtained from the frame formalism  upon 
gauge fixing, which in turn leads to deformed $O(D,D)$ transformations. 
Since a generalized metric formulation would imply an undeformed $O(D,D)$ symmetry, 
this result is in perfect agreement with the no-go result of sec.~3. 
In the first two subsections we discuss the frame formalism in terms of ${\cal E}_{ij}$ and 
additional pure gauge modes and then perform the gauge fixing. 
The resulting gauge algebra in terms of ${\cal E}_{ij}$ is discussed in the third subsection, 
while some aspects of the deformed $O(D,D)$ invariance are discussed in the final subsection.

\subsection{Frame field in terms of physical and gauge degrees of freedom}

We begin by giving an explicit parametrization of the frame field that solves 
the constraint ${\cal G}_{a\bar{b}}=0$, c.f.~(\ref{TANGENTmetric}), 
without having to fix a gauge: 
\be\label{generalFRAMEcomp}
  E_{A}{}^{M} \ = \  \begin{pmatrix}    E_{ai} & E_{a}{}^{i}  \\[0.5ex]
  E_{\bar{a}i} & E_{\bar{a}}{}^{i} \end{pmatrix} 
  \ = \  \begin{pmatrix}    -{\cal E}_{ji} \, e_{a}{}^{j}  & e_{a}{}^{i}  \\[0.5ex]
  {\cal E}_{ij} \, \bar{e}_{\bar{a}}{}^{j} & \bar{e}_{\bar{a}}{}^{i} \end{pmatrix} \;. 
 \ee 
It is written in terms of the three \textit{independent} tensors with $D^2$ components each: 
 \be
  {\cal E}_{ij} \ = \ g_{ij}+b_{ij}\;, \qquad e_{a}{}^{i}\;, \qquad \bar{e}_{\bar{a}}{}^{i}\,:\qquad 3D^2\quad  \text{components} \;. 
 \ee 
Thus, we have the right number of degrees of freedom, and indeed the above form 
identically solves the constraint on ${\cal G}$ in that 
 \be\label{flatMETRIC}
  {\cal G}_{AB} \ = \  \begin{pmatrix}   -2\,e_{a}{}^{i}\,e_{b}{}^{j}\, g_{ij}  & 0  \\[0.5ex]
  0 & 2\,\bar{e}_{\bar{a}}{}^{i}\, \bar{e}_{\bar{b}}{}^{j} \, g_{ij} \end{pmatrix} \;. 
 \ee 
 
In the remainder of this subsection we discuss how the symmetries of DFT, 
i.e.~generalized diffeomorphisms, the global $O(D,D)$ and local $GL(D)\times GL(D)$
frame transformations, are realized on the component fields.  
We begin with the $GL(D)\times GL(D)$ gauge transformations, first without $\alpha'$ corrections. 
Under finite transformations we have 
 \be 
   E^{\prime}_{A}{}^{M} \ = \ {\bf \Lambda}_A{}^{B} E_{B}{}^{M}\;, \qquad
   {\bf \Lambda}_A{}^{B} \ = \ \begin{pmatrix}    {\bf \Lambda}_a{}^b & 0 \\[0.5ex]
  0 & \bar{\bf \Lambda}_{\bar{a}}{}^{\bar{b}}\end{pmatrix} \;, 
 \ee
using boldface greek letters to denote the group element ${\bf \Lambda}\in GL(D)\times GL(D)$.    
In terms of the component fields in (\ref{generalFRAMEcomp}), this symmetry acts as
 \be\label{LDlowest}
  e' \ = \ {\bf \Lambda}\, e\;, \qquad \bar{e}^{\prime} \ = \ \bar{\bf \Lambda}\,\bar{e}\;, \qquad 
  {\cal E}' \ = \ {\cal E}\;, 
 \ee 
where we used matrix notation, with $e$ denoting $e_{a}{}^{i}$, etc. 
The component fields transform under frame transformations 
as indicated by their flat and curved indices. In particular, ${\cal E}$ is invariant. 
Let us also mention in passing that acting with the generalized Lie derivative on the frame  (\ref{generalFRAMEcomp}) 
allows one to determine straightforwardly the generalized diffeomorphism transformations of $e_{a}{}^{i}$, 
$\bar{e}_{\bar{a}}{}^{i}$ and ${\cal E}_{ij}$, which for the latter reproduces the form given in \cite{Hohm:2010jy}, 
see eq.~(\ref{usualdeltaE}) below.

Let us now turn to the $O(D,D)$ transformations. Consider the general $O(D,D)$ matrix 
 \be 
  h^{M}{}_{N} \ = \   \begin{pmatrix}   a & b  \\[0.5ex]
  c & d \end{pmatrix} \ \in \ O(D,D)\;, 
 \ee
which acts on coordinates as $X\rightarrow X'=hX$ and on the frame field as 
 \be
  E^{\prime}_{A}{}^{M}(X') \ = \ h^{M}{}_{N}\,E_{A}{}^{N}(X)
  \;. 
 \ee  
Using matrix notation, we have 
 \be
   E^{\prime}(X') \ = \ E(X)h^{t} \ = \   
   \begin{pmatrix}   -e{\cal E} & e  \\[0.5ex] \bar{e}{\cal E}^{t} & \bar{e} \end{pmatrix}
    \begin{pmatrix}   a^t & c^t  \\[0.5ex] b^t & d^t \end{pmatrix}\;, 
 \ee
which yields      
 \be\label{MbarM}
 \begin{split}
  e' \ &= \ eM\;, \qquad  M \ \equiv \ d^t -{\cal E}c^t\;, \\
  \bar{e}' \ &= \ \bar{e}\bar{M}\;, \qquad \bar{M} \ \equiv \ d^t +{\cal E}^t c^t\;, 
 \end{split}
 \ee 
and 
 \be\label{eEPrime}
   e'{\cal E}' \ = \  e({\cal E}a^t-b^t)\;, \qquad
  \bar{e}'{\cal E}^{\prime t} \ = \ \bar{e}({\cal E}^ta^t +b^t)\;. 
 \ee  
Using (\ref{MbarM}) in the second equation in here  we obtain\footnote{Using the first 
equation of (\ref{eEPrime}) 
instead gives a differently looking result, but it is equivalent as follows from the general formalism.} 
 \be\label{nonlinODD}
  {\cal E}'(X') \ = \ (a{\cal E}(X)+b)(c{\cal E}(X)+d)^{-1}\;. 
 \ee   
This is the well-known fractional-linear form of the $O(D,D)$ transformation. 
Thus, in presence of the gauge degrees of freedom $e$ and $\bar{e}$, the $O(D,D)$ always acts 
on ${\cal E}$ in this form. However, when fixing a gauge, say by setting 
$e_{a}{}^{i}=\bar{e}_{\bar{a}}{}^{i}=\delta_{a}{}^{i}$, we have to take into account 
compensating frame transformations. From (\ref{MbarM}) we infer that for 
$e=\bar{e}=1$ the compensating $GL(D)\times GL(D)$ transformations 
are  ${\bf \Lambda}=M$, ${\bf \bar{\Lambda}}=\bar{M}$. In (\ref{LDlowest}) 
we saw that ${\cal E}$ is inert under these frame transformations to lowest order in $\alpha'$
and hence not affected by 
compensating gauge transformations. However, this changes when turning on the 
$\alpha'$ deformations, as we will do in the next subsection. 

We finally recall that the non-linear $O(D,D)$ action (\ref{nonlinODD}) can be linearized 
by introducing the generalized metric 
 \be
  {\cal H}^{MN} \ = \ E_{\bar{a}}{}^{M} E^{\bar{a}N} - E_{a}{}^{M} E^{aN}\;, 
 \ee
where the indices are contracted with ${\cal G}$. This is manifestly invariant 
under the (undeformed) local frame transformations (\ref{LDlowest}) and transforms 
covariantly under $O(D,D)$ as indicated by the free $O(D,D)$ indices. 
Moreover, inserting the explicit frame (\ref{generalFRAMEcomp}) and using (\ref{flatMETRIC}), 
we recover the familiar expression in terms of $g$ and $b$, with the gauge degrees of 
freedom $e$ and $\bar{e}$ dropping out, as it should be by gauge invariance.
Note that the generalized metric is no longer invariant under $\alpha'$-deformed 
frame transformations.

\subsection{Gauge fixing and $\alpha'$-deformed gauge structure}

We now consider the $\alpha'$-deformed frame transformations and investigate 
the effect of gauge fixings. Acting with the $\alpha'$-deformed transformation matrix 
(\ref{DeformedMAtrix}) on the frame (\ref{generalFRAMEcomp}), we obtain for the 
components $e$ and $\bar{e}$
 \be
  \delta_{\Lambda}^{(1)} e_{a}{}^{i} \ = \ \Sigma_{a}{}^{\bar{b}}\,\bar{e}_{\bar{b}}{}^{i}\;, \qquad
  \delta_{\Lambda}^{(1)} \bar{e}_{\bar{a}}{}^{i} \ = \ -\Sigma^{b}{}_{\bar{a}}\, e_{b}{}^{i}\;. 
 \ee 
Next, for the remaining components we compute from (\ref{generalFRAMEcomp}), for instance,  
 \be
 \begin{split}
  \delta_{\Lambda} E_{ai} \ &= \ \delta_{\Lambda}(-{\cal E}_{ji} e_{a}{}^{j})
   \ = \ -\delta_{\Lambda} {\cal E}_{ji}\,  e_{a}{}^{j}
    -{\cal E}_{ji} \, \Sigma_{a}{}^{\bar{b}}\,\bar{e}_{\bar{b}}{}^{i} \\
   \ &= \ \Sigma_{a}{}^{\bar{b}} E_{\bar{b}i}
  \ = \ \Sigma_{a}{}^{\bar{b}}\,{\cal E}_{ij}\,\bar{e}_{\bar{b}}{}^{j}\;, 
 \end{split}
 \ee 
from which we conclude 
 \be\label{DEformedEvar}
  \delta_{\Lambda}{\cal E}_{ij} \ = \ -e_{i}{}^{a}\,\bar{e}_{j}{}^{\bar{b}}\, \Sigma_{a\bar{b}}\;, 
 \ee
recalling that raising and lowering of flat indices is done with ${\cal G}$ and using (\ref{flatMETRIC}).
Thus, ${\cal E}_{ij}$ is no longer inert under frame transformations, and hence     
the compensating $GL(D)\times GL(D)$ transformations  do affect ${\cal E}$, 
leading in particular to $\alpha'$ deformed $O(D,D)$ transformations.

Let us now investigate the gauge structure of the generalized diffeomorphisms parameterized by 
$\xi^M=(\tilde{\xi}_i,\xi^i)$ after gauge fixing.  
Thus, from now on we take the frame field to take the gauge fixed form 
 \be\label{gaugeFIXEDframe}
  E_{A}{}^{M} \ = \  \begin{pmatrix}    E_{ai} & E_{a}{}^{i}  \\[0.5ex]
  E_{\bar{a}i} & E_{\bar{a}}{}^{i} \end{pmatrix} 
  \ = \  \begin{pmatrix}    -{\cal E}_{ai} & \delta_{a}{}^{i}  \\[0.5ex]
  {\cal E}_{i\bar{a}} & \delta_{\bar{a}}{}^{i} \end{pmatrix} \;, 
 \ee 
where after gauge fixing we can identity curved and flat indices.  
Acting on this frame field with the generalized Lie derivative as in (\ref{fullHGauge}),  
it is easy to see that the following compensating $GL(D)\times GL(D)$ 
transformations are required in order to preserve the gauge: 
 \be\label{compLambdaAAA}
  \begin{split}
   \Lambda_{j}{}^{i} \ &= \ \partial_j\xi^i -\tilde{\partial}^i\tilde{\xi}_j + {\cal E}_{jk}(\tilde{\partial}^i\xi^{k}
   -\tilde{\partial}^k\xi^i)\,, \\
   \bar{\Lambda}_{j}{}^{i} \ &= \ \partial_j\xi^i -\tilde{\partial}^i\tilde{\xi}_j 
   - {\cal E}_{kj}(\tilde{\partial}^i\xi^{k}
   -\tilde{\partial}^k\xi^i)\;. 
  \end{split}
 \ee 
Here we included only the terms to zeroth order in $\alpha'$, as these are sufficient for our purposes below, 
and we used that flat and curved indices can be identified. 
Acting then on the components of (\ref{gaugeFIXEDframe}) encoding ${\cal E}_{ij}$ and including 
these compensating frame transformations it is easy to check that to lowest order in $\alpha'$ 
 \be\label{usualdeltaE}
   \delta_{\xi}^{(0)}{\cal E}_{ij} \ = \  {\cal D}_i\tilde{\xi}_j - \bar{\cal D}_{j}\tilde{\xi}_i
  +\xi^N\partial_N{\cal E}_{ij}  + {\cal D}_{i}\xi^{k}{\cal E}_{kj} +\bar{\cal D}_{j}\xi^k {\cal E}_{ik}\;, 
 \ee
where   
 \be\label{CALder}
  {\cal D}_i \ = \ \partial_i -{\cal E}_{ik} \tilde{\partial}^k\;, \qquad
  \bar{\cal D}_i \ = \ \partial_i + {\cal E}_{ki} \tilde{\partial}^k\;. 
 \ee 
See the discussion in sec.~4 in \cite{Hohm:2010xe} for more details on this derivation. 
Now turning on $\alpha'$ corrections we infer with (\ref{DEformedEvar}) 
that the gauge transformations read 
 \be
  \begin{split}
  \delta_{\xi}{\cal E}_{ij} \ = \  \delta_{\xi}^{(0)}{\cal E}_{ij}  -\Sigma_{i\bar{j}}(E,\Lambda) \;, 
 \end{split} 
 \ee   
where $\Lambda$ has to be eliminated in terms of  (\ref{compLambdaAAA}), and 
  \be \label{COnnSTEP}
  \begin{split}
     \Sigma_{i\bar{j}}(E,\Lambda) \  = \ \tfrac{a}{2} \, {\cal D}_i\Lambda_{l}{}^{k}\,\omega_{\bar{j}k}{}^{l}
     +\tfrac{b}{2}\, \bar{\cal D}_{\bar{j}}\Lambda_{\bar{l}}{}^{\bar{k}}\, \omega_{i\bar{k}}{}^{\bar{l}}
     \ = \ -\tfrac{a}{2}\,  {\cal D}_i\Lambda_{l}{}^{k}\,\Gamma_{\bar{j}k}{}^{l}
     -\tfrac{b}{2}\, \bar{\cal D}_{\bar{j}}\Lambda_{\bar{l}}{}^{\bar{k}}\, \Gamma_{i\bar{k}}{}^{\bar{l}}\;, 
  \end{split}
 \ee    
with the `$O(D,D)$ connections' defined in \cite{Hohm:2010jy}: 
 \be\label{ODDCONNN}
 \begin{split}
  \Gamma_{i\bar{j}}{}^{\bar{k}} \ &= \ \tfrac{1}{2}g^{kl}({\cal D}_i{\cal E}_{lj}+\bar{\cal D}_{j}{\cal E}_{il}
  -\bar{\cal D}_{l}{\cal E}_{ij})\;,  \\
  \Gamma_{\bar{i}j}{}^{k} \ &= \ \tfrac{1}{2}g^{kl}(\bar{\cal D}_{i}{\cal E}_{jl} + {\cal D}_{j}{\cal E}_{li}
  -{\cal D}_{l}{\cal E}_{ji})\;. 
 \end{split} 
 \ee 
The expressions for the spin connections in terms of the $O(D,D)$ connections 
were obtained in sec.~4.2 in \cite{Hohm:2010xe}. 
Summarizing, the deformed generalized diffeomorphisms to first order in $\alpha'$ act on ${\cal E}_{ij}$
as 
  \be\label{finalcalEtrans}
   \begin{split}
     \delta_{\xi}^{(1)}{\cal E}_{ij} \ = \ &\,\tfrac{a}{2}\, {\cal D}_{i}
     \big(\partial_l\xi^k -\tilde{\partial}^k\tilde{\xi}_l + {\cal E}_{lp}(\tilde{\partial}^k\xi^{p}-\tilde{\partial}^p\xi^k)\big) 
     \,\Gamma_{\bar{j}k}{}^{l}\\
     &\, 
     +\tfrac{b}{2}\, \bar{\cal D}_{j}\big(\partial_l\xi^k -\tilde{\partial}^k\tilde{\xi}_l 
   - {\cal E}_{pl}(\tilde{\partial}^k\xi^{p}-\tilde{\partial}^p\xi^k)\big)\, \Gamma_{i\bar{k}}{}^{\bar{l}}\,.
    \end{split}
   \ee
Let us emphasize that although the above gauge transformations are written in terms of doubled 
derivatives, such as the operators in (\ref{CALder}), and the connections (\ref{ODDCONNN}), they 
are \textit{not} $O(D,D)$ covariant in the undeformed sense. However, they are 
$O(D,D)$ covariant in a suitably deformed sense, as is guaranteed by the construction and 
will be discussed further below.

Rather than analyzing the gauge transformations (\ref{finalcalEtrans}) directly and 
verifying their closure, we can read off the gauge algebra from (\ref{xiDEFor}),  
upon eliminating the gauge parameters by (\ref{compLambdaAAA}): 
  \be\label{COmpFRameALgebra}
   \begin{split}
    \xi_{12}^{M} \ = \ &\,\tfrac{a}{4}\, \Lambda_{2i}{}^{j}\partial^M\Lambda_{1j}{}^{i}
    -\tfrac{b}{4}\,  \bar{\Lambda}_{2i}{}^{j}\partial^M\bar{\Lambda}_{1j}{}^{i} \\
    \ = \  &\, \tfrac{a}{4}\, \big( \partial_i\xi_2^j -\tilde{\partial}^j\tilde{\xi}_{2i} + {\cal E}_{ik}(\tilde{\partial}^j\xi_2^{k}
   -\tilde{\partial}^k\xi_2^j)\big)
   \partial^M\big(\partial_j\xi_1^i -\tilde{\partial}^i\tilde{\xi}_{1j} + {\cal E}_{jk}(\tilde{\partial}^i\xi_1^{k}
   -\tilde{\partial}^k\xi_1^i)\big) \\
   &-\tfrac{b}{4}\, \big(\partial_i\xi_2^j -\tilde{\partial}^j\tilde{\xi}_{2i}  - {\cal E}_{ki}(\tilde{\partial}^j\xi_2^{k}
   -\tilde{\partial}^k\xi^{j}_2) \big) \partial^M 
   \big(\partial_j\xi_1^i -\tilde{\partial}^i\tilde{\xi}_{1j}  - {\cal E}_{kj}(\tilde{\partial}^i\xi_1^{k}
   -\tilde{\partial}^k\xi_1^i)\big) \;, 
   \end{split}
   \ee   
where we left again the antisymmetrization in $(1\leftrightarrow 2)$ implicit.    
Our goal here and in the next subsection is to relate this algebra to the unique 
field-independent gauge algebra found for the HSZ theory in \cite{Hohm:2013jaa}, plus field-dependent 
corrections. Thus, focusing on the field-independent part, we have 
 \be\label{fieldindependentALG}
 \begin{split}
   \xi_{12}^{M} \ = \ & \,  \tfrac{a-b}{4}\,\tilde{\partial}^j\tilde{\xi}_{2i}\,\partial^M\tilde{\partial}^i\tilde{\xi}_{1j}
    +\tfrac{a-b}{4}\,\partial_i\xi_{2}^{j}\,\partial^M\partial_j\xi_{1}^{i} \\
    & -\tfrac{a-b}{4}\,\partial_i\xi_{2}^{j}\,\partial^M\tilde{\partial}^i\tilde{\xi}_{1j}
    -\tfrac{a-b}{4}\,\tilde{\partial}^{j}\tilde{\xi}_{2i}\,\partial^M\partial_j\xi_1^i
   + {\cal O}({\cal E})\;, 
 \end{split}
 \ee  
while the HSZ algebra of \cite{Hohm:2013jaa} reads 
 \be\label{HSZalgebra}
  \tfrac{1}{2}\,\partial_K\xi_2^L\,\partial^M\partial_L\xi_1^K 
  \ = \ 
  \tfrac{1}{2}\,\partial_k\xi_2^l\,\partial^M\partial_l\xi_1^k
  +\tfrac{1}{2}\,\tilde{\partial}^k\tilde{\xi}_{2l}\,\partial^M\tilde{\partial}^l \tilde{\xi}_{1k}
  +\tfrac{1}{2}\,\partial_k\tilde{\xi}_{2l}\,\partial^M\tilde{\partial}^l \xi_1^k
  +\tfrac{1}{2}\,\tilde{\partial}^k\xi_2^l\,\partial^M\partial_l\tilde{\xi}_{1k}\;. 
  \ee
In order to bring this closer to the algebra above   
we use the strong constraint in the second line of (\ref{fieldindependentALG}) to obtain 
  \be
 \begin{split}
   \xi_{12}^{M} \ = \ & \,  \tfrac{a-b}{4}\,\tilde{\partial}^j\tilde{\xi}_{2i}\,\partial^M\tilde{\partial}^i\tilde{\xi}_{1j}
    +\tfrac{a-b}{4}\,\partial_i\xi_{2}^{j}\,\partial^M\partial_j\xi_{1}^{i} \\
    & +\tfrac{a-b}{4}\,\tilde{\partial}^i\xi_{2}^{j}\,\partial^M{\partial}_i\tilde{\xi}_{1j}
    +\tfrac{a-b}{4}\,{\partial}_{j}\tilde{\xi}_{2i}\,\partial^M\tilde{\partial}^j\xi_1^i
   + {\cal O}({\cal E})\;. 
 \end{split}
 \ee  
Comparing with (\ref{HSZalgebra}) we infer 
 \be\label{simplifiedALgebra}
   \xi_{12}^{M} \ = \   \tfrac{a-b}{4}\,\partial_K\xi_2^L\,\partial^M\partial_L\xi_1^K
   +\tfrac{a-b}{4}\,\tilde{\partial}^k\xi_2^l\,\partial^M(\partial_k\tilde{\xi}_{1l}-\partial_l\tilde{\xi}_{1k})
   +\tfrac{a-b}{4}\,\partial_k\tilde{\xi}_{2l}\,\partial^M(\tilde{\partial}^k\xi_1^l - \tilde{\partial}^l\xi_1^k)+ {\cal O}({\cal E})\;. 
 \ee  
Our goal is to prove that for $a=-b$ the gauge algebra is fully equivalent to (\ref{HSZalgebra}), 
up to trivial parameters and parameter redefinitions. Put differently, 
we want to prove that the gauge algebra can be brought into a form for which all field-dependent contributions
are proportional to $a+b$. We turn to this in the next subsection.

\subsection{Simplifying the gauge algebra}

We now search for the parameter redefinition that simplifies the gauge algebra (\ref{COmpFRameALgebra}) 
as envisioned above. 
To this end it is convenient to organize the structure of the lowest order gauge transformations and gauge 
algebra in powers of the field ${\cal E}$. From (\ref{usualdeltaE}) we infer that 
the gauge transformations of ${\cal E}_{ij}$ can be written as 
 \be
  \delta_{\xi}{\cal E}  \ = \  \ell_1(\xi) + \ell_2(\xi,{\cal E}) - \tfrac{1}{2} \ell_3(\xi, {\cal E},{\cal E})\;,  
 \ee 
where 
 \be\label{ellEproducts}
 \begin{split}
  [\ell_1(\xi)]_{ij} \ &= \ \partial_{i}\tilde{\xi}_{j} -  \partial_{j}\tilde{\xi}_{i} \;, \\
  [\ell_2(\xi,{\cal E})]_{ij} \ &= \ { L}_{\xi}{\cal E}_{ij}+\widetilde{ L}_{\tilde{\xi}}{\cal E}_{ij}\;, \\
  [\ell_3(\xi, {\cal E}_1,{\cal E}_{2})]_{ij} \ &= \ {\cal E}_{1ik}(\tilde{\partial}^k\xi^l -\tilde{\partial}^l\xi^k)
  {\cal E}_{2lj}+(1\leftrightarrow 2)\;, 
 \end{split}
 \ee 
using the notation of \cite{Hohm:2010jy} for Lie derivatives $L_{\xi}$ and 
dual Lie derivatives $\widetilde{ L}_{\tilde{\xi}}$. 
Here we use the language of $L_{\infty}$ algebras to deal with the various multi-products appearing in the 
gauge structure,  
but the reader unfamiliar with this framework can simply take the above as a convenient notation. 
This $L_{\infty}$ algebraic viewpoint will be elaborated on in \cite{HOHMZWIEBACHLINFTY}. 

Let us make explicit the constraints implied by closure of the gauge algebra: 
\be
  \begin{split}
   \big[\delta_{\xi_1},\delta_{\xi_2}\big]{\cal E} \ &= \ 
   \delta_{\xi_1}\Bigl(\ell_1(\xi_2)+\ell_2(\xi_2,{\cal E})
   -\tfrac{1}{2}\ell_3(\xi_2,{\cal E},{\cal E})\Bigr)-(1\leftrightarrow 2)  \\
    \ &= \ \ell_2\Bigl( \xi_2\, ,\ \ell_1(\xi_1)+\ell_2(\xi_1,{\cal E})-\tfrac{1}{2}\ell_3(\xi_1,{\cal E},{\cal E}) \Bigr) \\
    &\qquad -\ell_3\Bigl(\, \xi_2\, ,\  \ell_1(\xi_1)+\ell_2(\xi_1,{\cal E})-\tfrac{1}{2}
    \ell_3(\xi_1,{\cal E},{\cal E})\, , \ {\cal E}\, \Bigr) -(1\leftrightarrow 2)\\
    \ &= \ \ell_1(\xi_{12})+\ell_2(\xi_{12},{\cal E})-\tfrac{1}{2}\ell_3(\xi_{12},{\cal E},{\cal E})\;, 
  \end{split}
 \ee   
where $\xi_{12}  \equiv  [\xi_2,\xi_1]_{c}  \equiv \ell_2(\xi_2,\xi_1)$ and we used that $\ell_3$ is defined 
to be symmetric in its last two arguments. 
Thus, comparing orders in ${\cal E}$ closure requires 
 \be\label{Linftyidentities}
  \begin{split}
   \ell_1(\ell_2(\xi_2,\xi_1)) \ &= \ \ell_2(\ell_1(\xi_2),\xi_1) - (1\leftrightarrow 2)\;, \\
   \ell_2(\ell_2(\xi_2,\xi_1),{\cal E}) \ &= \ \ell_2(\xi_2,\ell_2(\xi_1,{\cal E})) + \ell_3(\xi_2,\ell_1(\xi_1),{\cal E}) 
   - (1\leftrightarrow 2)\;, \\
   \ell_3(\ell_2(\xi_2,\xi_1),{\cal E},{\cal E}) \ &= \ \ell_2(\xi_2,\ell_3(\xi_1,{\cal E},{\cal E}))
   +2\,\ell_3(\xi_2,\ell_2(\xi_1,{\cal E}),{\cal E}) - (1\leftrightarrow 2)\;, \\
   0 \ &= \ \ell_3(\xi_2,\ell_3(\xi_1,{\cal E},{\cal E}),{\cal E})- (1\leftrightarrow 2)\;. 
  \end{split}
 \ee  
These are precisely the relations defining an $L_{\infty}$ algebra to the desired order, 
with all higher products being identically zero. 
These relations can be verified by direct computations and are equivalent to 
 closure of the gauge algebra in DFT.

With these relations at hand we can next analyze the effect of a parameter redefinition 
of the form
 \be
  \xi \ \rightarrow \ \xi' \ =  \ \xi  \ + \ F(\xi,{\cal E})\;, 
 \ee
where $F$ is field-dependent and of first order in $\alpha'$.  
More precisely, we interpret this parameter redefinition as modifying the  gauge transformations at order $\alpha'$
according to 
 \be
 \begin{split}
  \delta_{\xi}^{(1)}{\cal E}  \ = \  \ell_1(F(\xi,{\cal E})) + \ell_2(F(\xi,{\cal E}),{\cal E}) 
  - \tfrac{1}{2} \ell_3(F(\xi,{\cal E}), {\cal E},{\cal E})\;. 
 \end{split}
 \ee 
Using the relations (\ref{Linftyidentities}) it is straightforward to verify  that these transformations close to 
first order in $\alpha'$, 
 \be
  \big[\delta_{\xi_1},\delta_{\xi_2}\big] \ = \ \delta^{(0)}_{\xi^{(0)}_{12}+\alpha'\xi_{12}^{(1)}}
  \ + \ \alpha' \delta^{(1)}_{\xi_{12}^{(0)}}\;, 
 \ee  
where $\xi_{12}^{(0)}=[\xi_2,\xi_1]_c$ is the lowest order C-bracket algebra, and  
 \be\label{algebraEDEF}
 \begin{split}
  \xi_{12}^{(1)} \ &= \ \delta_{\xi_1}^{(0)}F(\xi_2,{\cal E})+ \ell_2(\xi_2,F(\xi_1,{\cal E})) - (1\leftrightarrow 2)
  +F([\xi_1,\xi_2]_c,{\cal E})\\
  \ &= \ \delta_{\xi_1}^{(0)}F(\xi_2,{\cal E})+ [\xi_2,F(\xi_1,{\cal E})]_c - (1\leftrightarrow 2)
  +F([\xi_1,\xi_2]_c,{\cal E})\;. 
 \end{split}
 \ee 
 
We now have to find an appropriate $F$ that simplifies the algebra. With (\ref{ellEproducts}) it is 
easy to see that the second and third term in the field-independent part of the algebra 
(\ref{simplifiedALgebra}) are removed by putting  
 \be
  F^M(\xi,{\cal E}) \ = \ -\tfrac{a-b}{4}(\tilde{\partial}^k\xi^l -\tilde{\partial}^l\xi^k)\partial^M{\cal E}_{kl}\;. 
 \ee
We compute for the first term on the right-hand side of (\ref{algebraEDEF}) 
\be\label{CLOSURESTEPPPP}
 \begin{split}
  \delta_{\xi_1}^{(0)}F^M(\xi_2,{\cal E}) \ = \ &\,    
  -\tfrac{a-b}{4}(\tilde{\partial}^k\xi_2^l -\tilde{\partial}^l\xi_2^k)\,\partial^M(\partial_{k}\tilde{\xi}_{1l}
  -\partial_{l}\tilde{\xi}_{1k})\\
  &\,    
  -\tfrac{a-b}{4}(\tilde{\partial}^k\xi_2^l -\tilde{\partial}^l\xi_2^k)\,\partial^M\big[({ L}_{\xi_1}
  +\widetilde{ L}_{\tilde{\xi}_1}){\cal E}_{kl}\big] \\
   &\,    
  -\tfrac{a-b}{4}(\tilde{\partial}^k\xi_2^l -\tilde{\partial}^l\xi_2^k)\,\partial^M
  (-{\cal E}_{kp}(\tilde{\partial}^p\xi_1^q -\tilde{\partial}^q\xi_1^p){\cal E}_{ql})\;. 
 \end{split}
 \ee 
For the final term on the right-hand side of (\ref{algebraEDEF}) one finds 
 \be
 \begin{split}
  F^M([\xi_1,\xi_2]_c,{\cal E}) \ &= \ -\tfrac{a-b}{4}(\tilde{\partial}^k[\xi_1,\xi_2]_c^l 
  -\tilde{\partial}^l[\xi_1,\xi_2]_c^k)\partial^M{\cal E}_{kl}\;  \\
  \ &= \ -\tfrac{a-b}{4}[({ L}_{\xi_1}+\widetilde{L}_{\tilde{\xi}_1}) 
  (\tilde{\partial}^k\xi_2^l -\tilde{\partial}^l\xi_2^k)]\partial^M{\cal E}_{kl}\;, 
 \end{split}
 \ee
using the explicit expressions for the C-bracket and Lie derivatives. 
Finally, in order to compute the second term on the right-hand side of (\ref{algebraEDEF}), 
it is convenient to use that the free index on $F^M$ is carried by a derivative 
and that total derivative terms are trivial parameters that can be dropped in the gauge algebra. 
Indeed, by the relation (\ref{DCrelation}) 
we can then replace the C-bracket by the generalized Lie derivative ${\cal L}$ 
and compute 
 \be
 \begin{split}
  [\xi_2,F(\xi_1,{\cal E})]_c^M \ &= \ {\cal L}_{\xi_2}F(\xi_1,{\cal E})^M \ = \ 
  \xi_2^N\partial_NF(\xi_1,{\cal E})^M +\partial^M\xi_2^P F(\xi_1,{\cal E})_P \\
  \ &= \ ({ L}_{\xi_2}+\widetilde{L}_{\tilde{\xi}_2})F(\xi_1,{\cal E})^M
  +\partial^M\xi_2^P F(\xi_1,{\cal E})_P\;, 
 \end{split} 
 \ee 
 using the convention that ${L}$ and $\tilde{L}$ only acts on small latin indices, leaving 
 the index $M$ inert.  With the same convention we can rewrite the term in the second line of 
 (\ref{CLOSURESTEPPPP}) as 
  \be
  \begin{split}
    \partial^M\big[({L}_{\xi_1}+\widetilde{L}_{\tilde{\xi}_1}){\cal E}_{kl}\big] 
     \ = \ &\,({L}_{\xi_1}+\widetilde{L}_{\tilde{\xi}_1})\partial^M{\cal E}_{kl}
     +\partial^M\xi_1^P\partial_P{\cal E}_{kl} \\
     &\, +\partial^M\big(\partial_k\xi_1^p-\tilde{\partial}^p\tilde{\xi}_{1k}\big)\, {\cal E}_{pl}
     +\partial^M(\partial_l\xi_1^p-\tilde{\partial}^p\xi_{1l}){\cal E}_{kp}\;. 
  \end{split}
  \ee   
Combining all terms  in (\ref{algebraEDEF}) one may verify that the Lie derivative terms cancel 
as well as the term $\partial^M\xi F$, leaving 
 \be
   \begin{split} 
       \xi_{12}^{(1)} \ = \ &\, 
       -\tfrac{a-b}{4}(\tilde{\partial}^k\xi_2^l -\tilde{\partial}^l\xi_2^k)\,\partial^M(\partial_{k}\tilde{\xi}_{1l}
       -\partial_{l}\tilde{\xi}_{1k}) \\
       &\,-\tfrac{a-b}{4}(\tilde{\partial}^k\xi^l_2-\tilde{\partial}^l\xi_2^k)
       (\partial^M(\partial_k\xi_1^p-\tilde{\partial}^p\tilde{\xi}_{1k})\,{\cal E}_{pl}
       +\partial^M(\partial_l\xi_1^p-\tilde{\partial}^p\tilde{\xi}_{1l})\,{\cal E}_{kp})\\
       &\,    
  -\tfrac{a-b}{4}(\tilde{\partial}^k\xi_2^l -\tilde{\partial}^l\xi_2^k)\,\partial^M
  (-{\cal E}_{kp}(\tilde{\partial}^p\xi_1^q -\tilde{\partial}^q\xi_1^p){\cal E}_{ql})\;. 
   \end{split}
  \ee     
Combining with the terms of the compensating frame algebra (\ref{COmpFRameALgebra}) one finds 
significant simplifications, leaving for the total gauge algebra 
 \be\label{alpha'gaugealgebra}
 \begin{split}
  \xi_{12}^M \ = \  &\, \tfrac{a-b}{4}\,\partial_K\xi_2^L\,\partial^M\partial_L\xi_1^K\\
  &\,-\tfrac{a+b}{4}\Big\{({\cal E}_{kl}+{\cal E}_{lk})(\tilde{\partial}^k\xi_2^p-\tilde{\partial}^p\xi_2^k)\partial^M
  (\partial_p\xi_1^l-\tilde{\partial}^l\tilde{\xi}_{1p})\\[-0.5ex]
  &\qquad\;\;\;\;\;\;   -{\cal E}_{kp}\,\partial^M{\cal E}_{ql}\,(\tilde{\partial}^k\xi_2^l-\tilde{\partial}^l\xi_2^k)
  (\tilde{\partial}^p\xi_1^q-\tilde{\partial}^q\xi_1^p)\Big\} -(1\leftrightarrow 2) \;. 
 \end{split}
 \ee      
This form makes it manifest that for the HSZ case $a=-b$ the gauge algebra indeed reduces 
to the expected field-independent algebra structure in the first line.   
Moreover, only this part of the algebra is manifestly $O(D,D)$ invariant, with all indices 
transforming linearly and being properly contracted. 
In contrast, for the bosonic string case $a=b$ the algebra is not $O(D,D)$ covariant in the original 
sense, but only in a deformed sense, as we will discuss now.

\subsection{Remarks on $O(D,D)$ covariance} 

We close this section by making some general remarks on how $O(D,D)$ is realized in the formulation 
based only on the non-symmetric metric ${\cal E}_{ij}$. As discussed above, unless $a=-b$ and/or 
the pure gauge degrees of freedom are included, the familiar 
fractional-linear $O(D,D)$ transformation (\ref{nonlinODD}) of ${\cal E}_{ij}$ is no longer a symmetry to 
first order in $\alpha'$. Rather, the $O(D,D)$ action itself is $\alpha'$-deformed.  While these 
deformed transformations can always be obtained and understood from compensating gauge transformations 
in the frame formulation that makes all symmetries manifest, it is instructive to analyze some aspects 
of these transformations explicitly, which also corroborates independently the conclusions derived in this paper.

Let us begin by considering the infinitesimal $O(D,D)$ transformations with parameter 
$\tau^M{}_N$, satisfying  $\tau_{(MN)}=0$, 
  \be\label{infinitODD}
   \delta_{\tau} E_{A}{}^{M} \ = \ \xi_{\tau}^{N}\partial_{N}E_{A}{}^{M}  + \tau^{M}{}_{N} E_{A}{}^{N}\;, 
   \qquad
   \xi_{\tau}^{M} \ \equiv \ -\tau^{M}{}_{N} X^N\;. 
  \ee
We parameterize  the transformation matrix as   
  \be\label{TauMatrix}
   \tau^{M}{}_{N} \ = \   \begin{pmatrix}    \tau_i{}^{j} & \tau_{ij}  \\[0.5ex]
   \tau^{ij} &\tau^i{}_{j} \end{pmatrix} \ = \  \begin{pmatrix}    \alpha_i{}^{j} & \beta_{ij}  \\[0.5ex]
   \gamma^{ij} &- \alpha_j{}^{i}\end{pmatrix}\;, \qquad 
   \text{$\gamma$ and $\beta$ antisymmetric} \;, 
  \ee   
which is related to the $O(D,D)$ matrix as $h={\bf 1}+\tau$, and assume the 
gauge fixed form (\ref{gaugeFIXEDframe}) of the frame field. This, in turn, requires compensating 
gauge transformations in order to satisfy $0=\delta_{\tau}E_{a}{}^{i} = \delta_{\tau}E_{\bar{a}}{}^{i}$. 
A short computation yields for the compensating parameters 
\be\label{ODDcompensating}
\begin{split}
 \Lambda_{j}{}^{i} \ &= \ \alpha_{j}{}^{i} + {\cal E}_{jk}\,\gamma^{ik}- \alpha'\Sigma_{j}{}^{\bar{i}}\;, \\
 \bar{\Lambda}_{j}{}^{i} \ &=  \ \alpha_{j}{}^{i} - {\cal E}_{kj}\,\gamma^{ik}-\alpha' \Sigma_{\bar{j}}{}^{i}\;, 
\end{split} 
\ee  
where we identified flat and curved indices as usual. 
Including these compensating transformations  in  (\ref{infinitODD}) evaluated for, say, $E_{ai}$ one finds 
 \be
  \delta_{\tau}{\cal E}_{ij} \ = \ \xi_{\tau}^{N}\partial_{N}{\cal E}_{ij} 
  +\beta_{ij} +\alpha_{i}{}^{k}\, {\cal E}_{kj} + {\cal E}_{ik}\,\alpha_{j}{}^{k}
  -{\cal E}_{ik}\,\gamma^{kl}\,{\cal E}_{lj} -\alpha' \Sigma_{i\bar{j}}\;. 
 \ee 
To zeroth order in $\alpha'$ this result 
is equal to that obtained from the fractional-linear transformation (\ref{nonlinODD}) 
upon expanding to first order in the transformation parameters, which in matrix notation reads 
 \be
  {\cal E}'(X') \ = \ {\cal E}(X)+\beta + \alpha\, {\cal E}(X)+{\cal E}(X)\,\alpha^t 
  -{\cal E}(X)\, \gamma \, {\cal E}(X)\;. 
 \ee  
We infer that the antisymmetric $\beta$ encodes the `$b$-shifts' shifting the NS-NS 2-form
and $\alpha$ encodes the $\mathfrak{gl}(D)$ matrix rotating all $D$-dimensional indices. 
These symmetries are manifest in any diffeomorphism  and $b$-field gauge invariant theory. 
However, the full $O(D,D)$ also includes the symmetries parametrized by $\gamma^{ij}$, 
which act in a truly non-linear way on the physical fields, in this sense representing 
genuine T-duality transformations. 

We now turn to the $\alpha'$-deformed transformation, 
which is obtained by inserting the compensating parameters (\ref{ODDcompensating}) 
into $\Sigma_{i\bar{j}}$ as in (\ref{COnnSTEP}), (\ref{ODDCONNN}), 
 \be
  \delta_{\tau}^{(1)}{\cal E}_{ij} \ = \ \tfrac{a}{2}\,{\cal D}_{i}{\cal E}_{kp}\,\gamma^{pl}\,\Gamma_{\bar{j}l}{}^{k}
  -\tfrac{b}{2}\,\bar{\cal D}_{j}{\cal E}_{pk}\,\gamma^{lp}\,\Gamma_{i\bar{l}}{}^{\bar{k}}\;. 
 \ee 
We observe that only the $\gamma$ transformations are deformed.
This is as expected, because in a formulation based on ${\cal E}=g+b$ 
constant $b$-shifts  and global $GL(D)$ transformations 
can always be realized in the standard, undeformed sense. 
Since the $O(D,D)$ action is deformed, one may ask whether we still have a group action, in other words, 
whether these transformations still close by themselves. We can answer this question by recalling that 
the deformed transformations originated from compensating frame transformations. 
Since these frame transformations only close modulo a modification of the C-bracket of 
generalized diffeomorphisms it follows that also the deformed $O(D,D)$ transformations 
require generalized diffeomorphisms for closure. 
The effective parameter is again obtained by eliminating the gauge parameters in (\ref{xiDEFor}) by (\ref{ODDcompensating}),  
 \be
  \begin{split}
   \xi_{12}^M 
    \ &= \ 
    \tfrac{a}{4}\,(\alpha_{2i}{}^{j} + {\cal E}_{ik}\,\gamma_2^{jk})\,\partial^M(\alpha_{1j}{}^{i} 
    + {\cal E}_{jl}\,\gamma_1^{il})
   -\tfrac{b}{4}\, (\alpha_{2i}{}^{j} - {\cal E}_{ki}\,\gamma_2^{jk})\,\partial^M(\alpha_{1j}{}^{i} - {\cal E}_{lj}\,\gamma_1^{il})
      \\
   \ &= \ \tfrac{1}{4} \big(a\, {\cal E}_{ik}\,  \partial^M{\cal E}_{jl}
   -b\, {\cal E}_{ki}\, \partial^M{\cal E}_{lj} \big)  \gamma_2^{jk}\gamma_1^{il} \;, 
  \end{split}
 \ee  
ignoring in the last step total derivative terms, which is legal because they correspond to trivial parameters. 
Decomposing ${\cal E}_{ij}=g_{ij}+b_{ij}$ and using the symmetry of $g$ and the antisymmetry of $b$, 
the algebra reads  
 \be
   \xi_{12}^M \ = \ \tfrac{1}{4}\big((a-b)g_{ik}\,\partial^M g_{jl}+(a+b)g_{ik}\,\partial^M b_{jl}+(a+b)b_{ik}\,\partial^M g_{jl}
   +(a-b)b_{ik}\,\partial^M b_{jl}\big)\gamma_2^{jk}\gamma_1^{il}\;, 
  \ee 
where we recall that the antisymmetrization in $(1\leftrightarrow 2)$ is  implicit. 
Exchanging $i\leftrightarrow k$ and $j\leftrightarrow l$ leaves the terms proportional to $(a-b)$ invariant 
but changes the sign of the terms proportional to $(a+b)$, while $\gamma_2^{jk}\gamma_1^{il}$ 
changes sign under the antisymmetrization in $(1\leftrightarrow 2)$. 
Therefore, the terms proportional to $(a-b)$ drop out, and 
the gauge algebra reduces to 
 \be
  \xi_{12}^M \ = \ \tfrac{1}{2}\, (a+b)\, g_{ik}\, \partial^M b_{jl}\, \gamma_2^{jk}\gamma_1^{il}-(1\leftrightarrow 2)\;, 
  \ee 
where we restored the antisymmetrization in $(1\leftrightarrow 2)$ and used that by the antisymmetry under 
$i\leftrightarrow j$ and $k\leftrightarrow l$ the two remaining terms 
combine into one, up to a trivial parameter. 
We observe that the algebra is trivial for $a=-b$, confirming our above conclusion that in this case, and in this case only, 
the $O(D,D)$ remains undeformed in presence of $\alpha'$ corrections. 

Although for general parameters $a, b$ the $O(D,D)$ symmetry on the \textit{background independent} field 
${\cal E}_{ij}$ is deformed, we have seen that for the perturbative variables $h_{\,\nin{M}\bar{N}}$ 
the $O(D,D)$ symmetry is manifest in the original sense. The same holds for the string field theory variable 
$e_{ij}$, as both are related via background frame fields to the same frame-like variable $h_{a\bar{b}}$, 
  \be\label{hhhRElation}
   h_{\,\nin{M}\bar{N}} \ = \ 2\, \bar{E}_{M}{}^{a} \,  \bar{E}_{N}{}^{\bar{b}} \, h_{a\bar{b}}\;, \qquad
   h_{a\bar{b}} \ = \  \bar{E}_{a}{}^{i} \, \bar{E}_{\bar{b}}{}^{j}  \, e_{ij}\;. 
  \ee
It is instructive to pause for a moment and to explain why these two statements (deformed symmetries 
in the background independent formulation but undeformed symmetries in the perturbative 
formulation) are consistent. 
 We first note that 
 in the expansion (\ref{backgroundExp}) about flat space the background  transforms under $O(D,D)$  as 
 $\delta_{\tau} \bar{E}_{A}{}^{M}   =  \tau^{M}{}_{N} \bar{E}_{A}{}^{N}$, so that $h_{AB}$ is $O(D,D)$ invariant. 
 If we introduce the \textit{constant} 
 background $E_{ij}$ through a gauge fixed background frame field as above, however,  we need compensating 
 $GL(D)\times GL(D)$ transformations. The parameters take the same form as (\ref{ODDcompensating}), 
 but now depending only on the constant $E_{ij}$. Being constant, they drop out of  the 
 frame transformations at order $\alpha'$. Therefore, the $O(D,D)$ transformations of the fluctuations 
 are indeed undeformed.

 \medskip
 
We close this subsection by making some brief remarks about the covariance 
of the $\alpha'$ corrected gauge algebra (\ref{alpha'gaugealgebra}), which we recall here      
 \be\label{ALGEBRAAAAAAA}
 \begin{split}
  \xi_{12}^M \ = \  &\, \tfrac{a-b}{4}\,\partial_K\xi_2^L\,\partial^M\partial_L\xi_1^K\\
  &\,-\tfrac{a+b}{4}\Big\{({\cal E}_{kl}+{\cal E}_{lk})\,K_2^{kp}\, \partial^M K_{1p}{}^{l} 
  -{\cal E}_{kp}\,\partial^M{\cal E}_{ql}\,K_{2}^{kl}\, K_{1}^{pq}\Big\}  -(1\leftrightarrow 2) \;, 
 \end{split}
 \ee    
using the straightforward notation for matrices $K$.  Under a transformation $r\in GL(D)\subset O(D,D)$
all indices transform covariantly, e.g., for the derivatives $\partial_i'  =  r_i{}^j\,\partial_j$, 
$\tilde{\partial}^{i\prime} = (r^{-1})_j{}^i\, \tilde{\partial}^j$. As such, the algebra is manifestly convariant. 
However, already for constant $b$-shifts, which by our reasoning above should be realized in the undeformed sense, 
the check of covariance is a little subtle: 
We first note that the $O(D,D)$ transformation given by 
 \be
  h_{\beta} \ = \ \begin{pmatrix}    1 & \beta \\[0.5ex]
   0 & 1 \end{pmatrix} \;, \qquad \beta \ = \ -\beta^t\;, 
 \ee
transforms all relevant objects according to  
 \be
 \begin{split}
  \tilde{\xi}_i^{\prime} \ &= \ \tilde{\xi}_i + \beta_{ij}\xi^j \;, \qquad
  \xi^{i\prime} \ = \ \xi^i \;, \\
  \partial_i^{\prime} \ &= \ \partial_i + \beta_{ij}\tilde{\partial}^j\;, \qquad
  \tilde{\partial}^{i\prime} \ = \ \tilde{\partial}^i\;, \\
   {\cal E}' \ &= \ {\cal E}+\beta\;, 
 \end{split}
 \ee 
so that we find for the $K$ components: 
 \be
  \begin{split} 
   K^{\prime \, kl} \ = \ K^{kl} \;, \qquad 
   K^{\prime}{}_{k}{}^{l} \ = \ K_{k}{}^{l} + \beta_{kp}K^{pl}\;. 
  \end{split}
 \ee   
We then compute for the transformation of (\ref{ALGEBRAAAAAAA}) 
 \be
 \begin{split}
   \xi_{12}^{M\prime} \ &= \ \xi_{12}^{M} -\tfrac{a+b}{4}\big(
   ({\cal E}_{kl}+{\cal E}_{lk})\,K_2^{kp}\, \partial^M ( \beta_{pq}K_1^{ql})
  -\beta_{kp}\,\partial^M{\cal E}_{ql}\,K_{2}^{kl}\, K_{1}^{pq}\big)  \\
   \ &= \  \xi_{12}^{M} -\tfrac{a+b}{4}\big( g_{kl} \, \beta_{pq}\,\partial^M( K_2^{kp}\, K_1^{ql}) 
  -\beta_{pq}\,\partial^M g_{lk}\,K_{2}^{pk}\, K_{1}^{ql}\big) \\
   \ &= \ \xi_{12}^{M} -\tfrac{a+b}{4}\, \partial^M\big(g_{kl} \, \beta_{pq}\, K_2^{kp}\, K_1^{ql}\big)\;, 
  \end{split} 
 \ee   
where we used the implicit antisymmetry under $(1\leftrightarrow 2)$. 
We see that, curiously, the expression for the algebra is not invariant, but rather transforms into a  total 
derivative. This is a trivial parameter that does not transform fields, however, and so in this sense 
the algebra really is invariant. We also note that it is not possible to rewrite the algebra using trivial 
parameters so that the algebra expression is strictly invariant, as opposed to invariant up to trivial parameters. 
 
Finally, let us mention that the $\gamma$ transformations in  (\ref{TauMatrix})
are not expected to be a symmetry of (\ref{ALGEBRAAAAAAA}) for $a+b\neq 0$, for these 
transformations where already deformed at the infinitesimal level. 
However, by construction of this algebra, they must be a symmetry in some deformed sense, 
but here we will not attempt to identify these transformations explicitly.

\section{Background independence} 

In this section we confirm and elucidate the conclusion that there is no generalized metric formulation 
for $\alpha'$-deformed double field theory, unless $a=-b$, from the point of view of 
manifest background independence. Let us recall the expansion (\ref{fullHexpan})  of the generalized metric
around a constant background, 
 \be\label{secondHEXPNAD}
 \begin{split}
  {\cal H}_{MN} \ = \ \bar{\cal H}_{MN} &+ h_{\,\nin{M}\bar{N}}+ h_{\,\nin{N}\bar{M}}
  -\tfrac{1}{2} h^{\,\nin{K}}{}_{\bar{M}}\, h_{\,\nin{K}\bar{N}} + \tfrac{1}{2} h_{\,\nin{M}}{}^{\bar{K}}\,
  h_{\,\nin{N}\bar{K}}\\
  &-\tfrac{1}{8} h^{\,\nin{K}}{}_{\bar{M}}\, h_{\,\nin{K}}{}^{\bar{L}} \,h^{\,\nin{P}}{}_{\bar{L}} \,h_{\,\nin{P}\bar{N}}
  +\tfrac{1}{8} h_{\,\nin{M}}{}^{\bar{K}}\, h^{\,\nin{L}}{}_{\bar{K}}\, h_{\,\nin{L}}{}^{\bar{P}}\, h_{\,\nin{N}\bar{P}}
  \  + \  {\cal O}(h^6)\;, 
 \end{split}
 \ee 
and let us also recall that, projecting this equation with background projectors $P$ and $\bar{P}$ based on $\bar{\cal H}$, 
we have  
   \be\label{hthroughH}
  h_{\,\nin{M}\bar{N}} \ = \  P_{M}{}^{K} \,   \bar{P}_{N}{}^{L} \,  {\cal H}_{KL}\;. 
 \ee  
This follows because in the expansion (\ref{secondHEXPNAD}) all higher order terms in $h$ carry index projections 
of the same type, either unbarred-unbarred or barred-barred. 
This relation allows one to translate any action written in terms of $h_{\,\nin{M}\bar{N}}$ into an 
action in terms of the generalized metric ${\cal H}_{MN}$. In this sense there is always a generalized metric 
formulation, but the point is that the resulting theory still depends on the background through the 
projectors in (\ref{hthroughH}).\footnote{I would like to thank Ashoke Sen for discussions on this point.} 
It is not guaranteed that the background structures will drop out. 
Therefore, the more precise statement of the no-go result for a generalized metric formulation is that there 
is no \textit{background independent} generalized metric formulation, which we confirm in 
the remainder of this section.

Let us begin by recalling that the property of background independence has two different but closely related meanings: 
First, one may have \textit{manifest} background independence, which plainly means that the theory does not 
depend on a background. An example is Einstein gravity written in terms of the full metric tensor. 
Second, even if a theory is written with explicit background structures it may still be secretly background independent 
in that  any shift of the background can be absorbed into a shift of the fields, 
possibly up to field redefinitions.\footnote{String field theory is actually background independent 
in this sense \cite{Sen:1993mh}, see also \cite{Erler:2014eqa}.} 
An example is Einstein gravity expanded around a background solution. 
Another example is any double field theory based on a generalized metric but expanded 
as in (\ref{secondHEXPNAD}).
Given such a theory written in terms of background structures and 
fluctuations $h_{\,\nin{M}\bar{N}}$ we can test for background independence 
by performing a background shift given by the variation  
$\delta_{\chi}\bar{\cal H}_{MN}  =  -\chi_{\,\nin{M}\bar{N}}-\chi_{\,\nin{N}\bar{M}}$, with constant parameter 
${\chi}$, or equivalently for the projectors,  
 \be\label{projectorShift}
 \begin{split}
  \delta_{\chi}P_{MN} \ = 
  \ \tfrac{1}{2}(\chi_{\,\nin{M}\bar{N}}+\chi_{\,\nin{N}\bar{M}}) 
  \ = \  - \delta_{\chi}\bar{P}_{MN} \;, 
 \end{split} 
 \ee
and asking whether we can find an opposite shift of the fluctuation $h_{\,\nin{M}\bar{N}}$ so that 
the expansion  (\ref{secondHEXPNAD}) and hence the corresponding action are invariant. 
Since the fluctuation is constrained by 
$\bar{P}_{M}{}^{K}  h_{\,\nin{K}\bar{N}} =  {P}_{N}{}^{K} h_{\,\nin{M}\bar{K}}  =  0$, 
its variation needs to preserve this condition: 
 \be\label{preSERVE}
  0 \ = \ \ \delta_{\chi} \bar{P}_{M}{}^{K} \,  h_{\,\nin{K}\bar{N}}  +\bar{P}_{M}{}^{K} \delta_{\chi} h_{\,\nin{K}\bar{N}} \;, 
  \quad {\rm etc.} 
 \ee 
 It is straightforward to verify with  (\ref{secondHEXPNAD}) that the variation satisfying 
 this constraint and $\delta_{\chi}{\cal H}_{MN}  =  0$ is given by 
 \be\label{firstchishift}
 \begin{split}
  \delta_{\chi}h_{\,\nin{M}\bar{N}} \ = \  &\, \chi_{\,\nin{M}\bar{N}} +\tfrac{1}{2}\,\chi^{\,\nin{K}}{}_{\bar{M}}\, 
  h_{\,\nin{K}\bar{N}} -\tfrac{1}{2}\,h_{\,\nin{M}}{}^{\bar{K}}\, \chi_{\,\nin{N}\bar{K}}\\
  &\,-\tfrac{1}{4}\,\chi_{\,\nin{M}}{}^{\bar{K}}\,h^{\,\nin{L}}{}_{\bar{K}}\, h_{\,\nin{L}\bar{N}}
  -\tfrac{1}{4}\,\chi^{\,\nin{K}}{}_{\bar{N}}\, h_{\,\nin{M}}{}^{\bar{L}}\, h_{\,\nin{K}\bar{L}}+{\cal O}(h^3)\;. 
 \end{split} 
 \ee  
Note that the different index projections appearing here on the left-hand and right-hand sides 
are as needed in order to satisfy (\ref{preSERVE}). 
This result can also be computed from 
$\delta_{\chi}h_{\,\nin{M}\bar{N}} = \delta_{\chi}(P_{M}{}^{K} \bar{P}_{N}{}^{L}){\cal H}_{KL}$
by using (\ref{projectorShift}). 

We conclude that a manifestly background independent formulation in terms 
of a generalized metric ${\cal H}_{MN}$ exists  for a given theory written in terms 
$h_{\,\nin{M}\bar{N}}$ if and only if the theory is invariant, possibly up to 
$O(D,D)$ covariant field redefinitions of $h_{\,\nin{M}\bar{N}}$, under the variations (\ref{projectorShift}) 
and (\ref{firstchishift}) of background and fluctuation.
Starting from the frame formulation of sec.~4,  
we will use this criterion to prove that there is no manifestly background independent 
generalized metric formulation for the $\alpha'$-deformed double field theory unless $a=-b$.

We now return to the frame formalism and discuss background independence, following 
the discussion in \cite{Hohm:2015ugy}. As was proved there and can be easily verified, 
the background expansion ansatz (\ref{backgroundExp}) is invariant under 
 \be\label{FRAMebackgroundINV}
  \delta_{\Delta}h_{AB} \ = \ \Delta_{AB}+h_{A}{}^{C}\Delta_{BC}\;, \qquad
  \delta_{\Delta}\bar{E}_{A}{}^{M} \ = \ \Delta_{A}{}^{B}\bar{E}_{B}{}^{M}\;, 
 \ee 
with constant parameter $\Delta_{AB}$ satisfying $\Delta_{a\bar{b}}=-\Delta_{\bar{b}a}$. 
The background structures transform in agreement with (\ref{projectorShift}). To see this, 
we first note  
 \be
  \delta_{\Delta}\bar{E}_{M}{}^{a}  \ = \ -\bar{E}_{M}{}^{b}\Delta_{b}{}^{a} + \bar{E}_{M}{}^{\bar{b}}\Delta^{a}{}_{\bar{b}}\;, 
  \qquad
  \delta_{\Delta}\bar{E}_{N}{}^{\bar{b}} \ = \ -\bar{E}_{N}{}^{c}\Delta_{c}{}^{\bar{b}}-\bar{E}_{N}{}^{\bar{c}}
  \Delta_{\bar{c}}{}^{\bar{b}} \; . 
 \ee 
Using this and the identification 
 \be\label{chidelta}
  \chi_{\,\nin{M}\bar{N}} \ = \ 2\,\bar{E}_{M}{}^{a}\,\bar{E}_{N}{}^{\bar{b}} \, \Delta_{a\bar{b}}\;, 
 \ee
we compute 
 \be
  \delta_{\chi}P_M{}^{N} \ = \ \delta_{\chi}(\bar{E}_{M}{}^{a} \bar{E}_{a}{}^{N}) \ = \ 
   \tfrac{1}{2}\big(\chi_{\,\nin{M}}{}^{\bar{N}}+\chi^{\,\nin{N}}{}_{\bar{M}}\big)\;, \quad {\rm etc.} 
 \ee 
As in previous sections, we will impose the gauge fixing condition $h_{ab}=h_{\bar{a}\bar{b}}=0$, 
so that the physical fields are encoded in $h_{a\bar{b}}$. The compensating local frame transformations
needed to preserve this gauge choice under the background shifts (\ref{FRAMebackgroundINV}) are given by 
 \be\label{COMpLambda}
  \Lambda_{ab} \ = \ \Delta_{ab}+h_{a}{}^{\bar{c}}\Delta_{b\bar{c}}\;, \qquad
  \Lambda_{\bar{a}\bar{b}} \ = \ \Delta_{\bar{a}\bar{b}} + h^{c}{}_{\bar{a}}\Delta_{c\bar{b}}\;. 
 \ee  
Including these transformations in (\ref{FRAMebackgroundINV}) we find for the background 
transformation of $h_{a\bar{b}}$ 
 \be
  \delta_{\Delta} h_{a\bar{b}} \ = \ \Delta_{a\bar{b}} + h_{a}{}^{\bar{c}} \Delta_{\bar{b}\bar{c}}
  +\Delta_{a}{}^{c} h_{c\bar{b}} + h_{a\bar{c}}\, \Delta^{c\bar{c}}\, h_{c\bar{b}} - \alpha' \Sigma_{a\bar{b}}\;, 
 \ee 
where in $\Sigma$ we have to substitute for $\Lambda$ according to (\ref{COMpLambda}). 
 
Let us first establish the relation to (\ref{firstchishift})  to zeroth  order in $\alpha'$. 
This requires the identifications (\ref{hhdictionary}) and (\ref{chidelta}) in order 
to translate flat to curved indices.  
We then compute 
 \be
  \begin{split}
   \delta_{\chi} h_{\,\nin{M}\bar{N}} \ &= \ 2(\delta_{\Delta}\bar{E}_{M}{}^{a})\bar{E}_{N}{}^{\bar{b}} h_{a\bar{b}}
   +2\,\bar{E}_{M}{}^{a}(\delta_{\Delta}\bar{E}_{N}{}^{\bar{b}}) h_{a\bar{b}}
   +2\,\bar{E}_{M}{}^{a}\bar{E}_{N}{}^{\bar{b}}\delta_{\Delta}h_{a\bar{b}} \\
    \ &= \ \chi_{\,\nin{M}\bar{N}} + \tfrac{1}{2}\,\chi^{\,\nin{K}}{}_{\bar{M}} \,h_{\,\nin{K}\bar{N}}
   -\tfrac{1}{2}\, \chi_{\,\nin{N}}{}^{\bar{K}}\,h_{\,\nin{M}\bar{K}}
   +\tfrac{1}{4}\, h_{\,\nin{M}\bar{L}}\,\chi^{\,\nin{K}\bar{L}}\, h_{\,\nin{K}\bar{N}}\;, 
  \end{split}
 \ee
where in the second line various terms cancelled.  Note that this form, which is exact,      
does not yet agree with (\ref{firstchishift}). They differ in the terms quadratic in $h$ and, more importantly,  
while the above is exact, (\ref{firstchishift}) requires higher order terms. 
We have to find an $O(D,D)$ covariant field redefinition that relates both forms. 
Renaming  the transformation in (\ref{firstchishift}) 
as $\bar{\delta}_{\chi}$, it is easy to check that 
 \be
  \bar{\delta}_{\chi} h_{\,\nin{M}\bar{N}} \ = \ \delta_{\chi}h_{\,\nin{M}\bar{N}} 
  -\delta_{\chi}\big(\tfrac{1}{4} h_{\,\nin{M}\bar{L}} \, h^{\,\nin{K}\bar{L}} \, h_{\,\nin{K}\bar{N}}\big)
  +{\cal O}(h^3)\;, 
 \ee 
where to this order only the lowest-order variation in the second term enters. 
Thus, $\delta_{\chi}$ and $\bar{\delta}_{\chi}$ agree up to a field redefinition. 
Note that this 
field redefinition leads to higher order terms in the expansion of ${\cal H}_{MN}$ that have the 
projections $_{\,\nin{M}\bar{N}}$, which did not appear in the expansion scheme (\ref{secondHEXPNAD}) 
but are perfectly allowed.\footnote{Although such terms complicate the background expansion of ${\cal H}_{MN}$
they do simplify the gauge transformations since, as we have seen, for the variable $h_{\,\nin{M}\bar{N}}$
emerging from the frame formalism at most quadratic terms appear in the gauge transformations.} 

Let us now turn to the first $\alpha'$ correction. In this framework the background shifts 
in general receive $\alpha'$ corrections, because the compensating gauge parameters (\ref{COMpLambda}) 
need to be inserted into the deformed frame transformations (\ref{alpha'perturbativedeformed}). 
Using (\ref{ExplConne}) and the constancy of $\Delta$ we compute
 \be
 \begin{split}
  \delta_{\Delta}^{(1)}h_{a\bar{b}} \ &= \ -\tfrac{a}{2}\, {\cal D}_{a}\Lambda_{c}{}^d\,\omega_{\bar{b}d}{}^{c}
  -\tfrac{b}{2}\, {\cal D}_{\bar{b}}\Lambda_{\bar{c}}{}^{\bar{d}}\,\omega_{a\bar{d}}{}^{\bar{c}}\\
  \ &= \ -\tfrac{a}{2}(D_a-h_{a}{}^{\bar{c}}D_{\bar{c}})h_{c}{}^{\bar{d}}\, \Delta^{d}{}_{\bar{d}}
  (\Gamma_{\bar{b}d}{}^{c} - h^{c\bar{e}}\Gamma_{d\bar{b}\bar{e}} - h_{d}{}^{\bar{e}}D_{\bar{e}}h^{c}{}_{\bar{b}})\\
  &\quad\;\, -\tfrac{b}{2}\,(D_{\bar{b}}+h^{c}{}_{\bar{b}} D_{{c}})h^{d}{}_{\bar{c}}\,\Delta_d{}^{\bar{d}}
  (-\Gamma_{a\bar{d}}{}^{\bar{c}} - h^{e\bar{c}}\Gamma_{\bar{d}ae} -h^{e}{}_{\bar{d}} D_{e}h_{a}{}^{\bar{c}})\\
  \ &= \ -\tfrac{a}{2}\,D_ah_{c\bar{d}}\,\Delta^{d\bar{d}}\,\Gamma_{\bar{b}d}{}^{c}
  +\tfrac{b}{2}\, D_{\bar{b}}h_{d\bar{c}}\,\Delta^{d\bar{d}}\,\Gamma_{a\bar{d}}{}^{\bar{c}}\\
  &\quad \;\,+\tfrac{a}{2}\,D_ah_{c\bar{d}}\,\Delta^{d\bar{d}}\,h^{c\bar{e}}\,\Gamma_{d\bar{b}\bar{e}}
  +\tfrac{a}{2}\, D_ah_{c\bar{d}}\,\Delta^{d\bar{d}}\, h_{d}{}^{\bar{e}}D_{\bar{e}}h^{c}{}_{\bar{b}}
  +\tfrac{a}{2}\,h_{a}{}^{\bar{c}} D_{\bar{c}}h_{c\bar{d}}\,\Delta^{d\bar{d}}\,\Gamma_{\bar{b}d}{}^{c} \\
  &\quad \;\, + \tfrac{b}{2}\,D_{\bar{b}}h_{d\bar{c}}\,\Delta^{d\bar{d}}\, h^{e\bar{c}}\, \Gamma_{\bar{d}ae} 
  +\tfrac{b}{2}\,D_{\bar{b}}h_{d\bar{c}}\,\Delta^{d\bar{d}}\, h^{e}{}_{\bar{d}} D_{e}h_{a}{}^{\bar{c}}
  +\tfrac{b}{2}\, h^{c}{}_{\bar{b}} D_{c}h_{d\bar{c}}\,\Delta^{d\bar{d}}\,\Gamma_{a\bar{d}}{}^{\bar{c}}
  +\cdots 
 \end{split} 
 \ee 
where we omitted terms of ${\cal O}(h^4)$. This result 
can now be translated into doubled $O(D,D)$ indices by contracting with the 
background frame fields and using  the relations (\ref{hhdictionary}) and (\ref{chidelta}).  
Moreover, we use that the shift of the background frame does not receive an $\alpha'$ correction, 
because the background is assumed to be constant. We find 
 \be\label{firstBackgroundTrans}
  \begin{split}
   &\delta_{\chi}^{(1)}h_{\,\nin{M}\bar{N}} \ = \  -\tfrac{a}{8}\,\partial_{\,\nin{M}}h_{\,\nin{K}\bar{L}}\,\chi^{\,\nin{P}\bar{L}}\,
   \Gamma_{\bar{N}\,\nin{P}}{}^{\,\nin{K}}
   +\tfrac{b}{8}\, \partial_{\bar{N}}h_{\,\nin{K}\bar{L}}\,\chi^{\,\nin{K}\bar{P}}\,\Gamma_{\,\nin{M}\bar{P}}{}^{\bar{L}}
   \\
  &\; \;\,+\tfrac{a}{16}\,\partial_{\,\nin{M}} h_{\,\nin{K}\bar{L}}\,\chi^{\,\nin{P}\bar{L}}\,
  h^{\,\nin{K}\bar{Q}}\,\Gamma_{\,\nin{P}\bar{N}\bar{Q}}
  +\tfrac{a}{16}\, \partial_{\,\nin{M}} h_{\,\nin{K}\bar{L}}\,\chi^{\,\nin{P}\bar{L}}\, h_{\,\nin{P}}{}^{\bar{Q}}
  \partial_{\bar{Q}}h^{\,\nin{K}}{}_{\bar{N}}
  +\tfrac{a}{16}\,h_{\,\nin{M}}{}^{\bar{K}} \partial_{\bar{K}}h_{\,\nin{L}\bar{P}}\,\chi^{\,\nin{Q}\bar{P}}\,
  \Gamma_{\bar{N}\,\nin{Q}}{}^{\,\nin{L}} \\
  &\; \;\, + \tfrac{b}{16}\,\partial_{\bar{N}}h_{\,\nin{K}\bar{L}}\,\chi^{\,\nin{K}\bar{P}}\, h^{\,\nin{Q}\bar{L}}\,
   \Gamma_{\bar{P}\,\nin{M}\,\nin{Q}} 
  +\tfrac{b}{16}\,\partial_{\bar{N}}h_{\,\nin{K}\bar{L}}\,\chi^{\,\nin{K}\bar{P}}\, h^{\,\nin{Q}}{}_{\bar{P}} \partial_{\,\nin{Q}} 
  h_{\,\nin{M}}{}^{\bar{L}}
  +\tfrac{b}{16}\, h^{\,\nin{K}}{}_{\bar{N}} \partial_{\,\nin{K}} h_{\,\nin{P}\bar{Q}}\,\chi^{\,\nin{P}\bar{L}}\,
  \Gamma_{\,\nin{M}\bar{L}}{}^{\bar{Q}}\,.
    \end{split}
 \ee  

We observe that the background shifts receive $\alpha'$-deformations 
that would not be present in a pure generalized 
metric formulation with the expansion scheme (\ref{secondHEXPNAD}). 
Next we have to analyze to what extent these deformations are non-trivial
in the sense that they are not removable by field redefinitions. 
The terms in the first line are quadratic in $h$ and can be trivially removed, 
because we can replace the constant $\chi$ by a 
bare $h$, using that $\partial h$ is invariant under $\delta_{\chi}$. 
Thus, defining 
 \be
  h'_{\,\nin{M}\bar{N}} \ = \ h_{\,\nin{M}\bar{N}} + \Delta^{[3]}_{\,\nin{M}\bar{N}}
  + \Delta^{[4]}_{\,\nin{M}\bar{N}}\;,
 \ee
with the number in brackets denoting the order of $h$, and setting 
 \be\label{Delta3}
  \Delta_{\,\nin{M}\bar{N}}^{[3]} \ = \ \tfrac{a}{8}\,\partial_{\,\nin{M}}h_{\,\nin{K}\bar{L}}\,h^{\,\nin{P}\bar{L}}
  \,\Gamma_{\bar{N}\,\nin{P}}{}^{\,\nin{K}}
  -\tfrac{b}{8}\,\partial_{\bar{N}}h_{\,\nin{K}\bar{L}}\,h^{\,\nin{K}\bar{P}}\,\Gamma_{\,\nin{M}\bar{P}}{}^{\bar{L}}\;, 
 \ee   
produces terms that precisely cancel the terms in the first line of (\ref{firstBackgroundTrans}). 
Note that for this we do not have to take into account the transformation of the background projectors, 
because these only yield terms cubic in $h$. 
This result implies that there is no obstacle for background independence and a generalized metric 
formulation to second order in perturbation theory (to second order in fields), in agreement with our 
findings in sec.~2 and 3.\footnote{For instance, the field redefinition (\ref{Delta3}) induces in the gauge transformations 
terms of the form $\partial h\partial h \partial\xi$, relevant for second order perturbation theory.}

Next, we inspect the terms cubic in $h$, for which we will see that the higher derivative terms cannot be 
removed for general $a,b$. 
For this we have to compute 
 \be
 \begin{split}
  \delta_{\chi}h'_{\,\nin{M}\bar{N}} \ &= \ \chi_{\,\nin{M}\bar{N}} 
  + \tfrac{1}{2}\,\chi^{\,\nin{K}}{}_{\bar{M}} \,h_{\,\nin{K}\bar{N}}
   -\tfrac{1}{2}\, \chi_{\,\nin{N}}{}^{\bar{K}}\,h_{\,\nin{M}\bar{K}} +\delta_{\chi}\Delta_{\,\nin{M}\bar{N}}
   +\delta_{\chi}^{(1)}h_{\,\nin{M}\bar{N}}\\
  \ &= \ \chi_{\,\nin{M}\bar{N}} 
  + \tfrac{1}{2}\,\chi^{\,\nin{K}}{}_{\bar{M}} \,(h'_{\,\nin{K}\bar{N}}-\Delta^{[3]}_{\,\nin{K}\bar{N}})  
   -\tfrac{1}{2}\, \chi_{\,\nin{N}}{}^{\bar{K}}\,(h'_{\,\nin{M}\bar{K}}-\Delta^{[3]}_{\,\nin{M}\bar{K}}) 
   +\delta_{\chi}^{(1)}h_{\,\nin{M}\bar{N}}\\
   &\quad\;  +\delta_{\chi}^{[0]}\Delta^{[4]}_{\,\nin{M}\bar{N}}+\delta_{\chi}^{[1]}\Delta^{[3]}_{\,\nin{M}\bar{N}}\;, 
 \end{split}
\ee   
where again the numbers in square bracket denote the number of fields. Moreover, the $\delta_{\chi}^{[1]}$
in the last term includes the transformation of the background strcutures, which also leaves the number 
of $h$ unchanged. The challenge is now to simplify the $\chi$-variations by determining an appropriate
$\Delta^{[4]}$. After a somewhat tedious but straightforward computation one finds that with 
 \be
  \begin{split}
   \Delta^{[4]}_{\,\nin{M}\bar{N}} \ = \ &\, \tfrac{a}{16} \, \partial_{\,\nin{M}}h_{\,\nin{K}\bar{L}}\,h^{\,\nin{P}\bar{L}}
   \, h^{\,\nin{K}\bar{Q}}\,\partial_{\bar{Q}}h_{\,\nin{P}\bar{N}}
   +\tfrac{b}{16}\,\partial_{\bar{N}}h_{\,\nin{K}\bar{L}}\,h^{\,\nin{K}\bar{P}}\,h^{\,\nin{Q}\bar{L}}\,
   \partial_{\,\nin{Q}}h_{\,\nin{M}\bar{P}}\\
   &\, -\tfrac{a}{16}\,\partial_{\,\nin{M}}h_{\,\nin{K}\bar{L}}\,h^{\,\nin{P}\bar{L}}\, h_{\,\nin{P}}{}^{\bar{Q}}\,
   \partial_{\bar{Q}}h^{\,\nin{K}}{}_{\bar{N}}
   - \tfrac{b}{16}\,\partial_{\bar{N}}h_{\,\nin{K}\bar{L}}\,h^{\,\nin{K}\bar{P}}\, h^{\,\nin{Q}}{}_{\bar{P}}\,
   \partial_{\,\nin{Q}}h_{\,\nin{M}}{}^{\bar{L}}\\
   &\, -\tfrac{a}{16}\,h_{\,\nin{M}}{}^{\bar{K}}\partial_{\bar{K}} h_{\,\nin{L}\bar{P}}\,h^{\,\nin{Q}\bar{P}}\,
   \Gamma_{\bar{N}\,\nin{Q}}{}^{\,\nin{L}}
   -\tfrac{b}{16}\, h^{\,\nin{K}}{}_{\bar{N}}\,\partial_{\,\nin{K}} h_{\,\nin{P}\bar{Q}}\, h^{\,\nin{P}\bar{L}}\,
   \Gamma_{\,\nin{M}\bar{L}}{}^{\bar{Q}}\;, 
  \end{split}
 \ee   
the final $\chi$ transformation simplifies significantly. Dropping the prime, 
the background transformations read 
 \be\label{APLphaPRimeDEformed}
 \begin{split}
  \delta_{\chi}h_{\,\nin{M}\bar{N}} \ = \ &\, 
    \chi_{\,\nin{M}\bar{N}} + \tfrac{1}{2}\,\chi^{\,\nin{K}}{}_{\bar{M}} \,h_{\,\nin{K}\bar{N}}
   -\tfrac{1}{2}\, \chi_{\,\nin{N}}{}^{\bar{K}}\,h_{\,\nin{M}\bar{K}}
   +\tfrac{1}{4}\, h_{\,\nin{M}\bar{L}}\,\chi^{\,\nin{K}\bar{L}}\, h_{\,\nin{K}\bar{N}}\\
   &\, +\tfrac{1}{16}(a+b)\,\partial_{\,\nin{M}}h_{\,\nin{K}\bar{L}}\,\chi^{\,\nin{P}\bar{L}}\, h^{\,\nin{K}\bar{Q}}\,
   \partial_{\bar{N}}h_{\,\nin{P}\bar{Q}}\;. 
  \end{split}
 \ee  
The term in the last line vanishes for the case $a=-b$ of the HSZ theory, but otherwise cannot be removed, 
as is easy to see. 
Indeed, although one could change that term by performing another redefinition by adding a term 
proportional to 
$\partial_{\,\nin{M}}h_{\,\nin{K}\bar{L}}\,h^{\,\nin{P}\bar{L}}\, h^{\,\nin{K}\bar{Q}}\,
\partial_{\bar{N}}h_{\,\nin{P}\bar{Q}}$, this would just change the order of $\chi$ and $h$, 
but not remove that term. 

The above result proves that there is no background independent formulation based on the 
generalized metric for general $a, b$. Indeed, if one had such a formulation, without loss of generality 
one could expand around a background as in (\ref{secondHEXPNAD}), for which the 
background shifts receive no $\alpha'$ corrections. Let us also emphasize that this result is not in conflict with the fact that in sec.~5 we found a manifestly background independent formulation in terms ${\cal E}_{ij}$, because this field transformed 
in a non-standard way under $O(D,D)$, while the above background shifts, redefinitions, etc., 
are manifestly $O(D,D)$ covariant in the original sense.

\section{Discussion and Outlook} 

In this paper I proved that,  
upon including $\alpha'$ corrections, there is no formulation of classical string theory in terms of the usual fields 
that makes both background independence and duality invariance 
manifest. 
More precisely, the universal background independent massless fields --- metric, $b$-field and dilaton --- cannot be 
organized into $O(D,D)$ covariant objects (the generalized metric ${\cal H}_{MN}$ or 
the non-symmetric metric ${\cal E}_{ij}$ transforming in the fractional linear form) 
to first order in $\alpha'$. A manifestly background independent and $O(D,D)$ invariant formulation
can, however, be obtained by use of a frame formalism, thereby introducing 
pure gauge degrees of freedom under (generalized and $\alpha'$-deformed) frame transformations. 
As reviewed in the introduction, the mere existence of such DFT formulations guarantees 
compatibility with the dualities of string theory.

In order to elucidate this point, it is instructive to recall how these dualities 
are conventionally tested  in the spacetime theory. Here one performs a dimensional reduction 
on a torus $T^d$, taking the fields to be independent of the $d$ internal coordinates $y^m$,  
and then shows that the resulting theory admits an $O(d,d)$ invariance. 
While to lowest order in $\alpha'$, at the two-derivative level, this can be achieved 
in a relatively straightforward manner \cite{Maharana:1992my}, to higher order in $\alpha'$ 
it becomes exceedingly tedious
and so far has only been worked out in `cosmological reductions' to one dimension 
\cite{Meissner:1996sa,Hohm:2015doa} or subject to further truncations \cite{Godazgar:2013bja}. 
Part of the complication is that the expression for the generalized metric in terms of $g$ and $b$ 
receives $\alpha'$ corrections. Equivalently, the 
$O(d,d)$ action on the original fields is $\alpha'$ deformed, as suggested by the DFT results. 
Nevertheless, in dimensional reduction it \textit{is} possible 
to write the action in terms of a field ${\cal H}$ satisfying ${\cal H}\eta{\cal H}=\eta$. 
This is consistent with the DFT results presented here, because the $O(d,d)$ only acts 
on the $d$ coordinates on which the fields no longer depend, so that ${\cal O}(\alpha')$
terms as e.g.~in (\ref{APLphaPRimeDEformed}) drop out.

The results of $\alpha'$-deformed DFT 
imply that the combination of three basic principles (duality invariance, gauge invariance
and background independence) leads to a powerful framework that strongly constrains 
the $\alpha'$ corrections. Indeed, in the formulation of 
\cite{Marques:2015vua} it was shown that the $\alpha'$-deformed generalized Green-Schwarz transformations 
uniquely determine the action to first order in $\alpha'$, up to the two free parameters $a, b$ that are 
needed in order to encode bosonic string theory, heterotic string theory and the `interpolating' 
HSZ theory \cite{Hohm:2013jaa}. While in this paper we did not attempt to construct a gauge invariant 
action, it is clear that to first order in $\alpha'$ 
this could be done upon augmenting the construction of \cite{Marques:2015vua}, which is based on a 
doubled local Lorentz symmetry, by the 
gauge degrees of freedom  corresponding to the enlarged $GL(D)\times GL(D)$ symmetry. 
More importantly, it remains to develop a full geometry for these symmetries, 
with suitably generalized connections or, perhaps, even more novel concepts, which would 
allow one to define a manifestly gauge invariant action. 

An important aspect of the problem of finding a proper geometric framework for the $\alpha'$-deformed 
frame symmetries is to extend the closure condition beyond first order in $\alpha'$. 
Almost certainly this requires a further deformation of the gauge transformations and/or the 
gauge algebra. Ideally, one would like to have gauge transformations that close exactly
and hence be able to define an action that is exactly gauge and duality invariant, 
as is the case for the HSZ theory \cite{Hohm:2013jaa}. 
One may speculate that this is possible upon further enlarging the gauge structure and introducing `higher' 
gauge modes, but so far no concrete proposal has presented itself. 
If an exact formulation exists, most likely it would include 
an infinite number of higher-derivative $\alpha'$ corrections. This does not imply, however, that 
such a theory includes all $\alpha'$ corrections of string theory, for it could be expected that there are 
different invariants starting at a higher number of derivatives. Correspondingly, there 
may be new free parameters, analogous to the parameters $a, b$ encountered to first 
order in $\alpha'$. It thus remains to be seen to what extent the $\alpha'$-deformed gauge symmetries 
determine the higher derivative corrections of string theory uniquely, but 
the results obtained so far show that they are strongly constrained by the symmetry principles of DFT. 
Moreover, for type II strings and M-theory one has to invoke 
U-dualities, as realized in the formulation of \cite{Hohm:2013pua}, which will further constrain 
the higher-derivative corrections.

Let me finally comment on the possible significance of these results for the full string theory. 
So far we have considered classical string theory for the massless fields, which 
is known to realize the continuous $O(d,d;\mathbb{R})$ symmetry. I uncovered 
an obstacle that arises when trying to make both this duality invariance and background 
independence manifest before compactification.  
We have seen that the duality transformations of the standard fields $g$ and $b$ 
are no longer given by the expected group action, but rather these transformations receive non-trivial 
$\alpha'$ corrections. In the full string theory on toroidal backgrounds these fields 
depend on the doubled coordinates subject to the weak (level-matching) constraint 
that allows for a simultaneous non-trivial dependence on $x$ and $\tilde{x}$. 
The periodicity conditions on $x$ and $\tilde{x}$ following from  the torus topology 
then imply that the duality group is the discrete  $O(d,d;\mathbb{Z})$. 
Recent results by Sen suggest that the massless fields together with their massive Kaluza-Klein and winding modes
provide a consistent subsector of string theory \cite{Sen:2016qap}, and if so the `strongly constrained' double field theory investigated in this paper can in turn be viewed 
as a subsector hereof. 
Therefore, the $\alpha'$-deformations of the $O(d,d;\mathbb{R})$ 
symmetry identified here must be consistent with the $O(d,d;\mathbb{Z})$ symmetry of the 
full string theory. This may seem puzzling since discrete symmetries cannot receive 
small deformations, but since one is now restricting to toroidal backgrounds 
one may at best demand a restricted property of background independence
so that it is no longer clear in which form the obstacles discussed here may 
manifest themselves.  
Indeed, we have seen that for the fluctuation fields around such backgrounds the $O(d,d)$
symmetry remains undeformed and hence there is no obstacle for restricting to the discrete 
subgroup in  a \textit{background dependent} formulation. 
It has been speculated since the 1980s, however, that a doubling of coordinates in string theory 
may be important and fundamental more generally 
(see e.g.~\cite{Atick:1988si,Duff:1989tf,Tseytlin:1990nb,Tseytlin:1990va,Hull:2006va}). 
If such a proposal will turn out to be correct in one form or another, the issues identified in this paper 
will play a role  for any background independent formulation of the full string theory, 
which certainly will require very novel geometric concepts.

\section*{Acknowledgments}

%\vskip .7cm

I would like to thank Diego Marques and Warren Siegel for discussions.  I am especially 
indebted to  Ashoke Sen and Barton Zwiebach for early collaboration on this problem 
and numerous helpful discussions. 
This work was supported by a DFG Heisenberg Fellowship 
of the German Science Foundation (DFG).

\end{document}